\documentclass[11pt,a4paper,doublespace,oneside,dvips]{book}

\usepackage{./styles/setspace}		
\usepackage{./styles/fancyhdr}		
\usepackage{graphicx}				
\usepackage{./styles/subfigure}		
\usepackage{url}					
\usepackage{ifthen}					
\usepackage{./styles/geometry}		
\usepackage{./styles/lgrind}
\input{./styles/epsf}
\usepackage{flafter}		
\usepackage{amssymb} 
 \usepackage{multirow}

\makeatletter
	\if@twoside
	\fi
\makeatother

\makeatletter%
	\def\cleardoublepage{%
		\clearpage%
		\if@twoside%
			\ifodd%
				\c@page%
			\else%
				\hbox{}%
				\thispagestyle{empty}%
				\newpage%
				\if@twocolumn%
					\hbox{}%
					\newpage%
				\fi%
			\fi%
		\fi%
	}%
\makeatother				

\fancypagestyle{plain}{%
	\fancyhf{} 
	\fancyfoot[CO,CE]{\thepage} 
	\renewcommand{\headrulewidth}{0pt} 
	\renewcommand{\footrulewidth}{0pt} 
}

\usepackage{algorithm}
\usepackage{algorithmic}
\usepackage{amsmath}
\usepackage{cite}
\usepackage{multirow}
\usepackage{titlesec}

\graphicspath{{ch\thechapter/images/}}

\begin{document}

\bibliographystyle{IEEEtran}


\frontmatter

\renewcommand{\chaptermark}[1]{%
\markboth{#1}{#1}}

\begin{titlepage}

	\vspace*{\fill}

	\begin{center}
		\huge
		\textbf{Accurate and Robust Localization Techniques for Wireless Sensor Networks}
	\end{center}

	\vfill

	\begin{center}
		\LARGE Mohamed AlHajri \\
		\LARGE Abdulrahman Goian \\
		\LARGE Muna Darweesh \\
		\LARGE Rashid AlMemari 
	\end{center}

	\vfill

	\begin{center}
		\Large Advisors: Prof. Raed Shubair, Dr. Luis Weruaga, Dr. Ahmed AlTunaiji
	\end{center}

	\vfill

	\begin{center}
		\Large 2015 
	\end{center}

	\vfill

%
%

\end{titlepage}

\chapter{Abstract}
\begin{singlespace}
The concept of wireless sensor network revolves around a group of sensor nodes that utilize the radio signals in order to communicate among each other. These nodes are typically made, of sensors, a memory, a multi-controller, a transceiver, and a power source to supply the energy to these components. There are many factors that restrict the design of wireless sensor network such as the size, cost, and functionality. The field of wireless sensor network, witnessed a remarkable revolution special after the environmental awakening in the last two decades. Its importance emanates from its capability to monitor physical and environmental conditions (such as sound, temperature, and pressure) with minimal power consumption. The principle of passing data cooperatively though a network to a main location played a vital role in the success of the methodology of WSN.
 
This report explores the concept of wireless sensor network and how it is been made viable through the convergence of wireless communications and micro-electro-mechanical systems (MEMS) technology together digital electronics. The report addresses some of the factors that have to be considered when choosing the localization algorithm. It is very important to choose properly since the localization process may involves intensive computational load, based on many different criteria, as well as analysis method. The report also views the advantages and disadvantages of localization techniques. Nevertheless, it investigates the challenges associated with the Wireless Sensor Networks. The report categorizes the algorithms, depending on where the computational effort is carried out, into centralized and distributed algorithms. With minimal computational complexity and signaling overhead, the project aims to develop algorithms that can accurately localize sensor nodes in real-time with low computational requirements, and robustly adapt to channel and network dynamics. The report focuses on three areas in particular: the first is the Received Signal Strength indicator technique, Direction of Arrival technique, and the integration of two algorithms, RSS and DOA, in order to build a hybrid, more robust algorithms.
 
In the Received Signal Strength (RSS), the unknown node location is estimated using trilateration. This report examines the performance of different estimators such as Least Square, Weighted Least Square, and Huber robustness in order to obtain the most robust performance. In the direction of arrival (DOA) method, the estimation is carried out using Multiple Signal Classification (MUSIC), Root-MUSIC, and Estimation of Signal Parameters Via Rotational Invariance Technique (ESPRIT) algorithms. We investigate multiple signal scenarios utilizing various antenna geometries, which includes uniform linear array (ULA) and uniform circular array (UCA). Specific attention is given for multipath scenarios in which signals become spatially correlated (or coherent). This required the use of pre-processing techniques, which include phase mode excitation (PME), spatial smoothing (SS), and Toeplitz. 
 
Further improvements of existing localization techniques are demonstrated through the use of a hybrid approach in which various combinations of RSS and DOA are explored, simulated, and analyzed. This has led to two major contributions: the first contribution is a combined RSS/DOA method, based on UCA, which has the tolerance of detecting both uncorrelated and coherent signals simultaneously. The second major contribution is a combined Root-MUSIC/Toepltiz method, based on UCA, which is outperforms other techniques in terms of increased number of detected signals and reduced computationally load.
\vfill

\end{singlespace}

\addcontentsline{toc}{chapter}{Contents}
\tableofcontents

\chapter{List of Acronyms}

\begin{singlespace}

\begin{list}{}
{
\setlength{\labelwidth}{2.5cm}
\setlength{\labelsep}{0.5cm}
\setlength{\leftmargin}{3cm}
}

\item[\textbf{WSN}] \textbf{W}ireless \textbf{S}ensor \textbf{N}etwork
\item[\textbf{GPS}] \textbf{G}lobal \textbf{P}ositioning \textbf{S}ystem
\item[\textbf{AWGN}] \textbf{A}dditive \textbf{W}hite \textbf{G}aussian \textbf{N}oise
\item[\textbf{TOA}] \textbf{T}ime \textbf{o}f \textbf{A}rrival
\item[\textbf{TDoA}] \textbf{T}ime \textbf{D}ifference \textbf{o}f \textbf{A}rrival
\item[\textbf{RSS}] \textbf{R}eceived \textbf{S}ignal \textbf{S}trength
\item[\textbf{DOA}] \textbf{D}irection \textbf{o}f \textbf{A}rrival
\item[\textbf{SDP}] \textbf{S}emi-\textbf{D}efinite \textbf{P}rogramming
\item[\textbf{MDS}] \textbf{M}ultidimensional \textbf{S}caling
\item[\textbf{LS}] \textbf{L}east \textbf{S}quare 
\item[\textbf{WLS}] \textbf{W}eighted \textbf{L}east \textbf{S}quare
\item[\textbf{ULA}] \textbf{U}niform \textbf{L}inear \textbf{A}rray
\item[\textbf{UCA}] \textbf{U}niform \textbf{C}ircular \textbf{A}rray
\item[\textbf{MUSIC}] \textbf{MU}ltiple \textbf{SI}gnal \textbf{C}lassification
\item[\textbf{ESPRIT}] \textbf{E}stimation \textbf{S}ignal \textbf{P}arameters via \textbf{R}otational \textbf{I}nvariance \textbf{T}echniques
\item[\textbf{PME}] \textbf{P}hase \textbf{M}ode \textbf{E}xcitation
\item[\textbf{CCA}] \textbf{C}ontinuous \textbf{C}ircular \textbf{A}rray
\item[\textbf{SS}] \textbf{S}patial \textbf{S}moothing
\item[\textbf{FSS}] \textbf{F}orward \textbf{S}patial \textbf{S}moothing
\item[\textbf{FBSS}] \textbf{F}orward/\textbf{B}ackward \textbf{S}patial \textbf{S}moothing
\item[\textbf{VULA}] \textbf{V}irtual \textbf{U}niform \textbf{L}inear \textbf{A}rray
\item[\textbf{GUI}] \textbf{G}raphical \textbf{U}ser \textbf{I}nterface

\end{list}

\end{singlespace}

\chapter{List of Symbols}

\begin{singlespace}

\begin{list}{}
{
\setlength{\labelwidth}{3.5cm}
\setlength{\labelsep}{0.5cm}
\setlength{\leftmargin}{3cm}
}

\item[$P_r$] Received Power
\item[$P_t$] Transmitted Power
\item[$G_t$] Transmitter Gain
\item[$G_r$] Receiver Gain
\item[$D_i$] Distance
\item[\textbf{A}] System Coordinated Matrix
\item[$\boldsymbol{\mathrm{p}}_s$] Unknown node
\item[\textbf{b}] Distance Vector
\item[$N$] Number of Array Elements
\item[$M$] Number of Received Signals
\item[$s_m$] Narrowband Signal
\item[$\theta_m$] Azimuth Angle
\item[$\theta_e$] Elevation Angle
\item[$\boldsymbol{\mathrm{A}}_s$] Array Steering Matrix
\item[$\phi_m$] Phase Shift between the Elements of the Sensor Array
\item[$\theta_n$] Angular Location of each Elements of the Sensor Array 
\item[$\zeta=\dfrac{2\pi r}{\lambda}$] UCA delay Parameter
\item[$w$] Beamforming Weights Vector
\item[$f$] Beampattern of an Array
\item[$f^c_p$] Normalized far-field Pattern of the p$^{th}$ phase mode in CCA
\item[$f^s_p$] Normalized far-field Pattern of the p$^{th}$ phase mode in UCA
\item[$h$] Highest Excited Mode
\item[$\boldsymbol{\mathrm{a}}{_sv}$] Steering Vector of VULA
\item[$\boldsymbol{\mathrm{T}}_v$] Transformation Matrix based on PME to mat UCA into VULA
\item[$\boldsymbol{\mathrm{x}}_v$] VULA Output Vector
\item[$\boldsymbol{\mathrm{R}}$] Covariance Matrix
\item[$\boldsymbol{\mathrm{R}}_{s}$] Signal Covariance Matrix
\item[$\boldsymbol{\mathrm{R}}_c$] Coherent Covariance Matrix
\item[$\boldsymbol{\mathrm{R}}_T$] Toeplitz Covariance Matrix constructed from $R_c$
\item[$\sigma^2_n$] Noise Variance
\item[$\boldsymbol{\mathrm{V}}_n$] Noise Eigenvector
\item[$\boldsymbol{\mathrm{V}}_s$] Signal Eigenvector
\item[P] MUSIC Spatial Spectrum
\item[$Q_{Root-MUSIC}$] Root-MUSIC Polynomial
\item[z] Root value of the Root-MUSIC Polynomial
\item[$\boldsymbol{\mathrm{T}}_w$] Noise-prewhitened Transformation Matrix
\item[$Q_{UCA-Root-MUSIC}$] UCA-Root-MUSIC Polynomial
\item[$\Delta x$] Displacement of ESPRIT Subarrays
\item[$\Phi$] Diagonal Matrix Containing the phase shift between ESPRIT Subarrays
\item[\textbf{T}] Non-singular Matrix to relate Steering Vectors of ESPRIT Subarrays
\item[$(.)^H$] Hermitian Operation
\item[$(.)^T$] Tranpose Operation of a Matrix 
\item[$(.)^*$] Complex Conjugate of a Matrix

\end{list}

\end{singlespace}

\addcontentsline{toc}{chapter}{List of Figures}
\listoffigures

\clearpage      
\addcontentsline{toc}{chapter}{List of Tables}
\listoftables
\chapter{Acknowledgements}

\noindent We would like to thank our three academic supervisors: Prof. Raed Shubair, Dr. Luis Weruaga, and Dr. Ahmed AlTunaji, for their technical help, valuable advice, and constant support.
 
Our main project supervisor Prof. Raed Shubair has been a tremendous mentor and an exceptional motivator.  He kept pushing us beyond the limits of a normal senior design project expectation. His timeless help and relentless guidance were essential for achieving the new results and contributions of this project.  Prof. Raed Shubair is the one who moved our work to the international academic frontier when he pushed us and taught us how to write eloquent technical papers which subsequently got accepted for presentation at two prominent conferences namely IEEE ICTRC’2015 in Abu Dhabi, UAE and IEEE APS/URSI’2015 in Vancouver, Canada.  Furthermore, Prof. Raed Shubair guidance and coaching were essential for us in winning the 1st place in the IEEE ICCSPA’15 Poster Competition.
 
Our co-supervisor Dr. Luis Weruaga has been instrumental in providing us with deep insights into statistical estimation theory and techniques.  We would like to thank him for his patience whenever we had any doubts on the theoretical aspects of the algorithms. 
 
Our co-supervisor Dr. Ahmed AlTunaiji helped us significantly at the early stages of the project.  He engaged with us in numerous discussions which made us grasp very well the fundamental concepts of wireless sensor network localization.  He coached us on the use of the professional technical writing software Latex, and helped us in developing various building blocks within our Matlab code.
 
This project would not have been completed without the generous support from Khalifa University which provided us with a distinctive educational experience. Special thanks are extended to the senior design project coordinator Dr. Kahtan Mezher, as well as to our examiners Dr. Shihab Jimaa  and Dr. Nazar Ali.
 
Finally, we would like to thank our own families, and especially our parents each.  Their endless love and countless sacrifices were the drive for all the successes and distinctions we achieved in this project.	

\mainmatter

\renewcommand{\chaptermark}[1]{%
\markboth{\chaptername\ \thechapter.\ #1}{}}

\chapter{Introduction}
\label{Introduction}

\section{Goals and Objectives}
\label{Goals and Objectives}

\subsection{Goals}
\begin{enumerate}
\item Survey on recent advances in localization for WSNs.
\item Implementation of selected algorithms in MATLAB. 
\item Proposed changes to selected algorithm for an improved performance.
\item Testing, verification, and validation of results.
\end{enumerate}
\subsection{Objectives}
\begin{enumerate}
\item Learn about recent advances in localization in WSNs.
\item Understand different practical applications and limitations for localization. 
\item Learn about simulation programs such as MATLAB.
\item Simulate various RSS and DOA techniques.
\item Enhance WSNs localization accuracy and robustness using RSS and DOA technique.
\item Investigate different pre-processing schemes to de-correlate coherent signals in DOA model.
\item Design a hybrid system by combining RSS with DOA model.
\item Improve the total performance by providing different modification for the algorithm.
\item Build a MATLAB-based GUI to facilitate the simulation of the developed localization system. 
\end{enumerate}

\section{Motivation}
\label{Motivation}
WSNs refer to networks involving spastically dispersed sensor nodes that communicate between each other through wireless channels.  Generally, the senor nodes have limited capability in terms of low power consumption, radio transceiver and small-size memory. Due to the continuous decrease of both cost and size of senor nodes, it becomes efficient to utilize such nodes to construct a large-scale WSN.  A WSN can be used for area monitoring like in military observation where the senor nodes are spread over a wide area to detect any possible intrusion from enemy. In such a scenario, the spatial information obtained from sensors nodes will be interpreted wrongly unless we know the respective locations of these nodes. For this reason, the localization process becomes an important aspect of a WSN and, over the recent years, researchers’ attention is focused on how we can improve the accuracy of the localization process.  Applications of WSN include healthcare monitoring, water quality monitoring and forest fire detection where these applications are unattainable unless we are capable of routing the information in the network \cite{kulkarni_computational_2011}.

Specifically, localization is defined as the ability to assign physical coordinates to unknown nodes in a network. The basic idea behind localization is that some nodes, so-called anchor nodes, in the network must have known coordinates. The anchor nodes transmit their coordinates to unknown nodes to assist them in localizing themselves. The assignment of coordinates to anchor nodes can be done through either hard coding or equipping them with Global Positioning System (GPS). The former method is impractical as it prevent the random disperse of nodes in an area and does not take in to account that a sensor can change its location with time as in military area where wind can move a sensor. The latter method is expensive and is inaccurate in indoor areas where GPS signals become inaccurate due to obstacles like trees and building. Besides, GPS demands significant power which may cause battery failure for nodes that does not rely on external power supply. Localization is carried out using various localization techniques such as trilateration and triangulation which utilized the neighbour anchor nodes’ locations to determine the locations of unknown nodes \cite{dargie_fundamentals_2010}. 

\section{Applications}
\label{Applications}
Recently, Wireless sensor networks have gathered the world's attention because of their diverse applications in military, environment, healthcare applications, habitat monitoring and industry \cite{kulaib_efficient_2014}. In these different applications, the need for the location information is essential.  The process of collecting information involves placing many different types of sensors on the sensor node such as mechanical, biological and thermal sensors to evaluate the environment’s properties \cite{yick_wireless_2008}. These sensors should be placed accurately to get the correct information to be sensed \cite{kulaib_efficient_2014}. 

\subsection{Health Systems}
Health systems based on WSNs are meant to provide accurate information about emergency situations and send records about the desired data. This is useful especially for elderly as sensors can send information about their health condition and thus monitor their activities, which would help them in their daily life \cite{puccinelli_wireless_2005}. Caregiver’s Assistant and CareNet Display project is an example to monitor the elderly. In this project, sensors are placed on each home object, thus, a record of the elderly’s movements and actions is recorded \cite{kulaib_efficient_2014}. 
For the patients in the hospitals, sensors give continuous information about their body, which help doctors to cure the illness in its early stage and prevent serious health issues. For instance, CodeBlue project has many sensors that sense the blood pressure and heart rate \cite{kulaib_efficient_2014}. Its configuration is shown in Figure 1.1. Also, this project, hopefully, will help to cure patients with different diseases \cite{puccinelli_wireless_2005}.  Moreover, doctors will be able to locate their patients and their colleagues in the hospital by using body sensors network. In this way, patients will be under full control and other doctors will be able to deliver the required assistance quicker \cite{kulaib_efficient_2014}. 
These health systems are not limited to patients only; infants, also, have special systems. For example, Sleep Safe is designed to monitor the sleeping position for the infant and alert the parent if their baby sleeps on his stomach. This system has two sensors, one is placed on the baby's clothing to detect the baby's position with respect to gravity and the other one is connected to computer. After collecting the required data, the computer processes these data upon standards that are set by the user and, consequently, the baby’s position will be known.   Moreover, the Baby Glove is a system that will measure the baby’s temperature, hydration and pulse rate. A sensor is attached to the swaddling wrap to send these data to a computer. This computer processes this data and determines if there is anything, which is abnormal and alerts the parent \cite{yick_wireless_2008}.   

\begin{figure} []
        \centering
        \includegraphics[width=0.6\textwidth,clip]{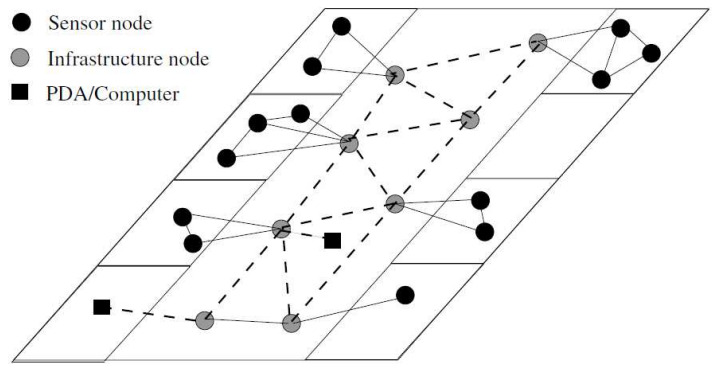}
        \caption[CodeBlue project configuration]{CodeBlue project configuration}
		 \label{Fig:CodeBlue}
\end{figure}
\subsection{Environmental Monitoring}
WSNs have an important role in the environmental side. It can be used to monitor environmental properties such as temperature and pressure. Many applications are in monitoring plants and animals, fire detection and natural changes such as volcanic eruption and floods are based on WSNs \cite{kulaib_efficient_2014}.     
Rare species of sea birds can be monitored by inserting sensors in their nets. These sensors can measure the humidity, temperature, and the pressure. Thus, the researchers will know the specific levels that affect these birds. This is the aim of Great Duck Island Project. Animals such as Zebras can be tracked. Their moves and behaviors are monitored to study the effects of human interference and how the Zebras interact with other species \cite{kulaib_efficient_2014,puccinelli_wireless_2005}.   Hence, around the zebra neck, a sensor node is placed \cite{puccinelli_wireless_2005}.  This node consists from several components GPS unit, long and short range radio transmitters, microcontroller, solar array and battery \cite{kulaib_efficient_2014}.  This is the main purpose of ZebraNet project \cite{kulaib_efficient_2014,puccinelli_wireless_2005}.  For monitoring the plants, combinations of sensors are used to sense the humidity, temperature, rainfall and sunlight. Also, all these nodes are equipped with camera to take images. After that these images are inserted in the rare plant species, that to be monitored, and in the surrounding area where these species does not exist. All these data will help the researchers in PODs project to understand the conditions that affect these rare plants species \cite{kulaib_efficient_2014}.  
In detecting forests fire, the nodes contain environmental sensor that can measure temperature, light intensity, barometric pressure and humidity. This will give indications about the possibility of having fire in the real time. Thus, this will save people's life and their properties \cite{kulaib_efficient_2014}.  
For volcanic eruptions, WSNs plays an important role because it does not require risking people’s life. The sensors in this application should provide reliable data at the event time and with high data rates \cite{yick_wireless_2008}. Combinations of microphone and seismometer sensors that collect seismic and acoustic data are used to monitor the volcano \cite{kulaib_efficient_2014}. An example of these sensors is shown in Figure 1.2.
Floods can be predicated using WSNs. This will help to save people and reduce losses. Sensors that can measure the water flow, rainfall and air temperature is used give information about the environment \cite{kulaib_efficient_2014}.   
\begin{figure} []
        \centering
        \includegraphics[width=0.6\textwidth,clip]{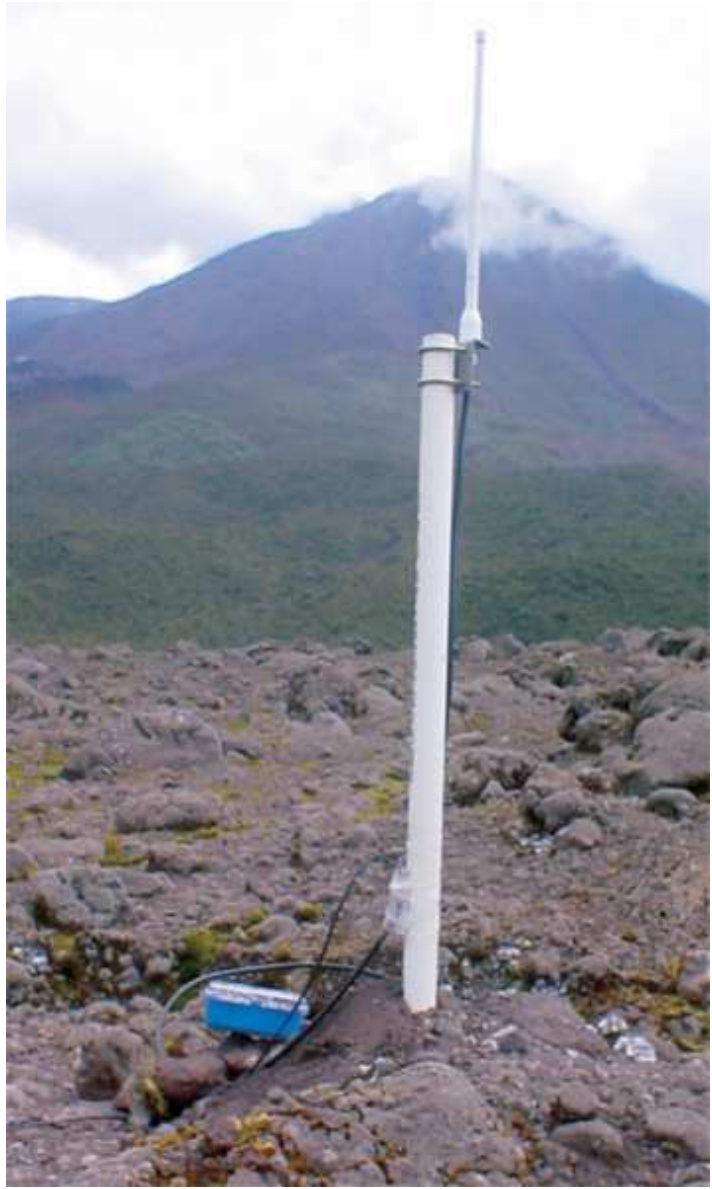}
        \caption[Sensors for detecting volcanic eruptions]{Sensors for detecting volcanic eruptions}
		 \label{Fig:Volcano}
\end{figure}
\subsection{Home Applications}
WSNs in home applications have systems that can spot human presence, control the air conditioning and the lights without the need for the human intervention. Thus, it is essential in these applications to have small size sensor node, so it can be placed easily on home objects \cite{kulaib_efficient_2014}.   
Non-intrusive Autonomous Water Monitoring System (NAWMS) is a project that aims to identify the wastage in water usage for each source alone.  The benefit is that each pipe in the home will be monitored properly and with minimum cost. The project works on the proportionality between the vibration of the pipe and the water flow. This vibration is determined by inserting accelerometers to the water pipes. Figure 1.3 shows the configuration for NAWMS. The sensor nodes are connected to the main water meter to compute the information about the water flow and send it to every node. The main purpose of these sensors is to measure the vibration for each pipe. Then, this vibration is routed to the computation node, so the sensor will be calibrated and the wastage will be specified. The computation node uses an optimized method to calibrate the sensors, in the sense that the addition of the water flows in every non-calibrated sensor should be equal to what in the water meter \cite{kulaib_efficient_2014}.   
Also, Smart building is one of the applications for WSNs. Inside this building, sensors and actuators are placed. These configurations control, monitor, and improve the living conditions and allow the reduction in the energy consumption, which can be done by controlling the airflow and temperature \cite{puccinelli_wireless_2005}.   
\begin{figure} []
        \centering
        \includegraphics[width=0.6\textwidth,clip]{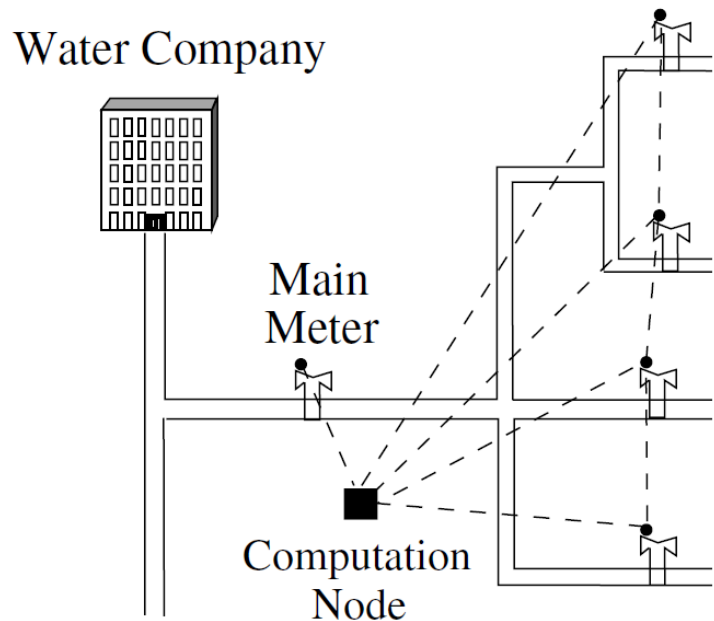}
        \caption[The configuration for NAWMS]{The configuration for NAWMS}
		 \label{Fig:NAWMS}
\end{figure}
\subsection{Infrastructure Health Monitoring}
WSNs are used to predict the health of different infrastructure such as bridge. Sensors measure the ambient structural vibrations without affecting the bridge’s operation. This monitors the bridge and, hence, the appropriate procedures are taken when any problem occurs \cite{kulaib_efficient_2014}.
\subsection{Intelligent Transportation}
In this application, a traffic monitoring system provides better safety for road users and prevents accidents. The sensors gather data about the vehicle’s direction and speed, moreover, the number of the vehicles that are in the road. This information helps to provide safety warnings on the roads \cite{kulaib_efficient_2014}. 

\subsection{Search and Rescue operation}
Search and rescue operation is one of many jobs that put the person life on the line.  WSN offers improvement to the Search and rescue operation that will reduce the risk faced by rescuers. WSN can help in finding the victim’s location and the area of the catastrophes \cite{kulaib_efficient_2014}. 

\subsection{Smart Campus}
An intelligent Campus is called iCampus that improves and transforms the end-to-end life cycle of the knowledge of ecosystem which can be done by using the WSNs. The intelligence can appear in two paths either in the sensor nodes itself or in the cooperative information network system or both of them.  This system does not only support the students but it provides with new teaching style and expands the management capabilities for the administration.  There are many examples for the iCampus, such as monitoring the students by the mean of sensory smart card. This card allows the student to use different services available in the university. Also, it helps the management to study their behaviors and actions \cite{kulaib_efficient_2014}. 

\subsection{Military applications} 
Since WSNs localize, identify and track the location of the desired target. Thus, it provides the soldiers or military troop with information about the enemy orientation and location. This is the main concept of the VigilNet which is surveillance system. This is used to tack the objects in the enemy's territories. Figure 1 shows the network for this system. Moreover, another application is the counter sniper system which can detect the shooters location.  This system has two different architectures. The first one is Boomerang system which is a group of microphones that detect the fire by processing the audio signal from the microphones. This problem in this system is that the lack of capability in the case of multi-path in sound detection. To solve that, a distributed network composed of acoustic sensors is used, shown in Figure 2. This works on the concept of the centralized concept, discussed later, which uses group of sensor to get information about the time and location to identify the bullet \cite{kulaib_efficient_2014}.  PinPrt is an ad hop acoustic sensor network that locates the sniper. This network senses the acoustic shock wave that results from the sound of gunfire. The arrival times at sensor nodes will estimate the snipers location \cite{puccinelli_wireless_2005}. 

Also, the minefield is monitored using a self-organizing sensor network in which there is a peer to peer communication between anti-tank mines, so react to any attacks and changes the place of the mines. This impedes the enemy movement \cite{puccinelli_wireless_2005}.  Large sensor network can be used to record any attempt from invaders to pass the country boarders illegally. These sensors must have camera to record any motion \cite{kulaib_efficient_2014}.  

\section{Challenges}
\label{Challenges}
Despite the variety of applications that are based on WSNs, researchers encounter many issues and challenges such as physical layer measurement errors, computation constraints, low-end sensor nodes, network shape, node mobility, and lack of GPS data.

There are different techniques used in WSNs to identify the position of the unknown node based on the signal strength and signal direction, which can be converted to distance and orientation measurements, respectively. These measurements have some errors due to multipath and shadowing, as shown in Figure 1.4. For example, if the signal is transmitted from the anchor A to the node B directly, the node B will be able to compute its location by estimating distance between A and B and this will be A, ̂ as shown in Figure 1.4 (a).  However, the real path between the transmitter and the receiver will not be clear. Many problems arise across this path such as multipath and shadowing \cite{kulaib_efficient_2014}.  In multipath, multiple copies of the transmitted signal from variety of paths combines in either constructive or destructive manner which affect the received signal. The receiver must specify the “first-arriving” peak. This is implemented by measuring the time that the cross-correlation first crosses the threshold \cite{alsindi_cooperative_2008}. In the shadowing, the signal is affected by different obstacles from the surrounding environment such as walls, buildings and trees. The signal diffracts along different paths between the transmitter and receiver \cite{akyildiz_wireless_2010}.  Therefore, in the case of multipath and shadowing, node B will not be able to get the accurate estimation with respect to node A, this case is shown in Figure 1.4 (b).  This difference will result in high errors in the estimated position \cite{akyildiz_wireless_2010}.  Hence, it is required to apply different techniques to compute the distance measurements and reduce the errors \cite{alsindi_cooperative_2008}.   Other attempt is to use averaging techniques to correct the errors, but it is not desired, so it is better to not take into account these measurements with errors \cite{kulaib_efficient_2014}.
The computation constrains are the source limitations. Nodes have restricted amount of power, small memory and small computations processors. Thus, most of the ranging techniques may not be done accurately. An attempt In order to solve this problem is to use centralized system because it can do large computation. On the other hand, these systems increase the communication overhead. Alternative option is to use simple and distributed algorithm. This algorithm shows reduction in the energy power and processing \cite{kulaib_efficient_2014}. 
\begin{figure} []
        \centering
        \includegraphics[width=0.6\textwidth,clip]{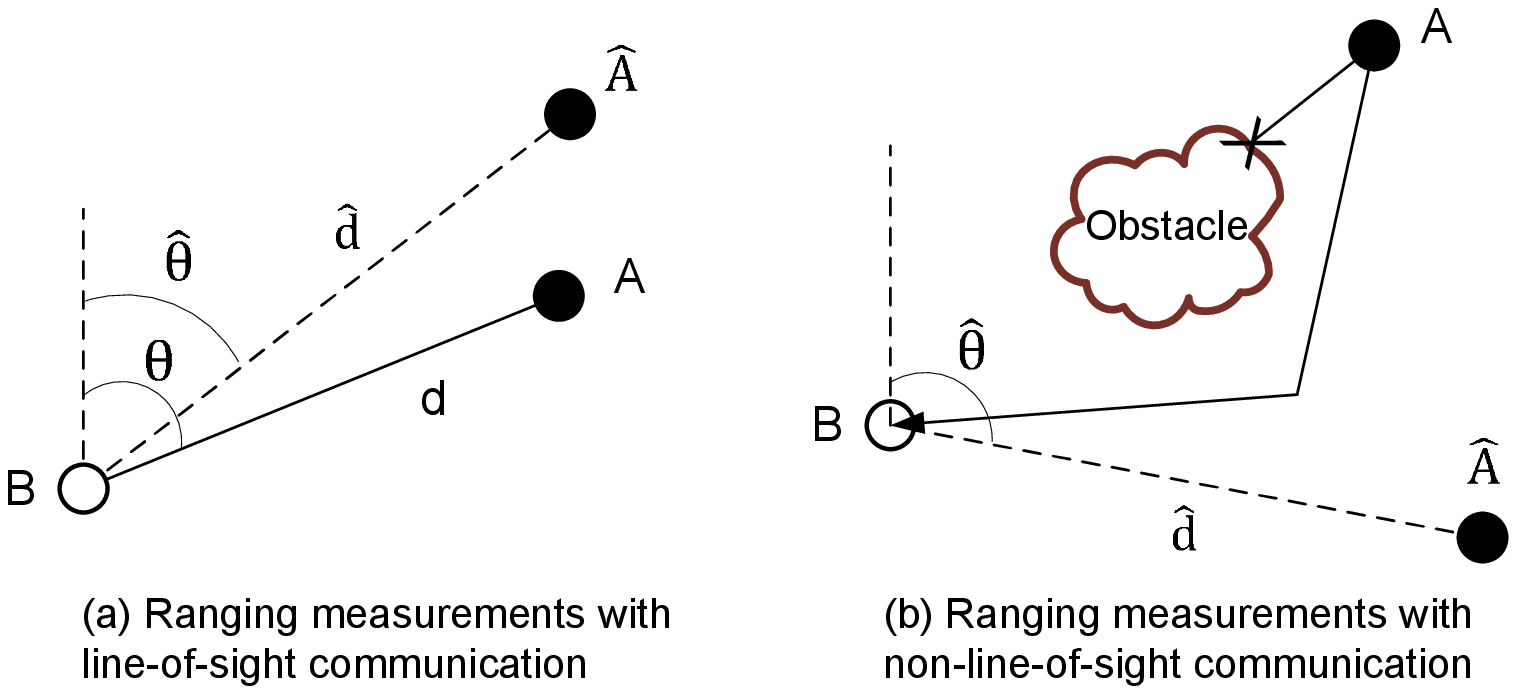}
        \caption[The effect of non-line-of-sight and multi-path on ranging technique]{The effect of non-line-of-sight and multi-path on ranging techniques}
		 \label{Fig:NLOS}
\end{figure} 
To get the location information of the node the GPS is used.  Actually, the GPS is expensive and has high energy consumption. Thus, it is used in the anchor nodes only \cite{kulaib_efficient_2014}.  
 
Moreover, the network shape plays an important role in the localization process.  For example, if the node is at the border of the network or not in the convex body, there will be few measurements; thus, the position will not be estimated correctly \cite{kulaib_efficient_2014}.   

Node mobility is one issue that should be considered in the localization process. The first scenario is when the anchor nodes are moving and the sensor nodes are in fixed position. This is an example in military in which the anchor nodes are placed to soldiers who scan the place and sensor nodes are static and distributed in the battlefield. This will improve the location accuracy of the sensor nodes. The second scenario is when the anchor nodes are in fixed positions while the sensor nodes are moving. For example, the anchor nodes are at the two sides of the river and sensor nodes are in the river. The last scenario is when both types of nodes are moving such as in the wind \cite{kulaib_efficient_2014}.  
  
The final challenge is for the wireless sensor nodes components is the quality of sensor node components. In the sensor nodes, low-end components or hardware measurement devices are placed because they reduce the system cost. These components allow error in distance or orientation estimation, which is added to the error produced by the channel. To solve this issue, different measurements can be taken from various neighbors in different time slots to get good localization accuracy.  In addition, these low-end components may introduce temporary or permanent node failure \cite{kulaib_efficient_2014}.     

\section{Achievements and Publications}
\label{Achievements and Publications}

\subsection{Achievements}
\begin{itemize}
\item Participation in RTA innovation week
\item Winning the first place in the IEEE International Conference on Communications, Signal Processing, and their Applications (ICCSPA'15) student poster competition
\item Participation in the IEEE UAE student day 
\end{itemize}
\subsection{Publications}
\begin{itemize}
\item M. I. AlHajri, R. M. Shubair, L. Weruaga, A. R. Kulaib, A. Goian, M. Darweesh, R. AlMemari,Hybrid Method for Enhanced Detection of Coherent Signals using Circular Antenna Arrays, IEEE International Symposium on Antennas and Propagation (APS'15), 2015.  
\item M. I. AlHajri, A. Goian, M. Darweesh, R. AlMemari, R. M. Shubair, L. Weruaga, A. R. Kulaib, Hybrid RSS-DOA Technique for Enhance WSN Localization in a Correlated Environment, IEEE International Conference on Information and Communication Technology Research, 2015.
\end{itemize}
\section{Overview of the Report}
\label{Overview of the Report}
The rest of report is arranged as follows. A generalized survey about the localization techniques is explained in Chapter 2. Also, Chapter 2 discusses the localization algorithms which is divided into centralized and distributed algorithms. Chapter 3 tackles in depth the formulation of RSS method and its simulation results. Chapter 4 studies the formulation of Direction of Arrival (DOA) method as theory and simulation of Uniform Linear Array (ULA) and Uniform Circular Array (UCA). In addition, Chapter 4 studies the formulation of Direction of Arrival method (DOA) in case of uncorrelated signals while introduces pre-processing techniques to deal with correlated signals. In addition, Chapter 4 compares the performance of different DOA algorithms based on environmental and array-related parameters. Chapter 5 explains in depth the formulation of different hybrid techniques and its simulation results. Chapter 6 describes the MATLAB-based GUI that facilitate the simulation of the developed localization system. Lastly, Chapter 7 concludes our report with summary of work done and the expected future work to be accomplished.   
\chapter{Survey of WSN Localization}
\label{Survey of WSN Localization}
In this chapter, several localization discovery and ranging techniques are discussed for estimating the unknown node \cite{kulaib_overview_2011}. In addition, a general classification for localization is presented.
\section{Location Discovery Techniques}
\label{Location Discovery Techniques}
There are three basic localization discovery technique; Triangulation, Trilateration, and Multilateration. 
\subsection{Triangulation}
\begin{figure} []
        \centering
        \includegraphics[width=0.6\textwidth,clip]{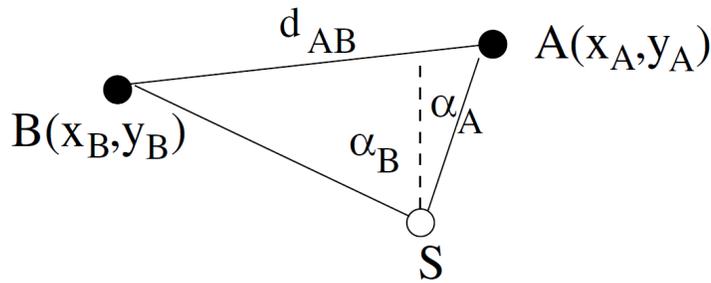}
        \caption[Triangulation]{Triangulation}
		 \label{Fig:Triangulation}
\end{figure}

Triangulation exploits the geometric properties of triangles in order to approximate the location of the unknown node. Precisely, triangulation requires at least two angles from the reference nodes and their respective locations to compute the unknown node position \cite{gentile_geolocation_2012}. Figure \ref{Fig:Triangulation} demonstrates the operation of triangulation where the unknown node will lie on the intersection of the three bearing lines (the line from the anchors to the node). 
Two anchor with locations $\textbf{p}_i=[x_i  , y_i]^T$  and measured angles $\boldsymbol{\theta}_i(x)=\left[ \boldsymbol{\theta}_1(x)~\boldsymbol{\theta}_2(x)\right]^T$  (expressed relative to a fixed vertical line as in the Figure 2.1), then, the location of the unknown node $\textbf{p}_s=[x_s  ,y_s]^T$   is determined from the relationships: 
\begin{equation}%
\tan\left( \boldsymbol{\theta}_i(x)\right)=\frac{y_i-y_s}{x_i-x_s}
\label{eqn}
\end{equation}%
From equation 2.1, we got a system of two equations and two unknowns which can be solved to obtain the unknown node position.  
In reality, the measured angles will not be the exact angles due to the present of noise. Thus, the equation of the real angles (\textbf{B}) is given as:
\begin{equation}%
\textbf{B}=\theta(x_s)+\boldsymbol{\delta}(\theta)
\label{eqn}
\end{equation}%
where $\boldsymbol{\delta}(\theta$) is a noise modeled as Additive Gaussian White Noise (AGWN). The error in the measured angles will prevent the bearing lines from intersecting into a single point leading to over-determined system. Solving such system will require the use of statistic approach like Maximum-likelihood or least-square method. 
\subsection{Trilateration}
\begin{figure} []
        \centering
        \includegraphics[width=0.6\textwidth,clip]{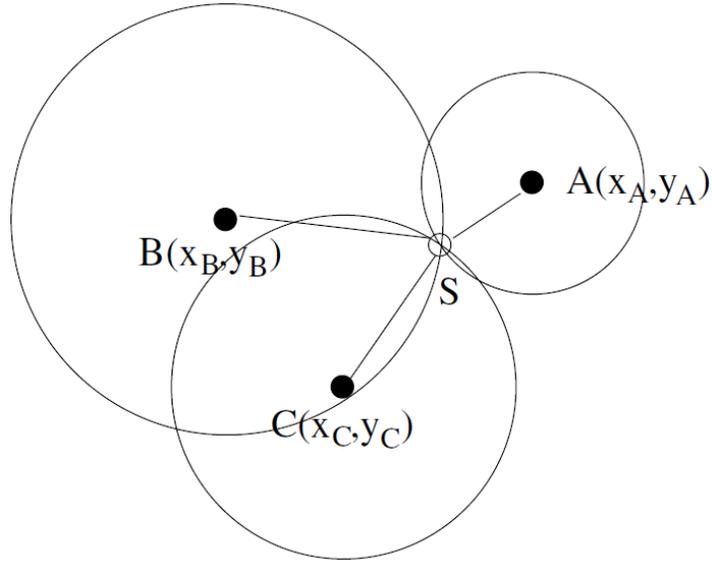}
        \caption[Trilateration]{Trilateration}
		 \label{Fig:Trilateration}
\end{figure}

Unlike triangulation, trilateration locates a node position based on the distances measured from three reference points with known locations. In two-dimensional space, trilateration requires at least three-distance measurement from non-collinear anchors to obtain a single location (the intersection of three circles) as showing in Figure \ref{Fig:Trilateration}. However, in reality 3-D space, we will demand at least four non-coplanar anchors to solve for the unknown location \cite{patwari_locating_2005}.
Assuming that we have $N$ anchor nodes with locations $\textbf{p}_i =[x_i  ,y_i]^T$  ($i=1$ $\cdot$ $N$) and distances between the unknown node’s position $\textbf{p}_s=[x_s  ,y_s]^T$ and the anchors are $\boldsymbol{d_i}$  ($i=1$ $\cdot$ $N$). Using these information, we can express the anchor-node relationship using this system of equations:
\begin{equation}%
\left[\begin{array}{c}(x_1-x_s)^2+(y_1-y_s)^2\\(x_2-x_s)^2+(y_2-y_s)^2\\\vdots\\(x_N-x_s)^2+(y_N-y_s)^2\end{array}\right]=\left[\begin{array}{c}d^2_1\\d^2_2\\\vdots\\d^2_N\end{array}\right]
\label{eqn}
\end{equation}
After simplifying the above system, we can express it as a linear system:
\begin{equation}%
\mathbf{Ap}_s=\textbf{b}
\label{eqn}
\end{equation}%
With coefficients:
\begin{equation}%
\boldsymbol{\mathrm{A}}=\left[\begin{array}{cc}2(x_N-x_1)^2&2(y_N-y_1)^2\\2(x_N-x_2)^2&2(y_N-y_2)^2\\\vdots&\vdots\\2(x_N-x_{N-1})^2&2(y_N-y_{N-1})^2\end{array}\right]
\label{eqnA}
\end{equation}%

\begin{equation}%
\boldsymbol{\mathrm{b}}=\left[\begin{array}{c}d^2_1-d^2_N-x^2_1-y^2_1+x^2_N+y^2_N\\d^2_1-d^2_N-x^2_2-y^2_2+x^2_N+y^2_N\\\vdots\\d^2_1-d^2_N-x^2_{N-1}-y^2_{N-1}+x^2_N+y^2_N\end{array}\right]
\label{eqnb}
\end{equation}%

In real-time, there is an error in the distance estimation which prevents the circles from intersecting in a single point. Therefore, in this scenario, we will have to solve the linear system to obtain the best fitting solution using estimation methods like least square approach. The node position $\boldsymbol{\mathrm{p}}_s=(x_s,y_s)^T$ from the least square is determined using:  
\begin{equation}%
\boldsymbol{\mathrm{p}}_s=(\boldsymbol{\mathrm{A}}^T\boldsymbol{\mathrm{A}})^-1\boldsymbol{\mathrm{A}}^T\boldsymbol{\mathrm{b}}
\label{eqn}
\end{equation}%
The error in the anchor position and location can be presented using Gaussian distribution with zero-mean as: 
\begin{equation}%
w_i=\dfrac{1}{\sqrt{\sigma^2_{position_i}+\sigma^2_{distance_i}}}
\label{eqn}
\end{equation}%
where $\sigma^2_{distance_i}$  represents the variance in  the distance between the node and the anchor while $\sigma^2_{position_i}$  , is the variance in the anchor locations will contains both variance as: 
\begin{equation}%
{\sigma^2_{position_i}}={\sigma^2_{x_i}}+{\sigma^2_{y_i}}
\label{eqn}
\end{equation}%
The new coefficient of the linear system $\mathbf{Ap}_s=\textbf{b}$  becomes:
\begin{equation}%
\boldsymbol{\mathrm{A}}=\left[\begin{array}{cc}2w_1(x_N-x_1)^2&2w_1(y_N-y_1)^2\\2w_1(x_N-x_2)^2&2w_1(y_N-y_2)^2\\\vdots&\vdots\\2w_1(x_N-x_{N-1})^2&2w_1(y_N-y_{N-1})^2\end{array}\right]
\label{eqnA1}
\end{equation}%

\begin{equation}%
\boldsymbol{\mathrm{b}}=\left[\begin{array}{c}w_1(d^2_1-d^2_N-x^2_1-y^2_1+x^2_N+y^2_N)\\w_1(d^2_2-d^2_N-x^2_2-y^2_2+x^2_N+y^2_N)\\\vdots\\w_1(d^2_{N-1}-d^2_N-x^2_{N-1}-y^2_{N-1}+x^2_N+y^2_N)\end{array}\right]
\label{eqnb1}
\end{equation}%

\subsection{Multilateration}
Both of two previous positioning techniques are limited to the need of the presence of at least 2 anchors (for triangulation) and 3 anchors (for trilateration) to compute the unknown node location. However, this problem can be overcome by turning the node that identified their locations into anchor broadcasting their locations to the all surrounding nodes. This technique is known as iterative multilateration and it continues until all nodes in a network have been localized \cite{tarrio_weighted_2011}. Figure 2.3 demonstrate the iterative multilateration where the grey node uses the black anchor to estimates its location. Then, the grey node becomes an anchor to assist the two white nodes to estimate their positions with the help of the original anchors. The disadvantage of this method is that the localization error accumulates with each iteration executed. In case a node is not surrounded by three anchors, then, it can still estimate its position by using a method called collaborative multilateration \cite{tarrio_weighted_2011}. Figure 2.4 illustrates this method where we have four anchors and two nodes participating all together to determine the two nodes locations.  A node will estimate its location by solving a system of over-determined quadratic equation, which relates its location to the other neighbors.
\begin{figure} []
        \centering
        \includegraphics[width=0.6\textwidth,clip]{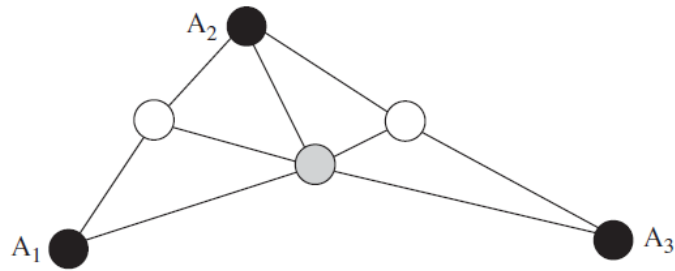}
        \caption[Iterative Multilateration]{Iterative Multilateration}
		 \label{Fig:Iterative}
\end{figure}
\begin{figure} []
        \centering
        \includegraphics[width=0.6\textwidth,clip]{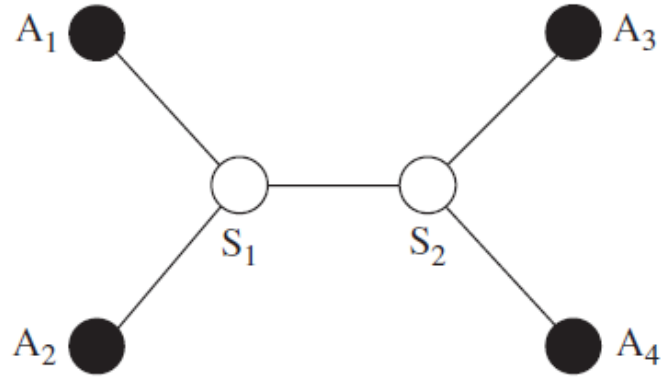}
        \caption[Collaborative Multilateration]{Collaborative Multilateration}
		 \label{Fig:Collaborative}
\end{figure}

\section{Ranging Techniques}
Several localization techniques that are used to estimate the distance or angle to localize the position of sensor nodes have been proposed in the literature. They include Time of Arrival (TOA), Received Signal Strength (RSS), Radio Hop Count, and Direction of Arrival (DOA). 
\label{Ranging Techniques}
\subsection{Time of Arrival (TOA) Technique}
One of the simplest techniques to measure the distance between nodes is TOA. TOA will utilize the proportionality relationship between the time and distance to estimate separation between the two nodes \cite{zitouni_simulated_2013}. A signal will be sent by a node at time $t_1$ and it reached by the receiver node at $t_2$, the distance between the two nodes is:
\begin{equation}%
d=s_r(t_2-t_1)
\label{eqn}
\end{equation}%
where $s_r$ is the propagation speed of the radio signal. However, for this technique to work accurately the two clocks at the transmitter and receiver needed to be synchronized precisely which is quite difficult to achieve in a practical environment. To overcome this limitation, Round-trip Time of Arrival (RToA) and Time Difference of Arrival (TDoA) are developed \cite{stojmenovic_handbook_2005}.

RTOA is quite similar to TOA but it does not require the synchronization between the sender and receiver nodes. It works by recording the time of transmission $t_t$ at node ‘A’ by its own clocks, then the signal is received by node ‘B’ which after a specific time period $t_{per}$ it will send the signal back and the reception time $t_r$ is recorded at node ‘A’. Then the distance is:
\begin{equation}%
d=s_r\dfrac{t_t-t_r-t_{per}}{2}
\label{eqn}
\end{equation}%
TDOA techniques can be classified into two main types: multi-node TDOA, and multi-signal TDOA \cite{boukerche_localization_2007}. The multi-node TDOA is where the difference in time at which the single signal from a single node arrives at three or more nodes resulting in two hyperboloids \cite{wymeersch_cooperative_2009}. These two hyperboloids will intersect at a single point, which is the location of the unknown node as shown in Figure 2.5. While the multi-signal node measures the difference in timings at which multiple signals from a single node arrives at another node as shown in Figure 2.6. To achieve this, nodes must be equipped with extra hardware to send two signals simultaneously. The first signal usually travels at speed of light ($3 \times 10^8 m/s$) and the second signal at the speed of sound ($340m/s$).The distance is: 
\begin{equation}%
d=(s_r-s_s)(t_2-t_1-t_{delay})
\label{eqn}
\end{equation}%
where $s_r$= radio signal speed, $s_s$= sound signal speed, $t_2$= arrival time of sound, $t_1$= arrival time of light
\begin{figure} []
        \centering
        \includegraphics[width=0.6\textwidth,clip]{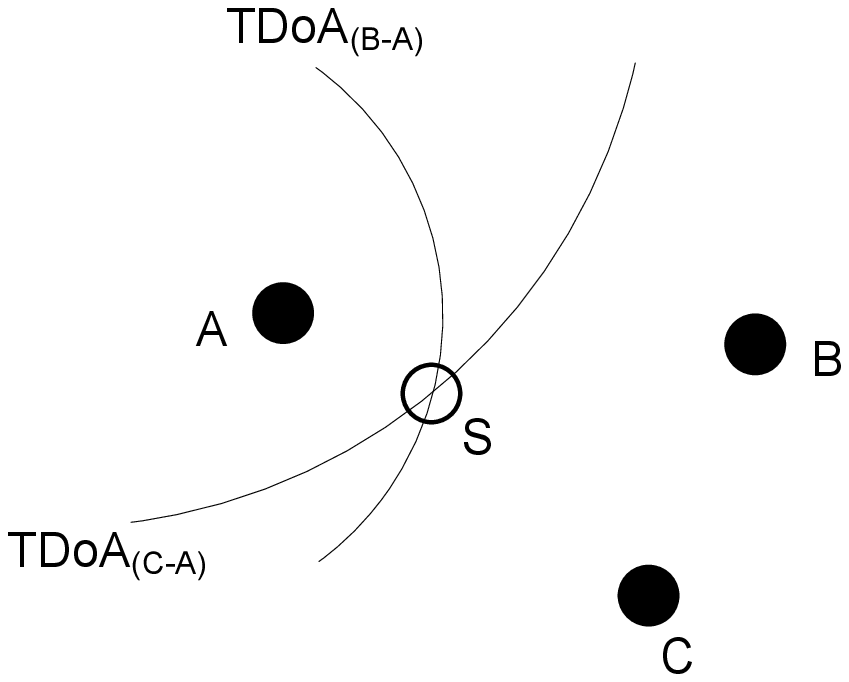}
        \caption[Multi-node TDOA]{Multi-node TDOA}
		 \label{Fig:Multinode-TDOA}
\end{figure}
\begin{figure} []
        \centering
        \includegraphics[width=0.6\textwidth,clip]{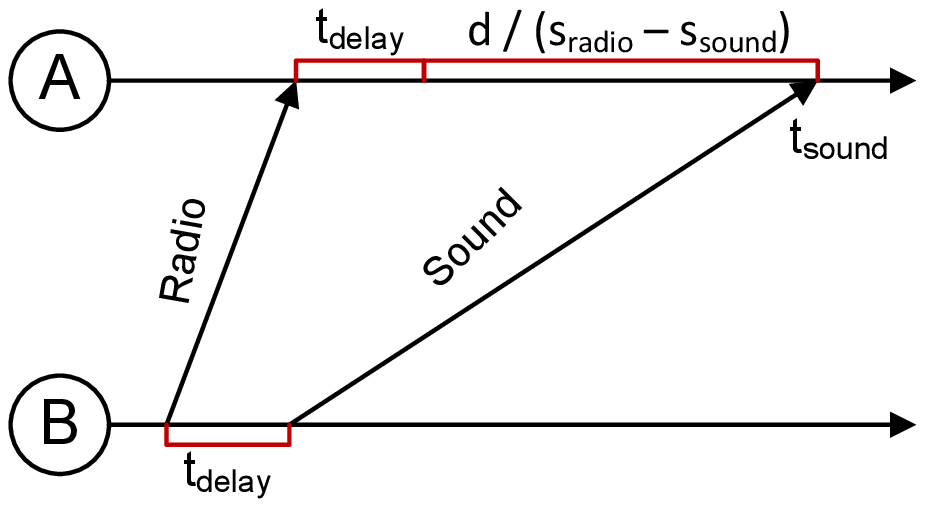}
        \caption[Multisignal-TDOA]{Multisignal-TDOA}
		 \label{Fig:Multisignal-TDOA}
\end{figure}
\subsection{Received Signal Strength (RSS) Technique}
RSS is a common technique in localizing sensor nodes; this is due to the fact that almost all nodes have the capability to measure the strength of the received signal. The distance is estimated between two nodes based on the strength of the signal received by the target node, transmitted power, and the path-loss model \cite{stojmenovic_handbook_2005}. The operation starts when an anchor node sends a signal that is received by the receiver and passed to the Received Signal Strength Indicator (RSSI) to determine the power of the received signal \cite{wymeersch_cooperative_2009}. Then the distance can be calculated using the equation:
\begin{equation}%
P_R=P_T-10\eta\log(d_{ij})+X_{ij}
\label{eqn}
\end{equation}%
where $P_T$ is a constant due to transmitted power and the antenna gains of the sensor nodes, $\eta$ is the attenuation constant, and $X_{ij}$ is the uncertainty factor due to multipath and shadowing.

The RSS measurement from three anchor nodes can be combined with their locations (trilateration) to estimate the location of the node. 
\subsection{Radio Hop Count Technique}
In Radio Hop Count Technique if two nodes can communicate with each other then the distance between them is less than R, where R is the maximum range of the radios. Then to calculate the distance between two nodes three main steps need to be done \cite{pal_localization_2010}. First, calculate the minimum number of hops between the unknown node and each anchor node. Then average the actual distance of one hop using the following equation:
\begin{equation}%
Hop Size_i=\frac{\sum \sqrt{(x_i-x_j)^2+(y_i-y_j)^2}}{\sum h_{ij}}
\label{eqn}
\end{equation}%
where $(x_i,y_i)$,$(x_j,y_j)$  anchor nodes, $h_{ij}$ number of hops between anchor nodes
Then the distance between the unknown nodes and anchor nodes is calculated using the following equation:
\begin{equation}%
d= Hop Size_i \times h_{ik}
\label{eqn}
\end{equation}%
\subsection{Direction of Arrival (DOA) Technique}
In this technique the directions of neighbouring sensors rather than the distance to neighbouring sensor is estimated. DOA can be classified into two techniques sensor array and directional antenna \cite{samhan_design_2006,al-ardi_investigation_2003,shubair_robust_2004,al-ardi_performance_2003,shubair_closed-form_1993,shubair_robust_2005,shubair_performance_2005,belhoul_modelling_2003,shubair_improved_2005,hakam_robust_2014,hakam_enhanced_2013,shubair_displaced_2008,alhajri_hybrid_2015,al-ardi_performance_2003-1,jimaa_convergence_2009,ardi_adaptive_2004,samahi_performance_2006,elayan_wireless_2017,lazovic_comparative_2013,moghaddam_novel_2010,hakam_novel_2016,shubair_adaptive_2013,kulaib_performance_2011,shubair_simple_1992,goian_fast_2015,alhajri_hybrid_2015-1,kulaib_accurate_2015,kulaib_improved_2015,kulaib_investigation_2013,jimaa_performance_2008,shubair_improved_2007,alayyan_mmse_2009,shubair_performance_2006,alkaf_improved_2012,shubair_detection_2011,shubair_enhanced_2009,shubair_robust_2006,hakam_accurate_2013,shubair_displaced_2007,shubair_improved_2006,al-ardi_direction_2006,al-nuaimi_direction_nodate,shubair_convergence_2005}. The sensor array is comprised of two or more individual sensors (microphone or antennas) as shown in Figure 2.7. The DOA will estimate the difference in the arrival timings. The angle of arrival is calculated using the following equation: 
\begin{equation}%
\Delta t=\dfrac{\delta\cos(\theta)}{y}
\label{eqn}
\end{equation}%
where $\Delta t$ is the difference in arrival time, $\delta$ is the antenna separation, $\theta$ is the angle of arrival, and $y$ is the velocity of the RF or acoustic signal
In the case of directional antenna, it operates by calculating the RSS ratio between several directional antennas.
\begin{figure} [h]
        \centering
        \includegraphics[width=0.6\textwidth,clip]{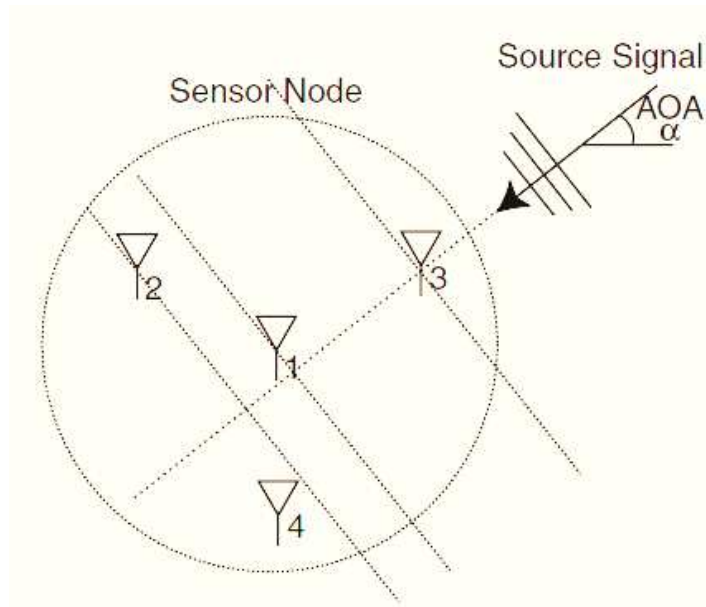}
        \caption[Sensor array processing]{Sensor array processing}
		 \label{Fig:SensorArrayProcessing}
\end{figure}
\section{Localization Classification}
A localization algorithm is a term that refers to the process or set of rules to be followed in calculations or other problem-solving operations, especially by a computer to establish location-based technology. The research community proposes many different classifications for the area of localization in WSNs. Depending on the needed criteria, localization algorithms can be categorized as single-hop and multi-hop or as anchor based and non-anchor based, or as centralized and decentralized (distributed). In this survey, the classification that shall be adopted is the centralized and decentralized (distributed), as shown in Figure 2.8 \cite{boukerche_localization_2007,kwon_anchor-free_2010,doherty_convex_2001,shang_localization_2003}:
\begin{figure} []
        \centering
        \includegraphics[width=0.8\textwidth,clip]{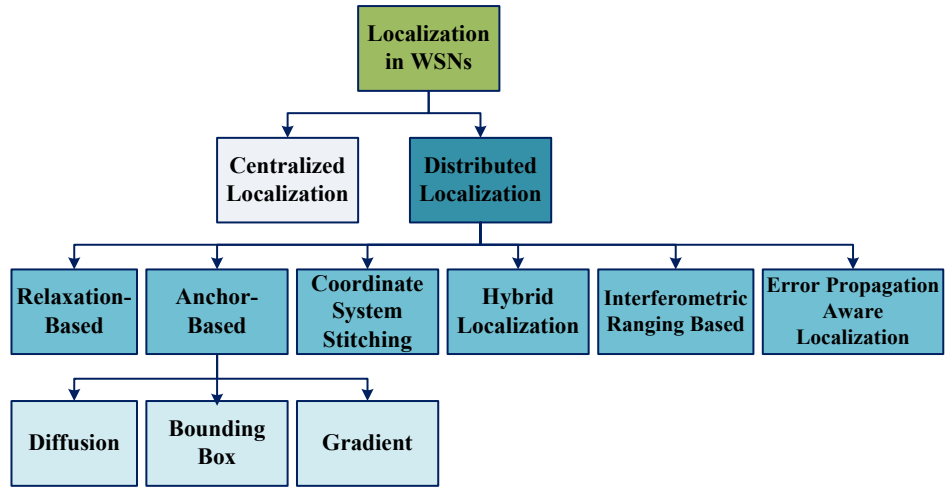}
        \caption[Classification of Localization Algorithms in WSNs]{Classification of Localization Algorithms in WSNs}
		 \label{Fig:Classification}
\end{figure}
\label{Localization Classification}
\subsection{Centralized Algorithm}
The first class of algorithm is referred to as centralized localization algorithm. Centralized localization is basically migration of inter-node ranging and connectivity data to a sufficiently powerful central base station and then the migration of resulting locations back to respective nodes. Centralized localization algorithm is characterized by its need for enormous computational power. The high amount of computational power gives the centralized localization its capability to execute complex mathematical operations. This superb advantage comes with the disadvantage of the high communication cost. This disadvantage is a result of the process itself. As all nodes of a network send their data to the central receiver, the computed positions are sent back to respective node; the communication cost, as a result, of such becomes considerably high. Centralized localization algorithm itself can be divided in different types, as shown in Figure 2.9. These types depend on the way they process data at the central receiver. There are two popular types of centralized algorithms \cite{kulaib_efficient_2014,boukerche_localization_2007}
\begin{enumerate}
\item Semi-definite Programming (SDP) 
\item Multidimensional Scaling (MDS)  
\end{enumerate}
\begin{figure} [h]
        \centering
        \includegraphics[width=0.6\textwidth,clip]{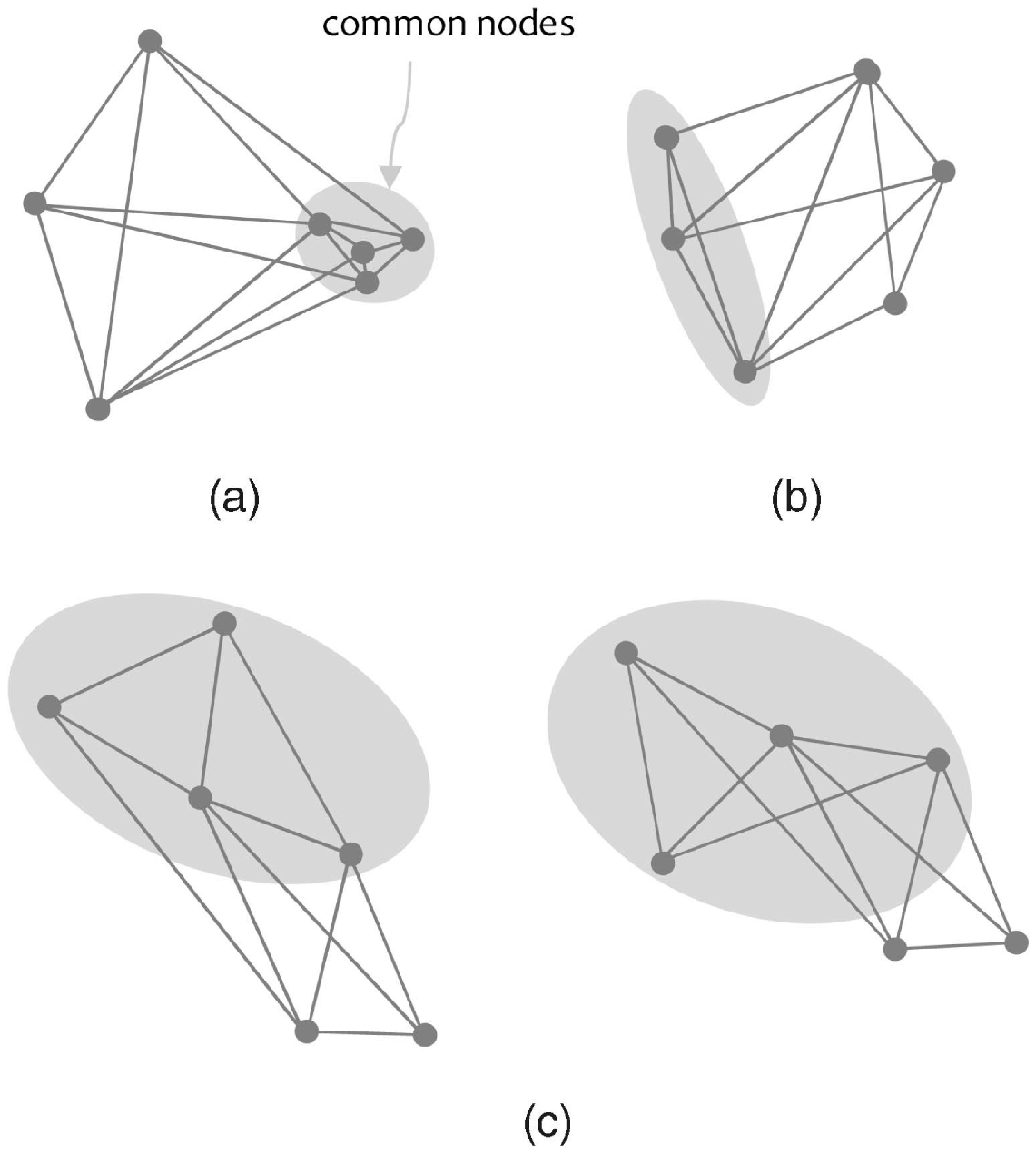}
        \caption[Configurations that represent Centralized Algorithms]{Configurations that represent Centralized Algorithms}
		 \label{Fig:Centralized}
\end{figure}
\subsubsection{Semi-definite Programming (SDP)}
One of the most popular approaches of centralized localization algorithms is known as Semi-definite Programming (SDP).  This approach uses Linear Matrix Inequalities (LMIs) to represent geometric constraints among nodes. After establishing the LMIs, they, all LMIs, get combined in order to form a single semi-definite program. Solving it produces a bounding region for every single node of the network. This technique can be used to execute enormously complex mathematical operations. However, since the SDP is based on the appropriate utility of LMIs, SDP can only work with those geometric that can be represented by LMIs. In other words, since LMI cannot be utilized to represent all geometric constraints (Precise range data, e.g. rings), by extension, SDP cannot be used to represent those configurations. The rule of thumb is that LMI can only be used to represent geometrical constraints that form convex regions, such as representing the hop count with a circle, and the angle of arrival with a triangle, Figure 2.10.
\begin{figure} [h]
        \centering
        \includegraphics[width=0.8\textwidth,clip]{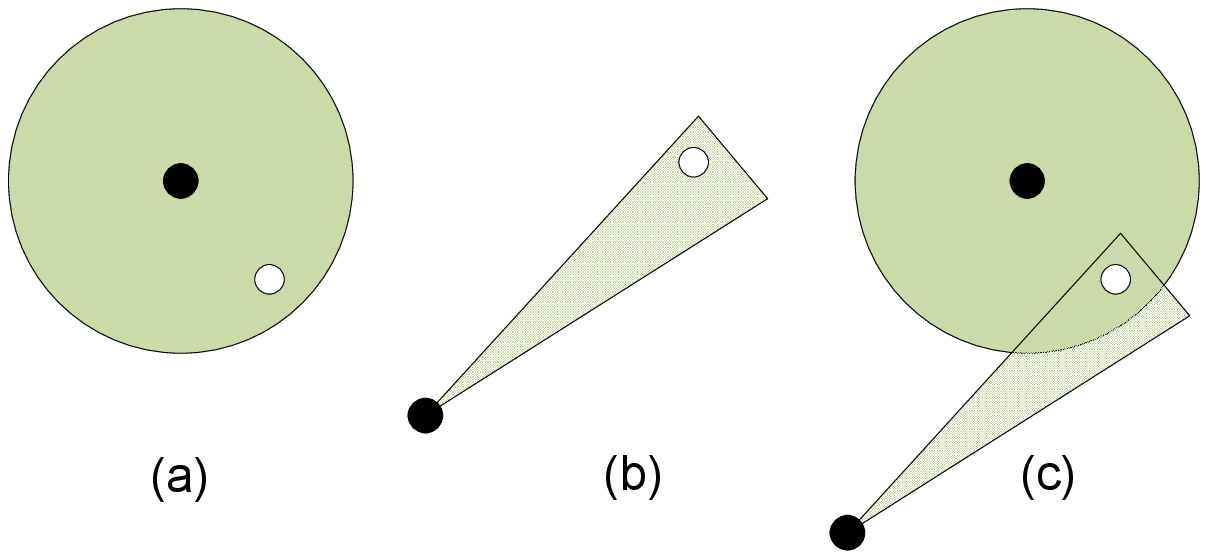}
        \caption[Illustrations of Semi-definite programming]{Illustrations of Semi-definite programming}
		 \label{Fig:SemiDe}
\end{figure}
\subsubsection{Multidimensional Scaling (MDS)}
Another popular approach of centralized localization algorithms is known as Multidimensional Scaling (MDS). This approach is also known as MDS-MAP. This approach utilizes the MDS that originates from mathematical psychology. MDS-MAP uses “Law of Cosines and linear algebra to reconstruct the relative positions of the points based on the pairwise distances”. This approach uses a set of well-known steps to be efficiently constricted.
\begin{enumerate}
\item The first step starts by the gathering of the ranging data. These data are being obtained from the network that used to obtain a matrix \textbf{R}, in such scenario $d_{ij}$ refers to the distance between nodes $j$ and $i$.
\item The second step is finished after using the gathered data for the sake of completing the matrix of the inter-node distances $D$, which is gotten by applying the shortest path algorithm such as Dijkstra’s or Floyd’s on \textbf{R}.
\item The third step starts after finishing the matrix of the inter-node distances. It starts by finding the estimates of node positions $p_s$ by administrating MDS technique on $D$, retaining the first 2 (or 3) largest eigenvectors and eigenvalues to construct a 2-D (or 3-D) map.
 \item 	The fourth and the final step is done as a transformation of the node positions $p_s$ take place into the global coordinates by utilizing 3 or more anchors in the case of 2-D, and 4 or more anchors in the case of 3-D.
\end{enumerate}
Comparing the above two centralized localization algorithms, MDS-MAP has a unique characteristic of that when any improvement administered in ranging accuracy will result in a noticeable improvement in MDS-MAP, which SDP lacks. Another advantage of the MDS-MAP on the SDP is that MDS-MAP does not need anchor nodes in the start of the process. It transforms into absolute locations by utilizing coordinates using 3 or more anchor nodes. A major drawback on MDS-MAP is that it costs way too high for higher order operations. Moreover, the cost can be reduced partially by using decentralized algorithms (i.e. coordinate system stitching), explained in the next section. 
\subsection{Decentralized (Distributed) Algorithm}
The second class of localization algorithms is referred to decentralized localization algorithm, which is also known as distributed localization. As centralized localization is a migration of inter-node ranging and connectivity data to a sufficiently powerful central base station and then the migration of resulting locations back to respective nodes, Distributed Localization Algorithm differs from it since in distributed localization algorithm all the computations relevant to the nodes are done by the nodes themselves; moreover, the nodes communicate with each other in order to obtain their positions in the network. In other words, the nodes of the distributed algorithms use each node’s computational power to perform its operations. This technique demands relatively high inner-node commination besides the parallelism for which it needs to run tasks comparable to centralized systems. Unlike the centralized localization algorithm, this technique does not require a centre node with significantly high computational power, which reduces the cost of implementation. Another advantage of the distributed localization algorithm is that it is, generally, faster than the centralized algorithm as each operation is done within the relevant node(s). Distributed algorithms can be divided into six main categories: 
\begin{enumerate}
\item Anchor-based
\item Relaxation-based
\item Coordinate system stitching
\item Hybrid Localization
\item Interferometric Ranging Based Localization
\item Error Propagation Aware Localization algorithms
\end{enumerate}
\subsubsection{Anchor-based}
The first category of distributed algorithms is better known as the anchor-based distributed algorithms. Anchor-based is a type of distributed algorithms that utilizes the anchors in order to find the location of unknown node(s). In such algorithms, the nodes obtaining a distance measurement of a few anchors, as a start. It then determines their location based on the found measurements. Anchor- based distributed algorithms has several algorithmic approaches:
\begin{itemize}
\item 	The first of these algorithmic approaches is diffusion. Diffusion algorithm depends sole on the radio connectivity data, which makes this algorithm fairly easy. This algorithm is based on the following working principle:
“Assuming that a node $p_s$ is most probably at a centroid of its neighbour’s positions. There are two different alternatives to this algorithm. The first option averages the positions of all anchors that can communicate with the node using radio, in order to localize the position of that node. This technique was developed by Bulusul \cite{savarese_location_2001}. The accuracy of this algorithm is low when a node is far away from the anchor nodes or anchor density is low. The second option considers both anchors and normal nodes in determining the position of the node at the centre. Also in this algorithm, the accuracy is low when a node is far away from the anchor nodes, node density varies across the network, or node density is low \cite{savvides_dynamic_2001}.” The first technique has an advantage when it comes to the needed number of node to start with, as it requires fewer number than the second technique.
\item •	The second anchor-based distributed algorithmic approach is known as the Bounding box algorithm. This algorithm calculates the node’s position based on the ranges of it to numerous anchors. Each anchor covers a range around it, which can be expressed as a box surrounding the anchor \cite{priyantha_poster_2003,moore_robust_2004}. The intersection between the boxes of the anchors determines the position (location) of an unknown node. An illustration is shown in Figure 2.11. 
\end{itemize}
\begin{figure} []
        \centering
        \includegraphics[width=0.6\textwidth,clip]{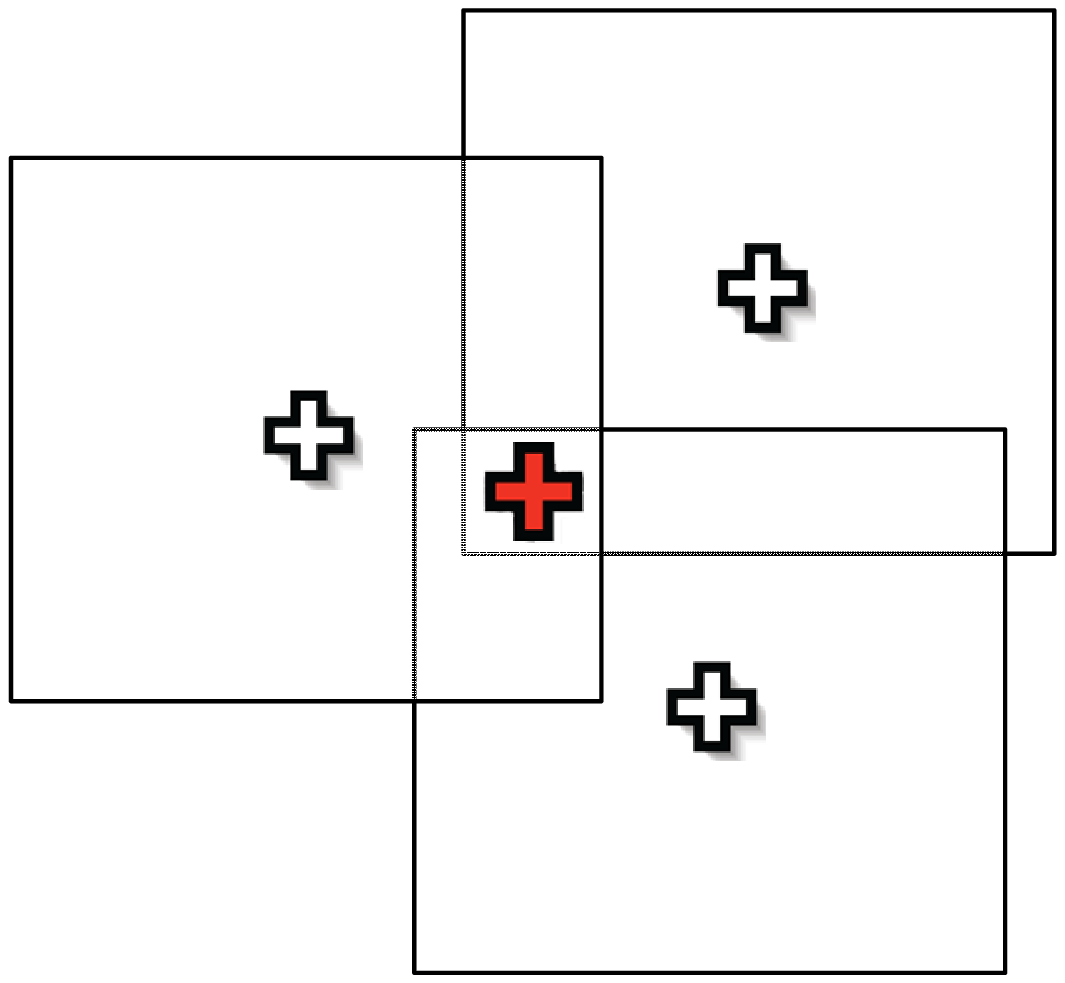}
        \caption[Illustrations of the bounding box interaction]{Illustrations of the bounding box interaction}
		 \label{Fig:Boundingbox}
\end{figure}
\subsubsection{Relaxation-based}
The second category of distributed algorithms is better known as the relaxation-based distributed algorithms. This algorithm combines the advantages of distributed algorithms scheme that lies in the computational power with the advantages of the centralized scheme that lies in the precision and accuracy. Such a fusion brings the algorithm to an utmost capability. 
The process of the relaxation-based distributed algorithm follows very specific steps, which are explained as below \cite{cheng_localization_2007,cho_mimo-ofdm_2010}:
\begin{enumerate}
\item The first step begins by giving an estimate of the position of the nodes. This step can be done using any of the distributed techniques mentioned earlier. 
\item The second step starts as the positions are refined using the position estimates of the neighbouring anchors. Those neighbouring anchors are considered temporary anchors. The refined step is usually obtained using local neighbourhood multilatertaion. In the case of this algorithm, neighbourhood multilatertaion is called “the spring model.” The terminology refers to the distances between the nodes with their resting springs.
\item Using an optimization technique will change the nodes position every single or a number of iteration(s). What determines the end of this step is that all the nodes involved must have zero forces subjected to them.  It is only then the process moves to the final step.
\item The final step involves forming a global minimum, which happens if and only if:
\begin{itemize}
\item The magnitude of all the acting-on-nodes forces is equal to zero.
\item The magnitude of the forces between nodes is also zero.
\end{itemize}
\end{enumerate}
By following the step, relaxation-based distributed algorithm can be obtained gracefully. Although this algorithm is quite an effective one, it has a serious drawback that lies in its sensitivity to the initial positions that it starts with. The problem manifests even more if the initial position in a local minima. However, this problem is solvable by utilizing a fold-free algorithm. This type of algorithm can position the nodes in a starting position such that they almost certainly never fall in local minima. 
\subsubsection{Coordinate System Stitching Algorithm}
The third category of distributed algorithms is known as Coordinate System Stitching algorithm. In this algorithm, the network is divided into smaller and overlapping regions, usually known as sub regions. Each of those sub regions creates its own optimum local map. There are many approaches for this technique, yet the general steps to applying it is as the following \cite{stutzman_antenna_2012}: 
\begin{enumerate}
\item The first step is the most crucial one, if not done correctly the algorithm can never work. This step involves splitting the network into areas. Those areas, sub regions, must be overlapping. This is usually used with a single node and its one-hop neighbour. This step differs from an algorithm to another.
\item The second step begins when the splitting of the sub regions is done. At this stage, a local map must be computed for each of the sub regions. This step differs from an algorithm to another.
\item The final step starts after finishing the computation of the optimum local map. The third step involves placing the entire sub regions into one single global coordinate system, thus the name. This step is done using registration procedure. This step common for all algorithms.
\end{enumerate}
\subsubsection{Hybrid Algorithm}
The fourth category of distributed algorithms is known as Hybrid Algorithm. In such scheme, two different localization techniques are being used. For example, MDS and the Proximity based map (PDM) or MDS with Ad- hoc positioning system (APS). Hybrid techniques and algorithms are usually used to make advantages and to overcome the limitations of both techniques in use. Overall, the performance of the hybrid algorithm is better than each of the techniques used individually.
Let us take an example in details; the author in \cite{tran-xuan_calibration_2009} combined the MDS with the PDM in order to establish a hybrid technique and so to localize the sensor nodes location. It starts by dividing the nodes into 3 classes:
\begin{itemize}
\item Primary anchors
\item Secondary anchors 
\item Normal sensors
\end{itemize}
The manifestation of the hybrid technique is in the following steps:
\begin{enumerate}
\item Initially, each primary anchor sends an invitation packer contains an ID that is unique to each invitation, which contains a value, ks, controlling the number of neighbouring secondary anchors. 
\item At this instant, a counter function is initialized and set to zero. The invitation sent is then received by the normal sensor, which by its turn performs a Bernoulli trial. 
\item If the output of the trail is true (p), the normal sensor increments the counter and thus becomes a secondary anchor.
\item Transmitting the packet from neighbouring sensors from one to another until all the counters are equal.
\item Following the steps with the primary and again with the secondary repeatedly can guarantee the success of the hybrid technique.
\end{enumerate}
The hybrid technique has an advantage of that it can be used in both indoor and outdoor localization.
Interferometric ranging based localization and error propagation aware localization algorithm is not discussed as it is beyond the scope of this project.

\chapter{Received Signal Strength}
\label{Received Signal Strength}
In this chapter, the log-normal channel model is derived and explained. In addition, a comparison between different estimators is presented.
\section{RSS Modeling}
\label{RSS modeling}
 “The free-space propagation model is used for predicting the received signal strength in the line-of-sight (LOS) environment where there is no obstacle between the transmitter and receiver” \cite{pozar_microwave_2000}. The RSS equation is derived from Friis Transmission formula \cite{kulaib_efficient_2014}
\begin{equation}%
P_r=P_tG_tG_r\dfrac{\lambda^2}{(4\pi \times d_{ij})^2}
\label{3.1}
\end{equation}%
where $P_r$ is the received signal strength from sensor node $i$ at node $j$ at time $t$,$P_t$ is the transmitted power,$G_t$ is the transmitter gain,$G_r$ is the receiver gain, $d_{ij}$ is the distance from sensor node $i$ at node $j$, and $\lambda$ is the wavelength of the signal. From the equation above, the received power $P_r$ attenuates exponentially with the distance $d_{ij}$ \cite{pozar_microwave_2000}.
The free-space path loss,$PL_F$, is derived from the equation above by 10 log the ratio of the transmitted power $P_t$ to the received power $P_r$ and setting $G_t$= $G_r$=1 because in most of the cases, the antennas that are used are isotropic antennas, which radiate equally in all direction, giving constant radiation pattern \cite{pozar_microwave_2000}. Thus, the formula is the following: 
\begin{equation}%
PL_F(d_{ij})[dB]=10\log \left(\frac{P_t}{P_r}\right)=20\log \left(\frac{4\pi \times d_{ij}}{\lambda}\right)
\label{3.2}
\end{equation}%

“In the free-space model, the average received signal in all the other actual environments decreases with the distance between the transmitter and receiver $d_{ij}$, in a logarithmic manner”. Therefore, path loss model generalized form can be obtained by changing the free-space path loss with the path loss exponent $n$ depending on the environment. This is known as the log-distance path loss model which will result in the following formula \cite{pozar_microwave_2000}:
\begin{equation}%
PL_{LD}(d_{ij})[dB]=PL_F(d_0)+10\eta\log \left(\dfrac{d_{ij}}{d_o}\right)
\label{3.3}
\end{equation}%

where $d_0$ is the reference distance at which the path loss inherit the characteristics of free-space in equation 3.2 \cite{pozar_microwave_2000}. This distance is set to different value depending on the propagation environment; for example, it is 1 km for a large coverage cellular system. In our case we will consider this value to be 1m. The value of $n$ is 2 which resembles the free-space. Different values for $n$ resembles different environment conditions \cite{pozar_microwave_2000}.

Every path between the sender and the receiver has different path loss since the environment characteristics changes as the location of the receiver changes. Moreover, the signal may not penetrate in the same way. For that reason, a more realistic modelling of the transmission is assumed which is the log-normal modelling.
\begin{equation}%
PL(d_{ij})[dB]=PL_F(d_0)+10\eta\log \left((\dfrac{d_{ij}}{d_o}\right)+X_\sigma
\label{3.4}
\end{equation}%
where $PL(d_ij)[dB]$ = $P_t[dB]$-$P_r[dB]$ is the path loss at distance $d_{ij}$ from the transmitter, $PL_F (d_0)$ is the path loss model at the reference distance $d_0$ which is constant. $X_\sigma$  is Gaussian random variable with a zero mean and standard deviation $\sigma$ \cite{tran-xuan_calibration_2009}.

In order to find the location of the blind node with respect to three reference nodes ($\boldsymbol{\mathrm{p}}_1, \boldsymbol{\mathrm{p}}_2, \boldsymbol{\mathrm{p}}_3$), three circles will be used to draw three lines. These lines passes  through two points at which the two circles intersect, and they are called the line of position (LOP), as shown in Figure 3.1. To find the equation that represents each line, we start with the distance equation that describes the length between the $i^{th}$ reference node and the blind node, which is given by \cite{kulaib_efficient_2014}:
\begin{equation}%
D_i=\left\Vert p_i-p_s\right\Vert=\sqrt{(x_i-x_s)^2+(y_i-y_j)^2}
\label{3.5}
\end{equation}%
where $\boldsymbol{\mathrm{p}}_i=(x_i,y_i)$ is the position of the reference node, $\boldsymbol{\mathrm{p}}_s=(x_s,y_s)$ is the position of the unknown node, and $\left\Vert x\right\Vert$  is the  norm of vector \textbf{x}.
\begin{figure} []
        \centering
        \includegraphics[width=0.6\textwidth,clip]{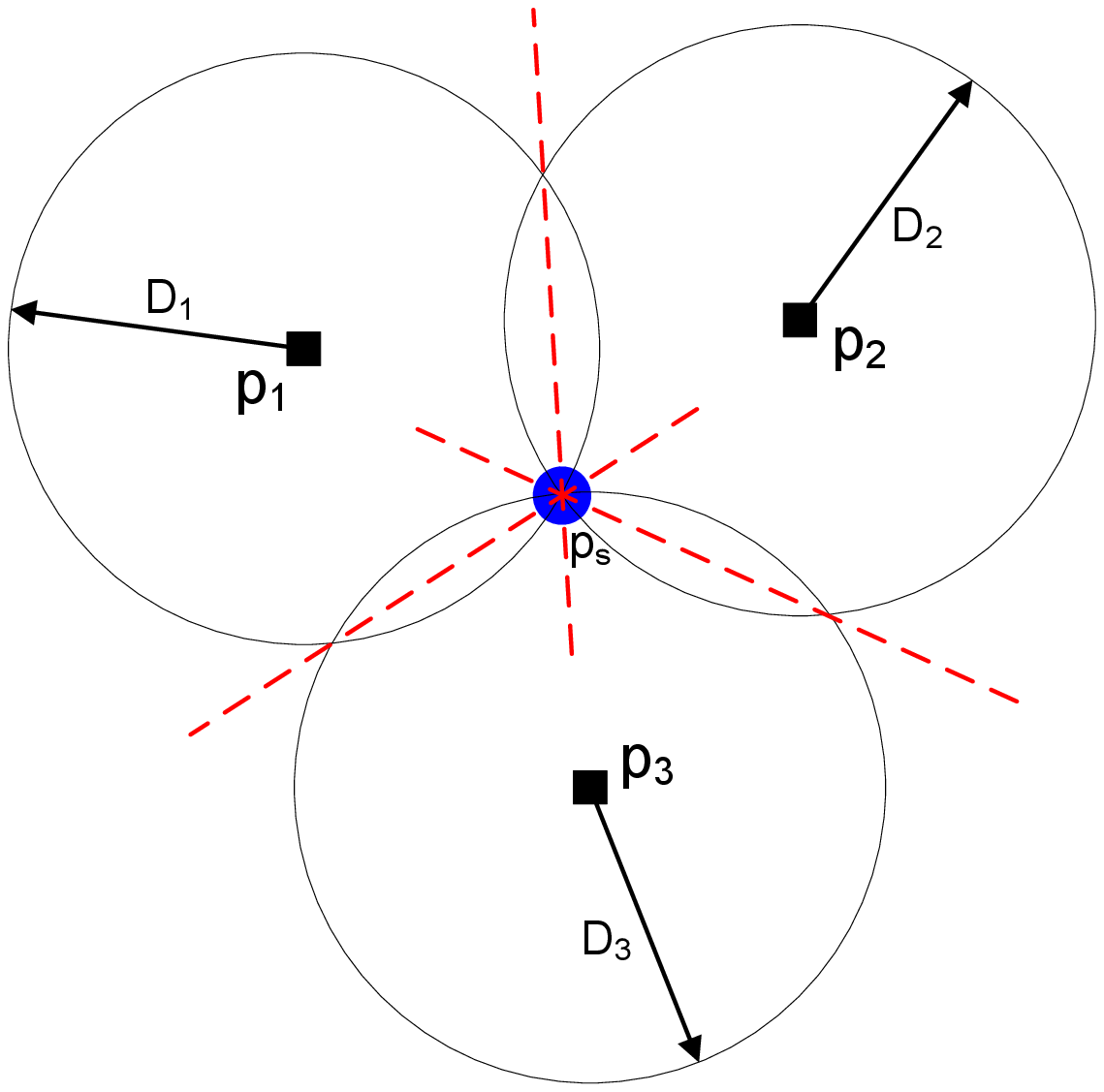}
        \caption[Node estimation in RSS]{Node estimation in RSS}
		 \label{Fig:RSS_Circles}
\end{figure}

Then, the LOP between $\boldsymbol{\mathrm{p}}_1$ and $\boldsymbol{\mathrm{p}}_2$, will be computed by squaring and taking the difference between $D_2$ and $D_1$ ,thus, the result will be as  follow:
\begin{equation}%
(x_2-x_1)x_s+(y_2-y_1)y_s=\frac{1}{2}(\left\Vert x_2\right\Vert^{2}-\left\Vert x _1\right\Vert^{2}+D_1^2-D_2^2)
\label{3.6}
\end{equation}%
The same scenario applies between $p_2$ and $p_3$ and between $p_1$ and $p_3$ , as follows \cite{kulkarni_computational_2011}

\begin{equation}%
(x_3-x_2)x_s+(y_3-y_2)y_s=\frac{1}{2}(\left\Vert x_3\right\Vert^{2}-\left\Vert x _2\right\Vert^{2}+D_2^2-D_3^2)
\label{3.7}
\end{equation}%

\begin{equation}%
(x_3-x_1)x_s+(y_3-y_1)y_s=\frac{1}{2}(\left\Vert x_3\right\Vert^{2}-\left\Vert x _1\right\Vert^{2}+D_1^2-D_3^2)
\label{3.8}
\end{equation}%

The blind node location will be estimated by solving for $x_s$ and $y_s$ in the above three equations.  However, increasing the number of the reference nodes will result in more intersection points. These lines will not intersect at the same point due to the error in the distance measurements. Hence, it is required to find another technique to provide better estimation for the location of the unknown node \cite{kulaib_efficient_2014}. 

For any localization system, with $N$ anchor nodes, there will be $N-1$ independent lines represented in the following formula:
\begin{equation}%
a_{i,1}x_s+a_{i,2}y_s=b_i
\label{3.9}
\end{equation}%
where the coefficients $\boldsymbol{\mathrm{a_i}}$ and $\boldsymbol{\mathrm{b_i}}$ are known.
The above equation, also, can be written as follows:
\begin{equation}%
\mathbf{a}_i\mathbf{p}_s=\mathbf{b}_i
\label{3.10}
\end{equation}%
where $\boldsymbol{\mathrm{a}}_i$ =[$a_{i,1}$  $a_{i,2}$] and $\boldsymbol{\mathrm{p}}_s$ = [$x_s$ $y_s$]$^T$. 
All lines equations can be described in matrix format: 
\begin{equation}%
\mathbf{Ap}_s=\mathbf{b}
\label{3.11}
\end{equation}%
where $\boldsymbol{\mathrm{A}}=[A_1~A_2............A_{N-1}]^T$,  	$\boldsymbol{\mathrm{b}}=[b_1~b_2............b_{N-1}]^T$,  
This equation has no solution, as the straight lines do not intersect at one point. For that reason, different estimation techniques have been used to find the position of $\mathbf{p_s}$.

\section{Norm Approximation}
The simplest norm approximation problem is an unconstrained problem of the form 
\begin{equation}%
\mathrm{minimize}~\left\Vert \boldsymbol{\mathrm{Ap}}_s-\boldsymbol{\mathrm{b}}\right\Vert^2_2 
\label{3.12}
\end{equation}%
where $\boldsymbol{\mathrm{A}} \in \boldsymbol{\mathrm{R}}^{ m_1 \times n_1}$ and $\boldsymbol{\mathrm{b}}\in\boldsymbol{\mathrm{R}}^{ m_1}$ are problem data, $\boldsymbol{\mathrm{p}}_s \in \boldsymbol{\mathrm{R}}^{ n_1}$ is the variable, and $\left\Vert .\right\Vert$ is a norm on $\boldsymbol{\mathrm{R}}^{ m_1}$. A solution of the norm approximation problem is sometimes called an approximate solution of $\boldsymbol{\mathrm{Ap_s}}\approx\boldsymbol{\mathrm{b}}$,in the norm $\left\Vert .\right\Vert$. The vector
\begin{equation}%
\mathbf{r_{residual}}=\mathbf{Ap}_s-\mathbf{b}
\label{3.13}
\end{equation}%
is called the residual for the problem; its components are sometimes called the individual residuals associated with $\boldsymbol{\mathrm{p}}_s$.

The norm approximation problem in \ref{3.12} is a convex problem, and is solvable. Its optimal value is zero if and only if $\boldsymbol{\mathrm{b}}$ $\in$ $\boldsymbol{\Re}{(\mathrm{A})}$. We can assume without loss of generality that the columns of $\boldsymbol{\mathrm{A}}$ are independent; in particular, that $m_1\geq n_1$. When $m_1=n_1$ the optimal point is simply $\boldsymbol{\mathrm{A^{-1}b}}$, so we can assume that $m_1 > n_1$ \cite{burden_numerical_2004}.
%
%
%
\subsection{Least-squares Approximation}
The most common norm approximation problem involves the $\ell_2$-norm \cite{chong_introduction_2008}. By squaring the objective, we obtain an equivalent problem which is called the least-squares approximation problem, 

\begin{equation}%
\mathrm{minimize}~\left\Vert \boldsymbol{\mathrm{Ap}}_s-\boldsymbol{\mathrm{b}}\right\Vert^2_2
\label{3.14}
\end{equation}%
where the objective is the sum of squares of the errors. This problem have an analytically solution by expressing the objective as a quadratic function:

\begin{equation}%
f(\boldsymbol{\mathrm{p}}_s)=\boldsymbol{\mathrm{p}}^T_s\boldsymbol{\mathrm{A}}^T\boldsymbol{\mathrm{Ap}}_s-2\boldsymbol{\mathrm{b}}^T\boldsymbol{\mathrm{Ap}}_s+\boldsymbol{\mathrm{b}}^T\boldsymbol{\mathrm{b}}
\label{3.15}
\end{equation}%
A point $\boldsymbol{\mathrm{p}}_s$ minimizes f if and only if

\begin{equation}%
\bigtriangledown f(\boldsymbol{\mathrm{p}}_s)=\boldsymbol{\mathrm{2A}}^T\boldsymbol{\mathrm{Ap}}_s-\boldsymbol{\mathrm{2A}}^T\boldsymbol{\mathrm{b}}=0
\label{3.16}
\end{equation}%

\begin{equation}%
\boldsymbol{\mathrm{A}}^T\boldsymbol{\mathrm{Ap}}_s=\boldsymbol{\mathrm{A}}^T\boldsymbol{\mathrm{b}}
\label{3.17}
\end{equation}%
which always have a solution. Since we assume the columns of \textbf{A} are independent, the least-squares approximation problem has the unique solution:

\begin{equation}%
\boldsymbol{\mathrm{p}}_s = \boldsymbol{\mathrm{(A}}^T\boldsymbol{\mathrm{A}})^{-1}(\boldsymbol{\mathrm{A}}^T\boldsymbol{\mathrm{b)}}
\label{3.18}
\end{equation}%

The least-squares estimator is a maximum likelihood estimator of a Gaussian distribution with a zero mean and variable variance (see Appendix A). However, in the least-squares all of the anchor nodes will be given the same weight in the estimation of the node regardless of the location of the node, which will result in a higher error in the distance estimation. To overcome this limitation, more weight will be given to those measurements corresponding to the closer distances and this will results in a higher accuracy. The Weighted Least Square (WLS) formuls is:
 
\begin{equation}%
\boldsymbol{\mathrm{p}}_s = \boldsymbol{\mathrm{(A}}^T\boldsymbol{\mathrm{WA}})^{-1}(\boldsymbol{\mathrm{A}}^T\boldsymbol{\mathrm{Wb)}}
\label{3.19}
\end{equation}%
where \textbf{W} is the weighting matrix

The computation complexity of the WLS is O(S$^3$) where S is the number of reference nodes inside the coverage area of the blind node but, not the total number of the anchor nodes.  On the other hand, the computational complexity in the LS is O(S) \cite{tarrio_weighted_2011}. However, in practical deployments the value of S is usually small, so WLS can be executed in a resource-constrained device.                                                 

\subsection{Huber Robustness}
In the Huber robustness the $\ell_1$-norm is used:

\begin{equation}%
\mathrm{minimize}~\left\Vert \boldsymbol{\mathrm{Ap_s-b}}\right\Vert_1
\label{3.20}
\end{equation}%

In the $\ell_1$-norm the absolute error is minimized, which means the size of the error will be smaller than the case of $\ell_2$-norm and so will be less sensitive to the outliers \cite{boyd_convex_2004}.However, in the case of Huber robustness we have an absolute value and so the derivative can not be applied and for that reason the problem is solved using the Iterative Reweighed Least Square (IRLS) technique.

In the first iteration the weights will be carried out using the WLS. After that, in the second iteration the absolute error ($e_i$) will be evaluated and the weights will be updated using the following equation:

\begin{equation}%
w_i=\sum\limits^{n}_{i=1}\frac{1}{\left| e_i\right|+\epsilon}\left| e_i\right|^2
\label{3.21}
\end{equation}%

After several iterations the answer will converge to the minimum. The value of $\epsilon$ is added to avoid having a discontinuity and it is critical to the performance of the estimator \cite{guitton_robust_2003}. This is because the value of $\epsilon$ resembles the size of the $\ell_2$-norm curve in the Huber robustness.

\section{Simulation Results}
\label{Simulation Results}

\subsection{Performance}
\begin{figure} [h]
        \centering
        \includegraphics[width=1\textwidth,clip]{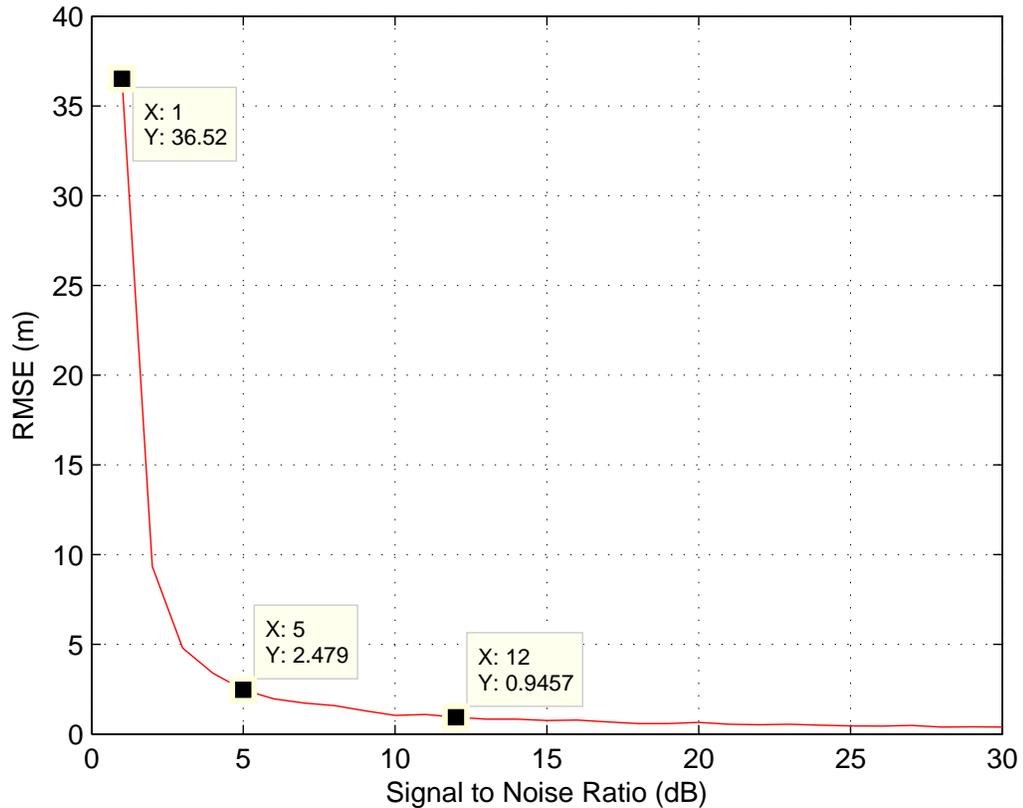}
        \caption[RMSE for different values of SNR using LS]{RMSE for different values of SNR using LS}
		 \label{Fig:Least Square}
\end{figure}

In order to compare the performance of the three RSS based localization approaches in terms of accuracy of the localization results, several nodes are placed in a two-dimensional region with the size of (100m $\times$ 100m). Each anchor node contains one antenna operating at a frequency of 1GHz. To evaluate the performance of the two techniques, Root Mean Square Error is evaluated 150 times for every SNR value. The performance of least square is plotted on Figure \ref{Fig:Least Square}.

It is evident that by increasing the SNR the RMSE decrease. This is because by increasing the SNR, the error in distance measurements will be reduced and therefore a better estimation of the unknown node location. When the SNR is 1 dB the RMSE was 36.52m and as the SNR increases the RMSE decreases reaching 0.9457m at 12 dB.
\begin{figure} [h]
        \centering
        \includegraphics[width=1\textwidth,clip]{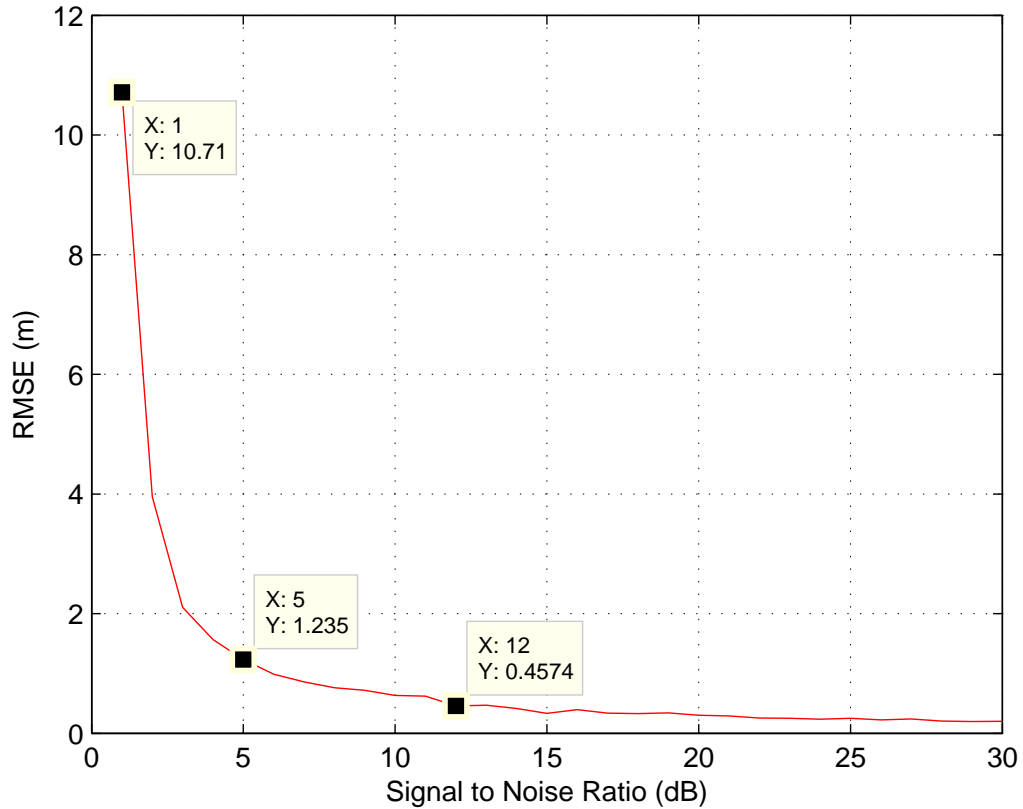}
        \caption[RMSE for different values of SNR using WLS]{RMSE for different values of SNR using WLS}
		 \label{Fig:Weighted Least Square}
\end{figure}

Comparing the two Figures \ref{Fig:Least Square} , \ref{Fig:Weighted Least Square} we noticed that at the same SNR value, WLS indicates lower RMSE value compared to that in LS, that is at SNR=12, RMSE in WLS is 0.4574m and in LS 0.9457m. Actually, a closer look in equation 3.4, we can notice that the RSS does not vary linearly on the distance between nodes, thus, the same error will result in larger errors in process of estimating the distance, especially, in long distance between the nodes. More precisely, the accuracy in the estimation relies on the distance between the nodes \cite{tarrio_weighted_2011}.  LS treats different distance lengths equally, which can be noticed from equation 3.18, so the error will be high. However, WLS takes into account the different lengths by using the weighting matrix, as in equation 3.19, which gives high weight for the short distances, thus better accuracy. This matches with the simulation results, which show the dramatic reduction in RMSE for WLS compared to that in LS at the same value for SNR.     

\begin{figure} [H]
        \centering
        \includegraphics[width=1\textwidth,clip]{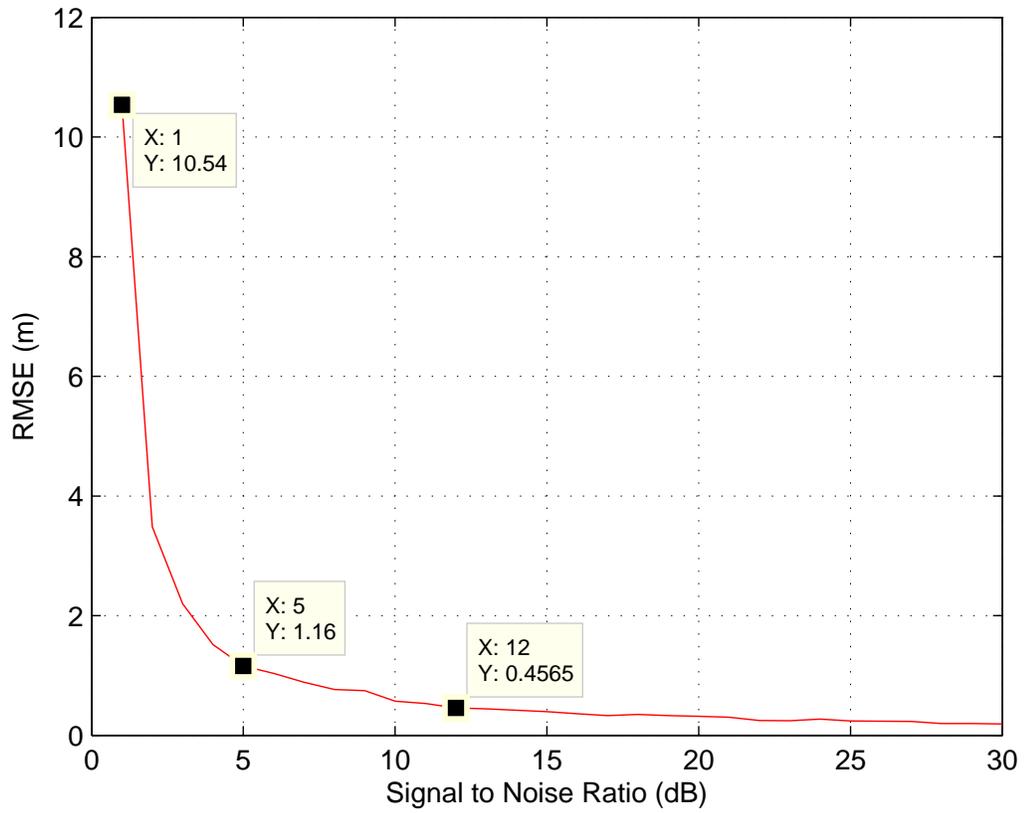}
        \caption[RMSE for different values of SNR using Huber Robustness]{RMSE for different values of SNR using Huber Robustness}
		 \label{Fig:IRLS_L1}
\end{figure}

The third estimator being used is the Huber robustness. From the simulation results in Figure \ref{Fig:IRLS_L1} it can be seen that the best performance achieved using the Huber robustness is the same as the WLS and can not achieve a better performance. This is because the noise is Gaussian and the best estimator to deal with a Gaussian noise is an $\ell_2$-norm and more specifically WLS.

\begin{figure} [H]
        \centering
        \includegraphics[width=1\textwidth,clip]{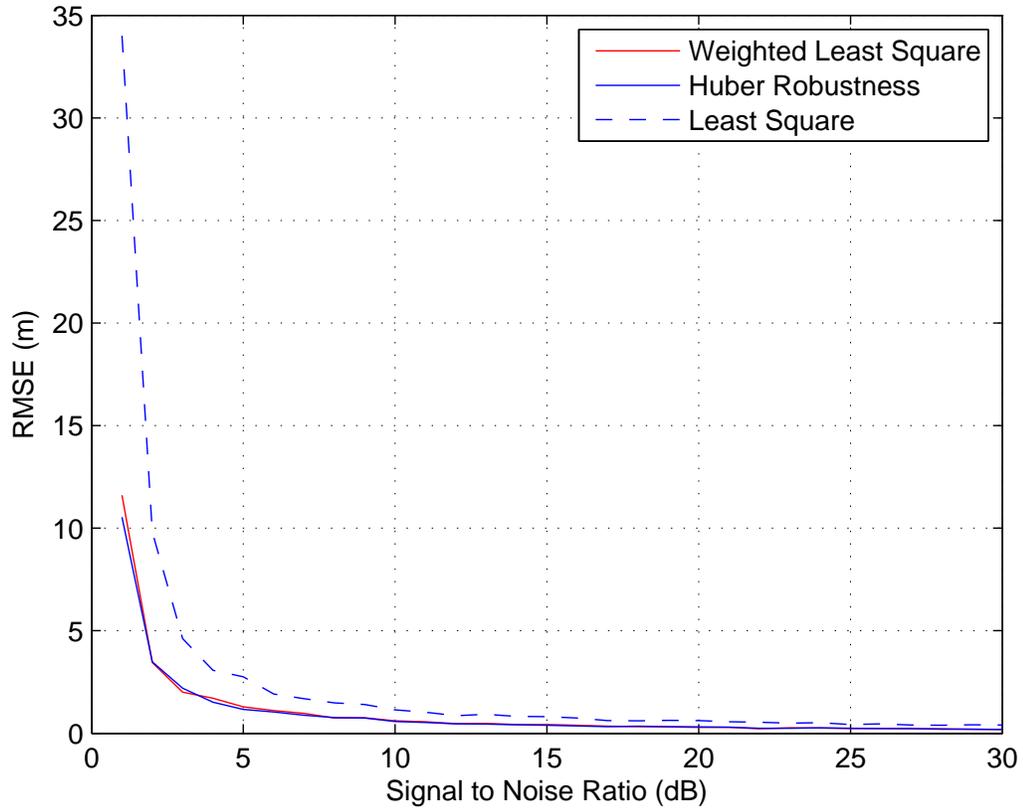}
        \caption[RMSE for different values of SNR (least,weighted least square, and huber robustness)]{RMSE for different values of SNR (least,weighted least square, and huber robustness)}
		 \label{Fig:Three_Envi1}
\end{figure}

It is clear that the WLS and Huber robustness achieves the optimal performance in comparison to the least square as shown in Figure \ref{Fig:Three_Envi1}. The performance of the WLS and Huber robustness is almost 5 times better than the least square. This is because small distance estimations have more significant effect on the final estimated location than large distance estimations.

\begin{figure} [H]
        \centering
        \includegraphics[width=1\textwidth,clip]{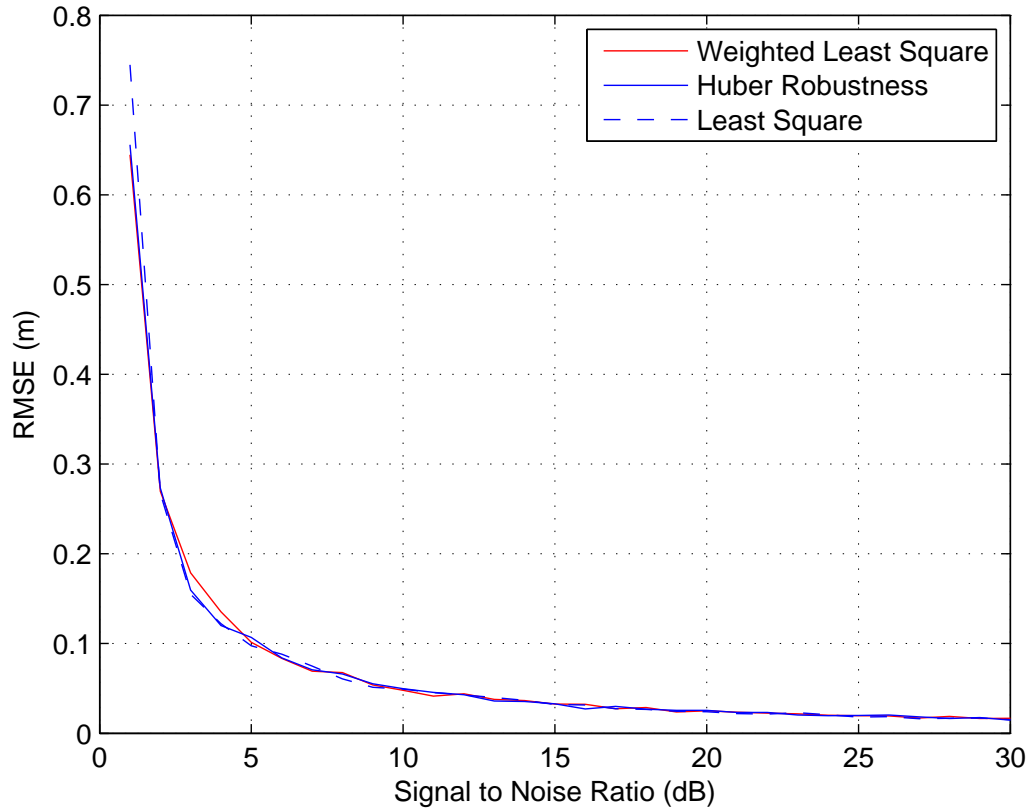}
        \caption[RMSE for different values of SNR (least,weighted least square, and huber robustness)]{RMSE for different values of SNR (least,weighted least square, and huber robustness)}
		 \label{Fig:Three_Envi2}
\end{figure}

However, by making the distance between the anchor nodes and unknown node almost equal the performance of the three estimators is almost the same as shown in Figure \ref{Fig:Three_Envi2}.

The performance is almost the same because the weight for each anchor node will be almost the same because the distance between nodes is almost equal. Therefore, the advantage of WLS and Huber over LS will vanish in this condition.

\subsection{Robustness}
\begin{figure} [h]
        \centering
        \includegraphics[width=1\textwidth,clip]{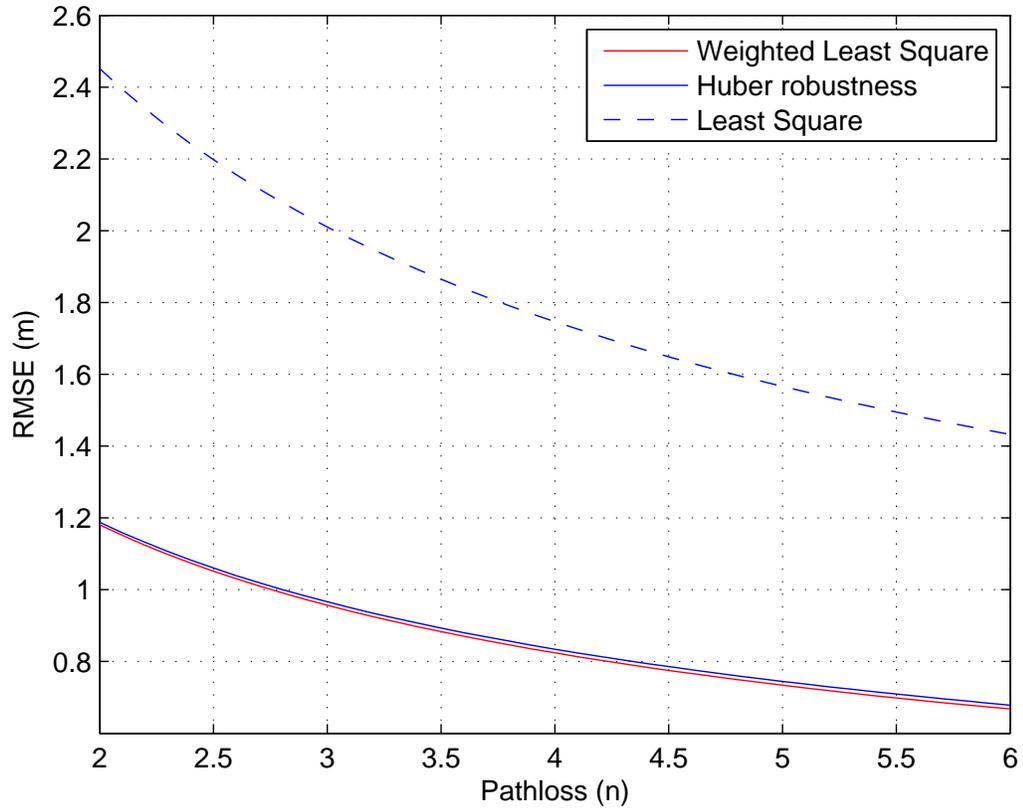}
        \caption[Robustness of the algorithm to changes to in the pathloss exponent]{Robustness of the algorithm to changes to in the pathloss exponent}
		 \label{Fig:Three_robustness}
\end{figure}

In the previous test, the environment was exactly the same as the environment used to develop the pathloss model. However, in a real situation it is very likely that there will be some changes in the environment. For example, changes in furniture locations or moving peeople in the area. When changes happen in the environment, the pathloss model does not resemble the real environment. Therefore, it is important to test the performance of the algorithm to changes in the pathloss exponent as shown in Figure \ref{Fig:Three_robustness}.

It can be seen clearly from Figure \ref{Fig:Three_robustness} that for a lower pathloss exponent the RMSE is larger and this is because the distance estimated will be larger and so a higher error is associated with it.However, for the higher pathloss exponent the distance estimated is smaller and so lower error associated with it.Moreover, it can be seen that the Huber robustness and the WLS is less sensitive to changes in the pathloss exponent compared to the least square.
\chapter{Direction of Arrival}
\label{Direction of Arrival}
In this chapter, two conventional array geometries namely Uniform Linear Array (ULA) and Uniform Circular Array (UCA) are analysed to be used with DOA-based algorithms. Then, the importance of received signals to be uncorrelated is highlighted in type of signal section as well as how to de-correlate correlated signals using Phase mode Excitation, Spatial Smoothing and Teopiltz algorithm.  After that, the DOA-based algorithm are introduced which includes MUSIC, Root-MUSIC, UCA-Root-MUSIC, ESPRIT and UCA-ESPRIT. Lastly, the numerical results are provided to show the performance of DOA-based algorithm and DOA-based localization.

\section{Conventional Sensor Array Configurations}
For the DOA, antenna arrays are recommended as they can detect multiple signals at the same time, which is a quality that directional antenna lacks. Based on this type of antennas, several geometrical conventional sensor array configurations can be used to perform the estimation of DOA. The most adopted configurations are known as Uniform Linear Array (ULA) and the Uniform Circular array (UCA). 
\subsection{Uniform Linear Array (ULA)}
\begin{figure} [h]
       \centering
        \includegraphics[width=0.8\textwidth,clip]{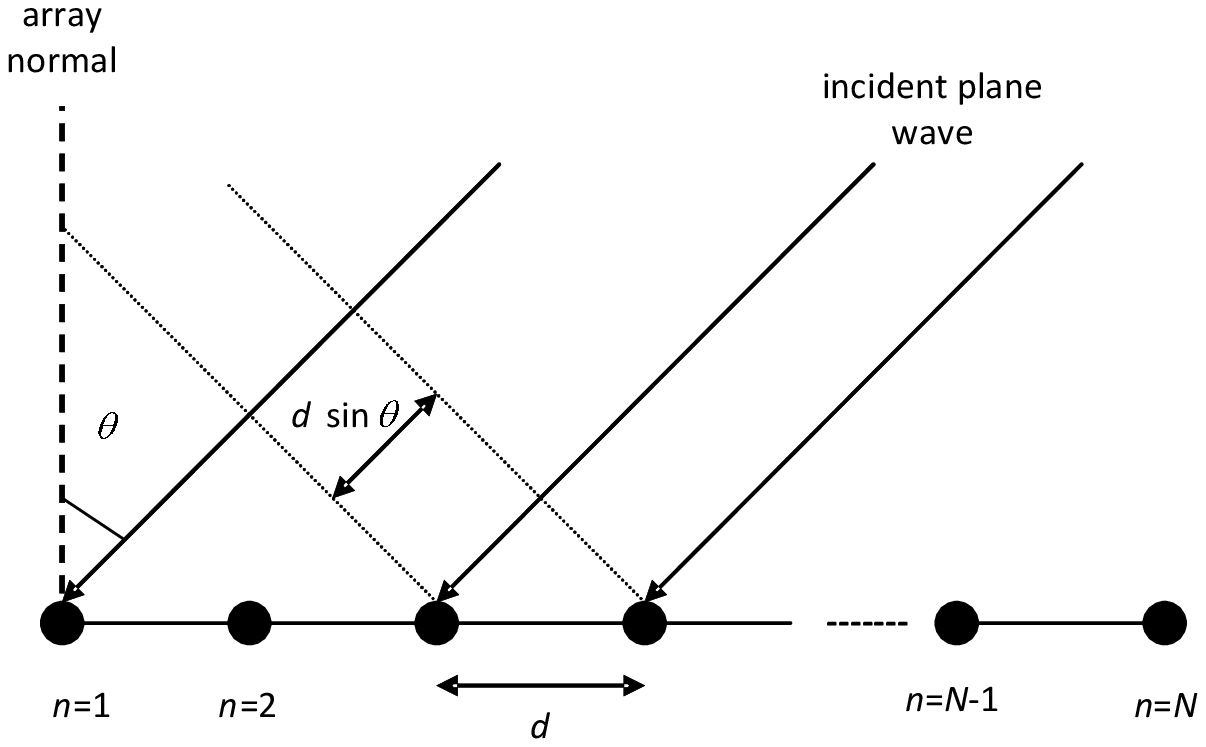}
        \caption[Geometry of N-elements (ULA)]{Geometry of N-elements (ULA)}
		 \label{Fig:linear_array}
\end{figure}

The Uniform Linear Array (ULA) is one of the adopted configurations of DOA, as shown in Figure \ref{Fig:linear_array}. In it, a set of $N$ sensors (antennas) is being scattered along a single dimensional line. The sensors must maintain uniform and equal distances between them, often, the sensors used in such configuration are known as omni-directional sensors. In this scenario, those sensors receive $M$ numbers of narrowband signals $s_m(t)$, where $1 \leq m \leq M$. The angle of which the signal originates from is different from the others. Those angles are known as azimuth angles $\theta_m$. The $N$-dimensional received data vector \textbf{x} at any time $t$ is given by \cite{oumar_comparison_2012}:

\begin{equation}%
\boldsymbol{x}(t)=\sum\limits^{M}_{m=1}\boldsymbol{\mathrm{a}}_s(\theta_m)s_m(t)+\boldsymbol{\mathrm{n}}(t)=\boldsymbol{\mathrm{A}}_s(\theta)\boldsymbol{s}(t)+\boldsymbol{\mathrm{n}}(t)
\label{4.1}
\end{equation}%
where “\textbf{n}” is a noise vector modelled as white and zero-mean complex Gaussian, “$\boldsymbol{\mathrm{A}}_s(\theta)$” is a matrix which consist of $M$ steering vectors, and “$\boldsymbol{\mathrm{{a}}_s}(\theta_m)$” is the steering vector, which corresponds to the DOA of the m$^{th}$ signal, defined as:

\begin{equation}%
\boldsymbol{\mathrm{a}}_s(\theta_m)=\left[\begin{array}{lllll}1&e^{-j\phi_m}&e^{-2j\phi_m}&...&e^{-j(N-1)\phi_m}\end{array}\right]^T
\label{4.2}
\end{equation}%
where $\phi_m$ represents the phase shift between the elements of the sensor array and is expressed as:

\begin{equation}%
\phi_m=2\pi \left(\frac{d}{\lambda})\sin(\theta_m)\right)
\label{4.3}
\end{equation}%

\subsection{Uniform Circular Array (UCA)}
\begin{figure} [h]
       \centering
        \includegraphics[width=0.6\textwidth,clip]{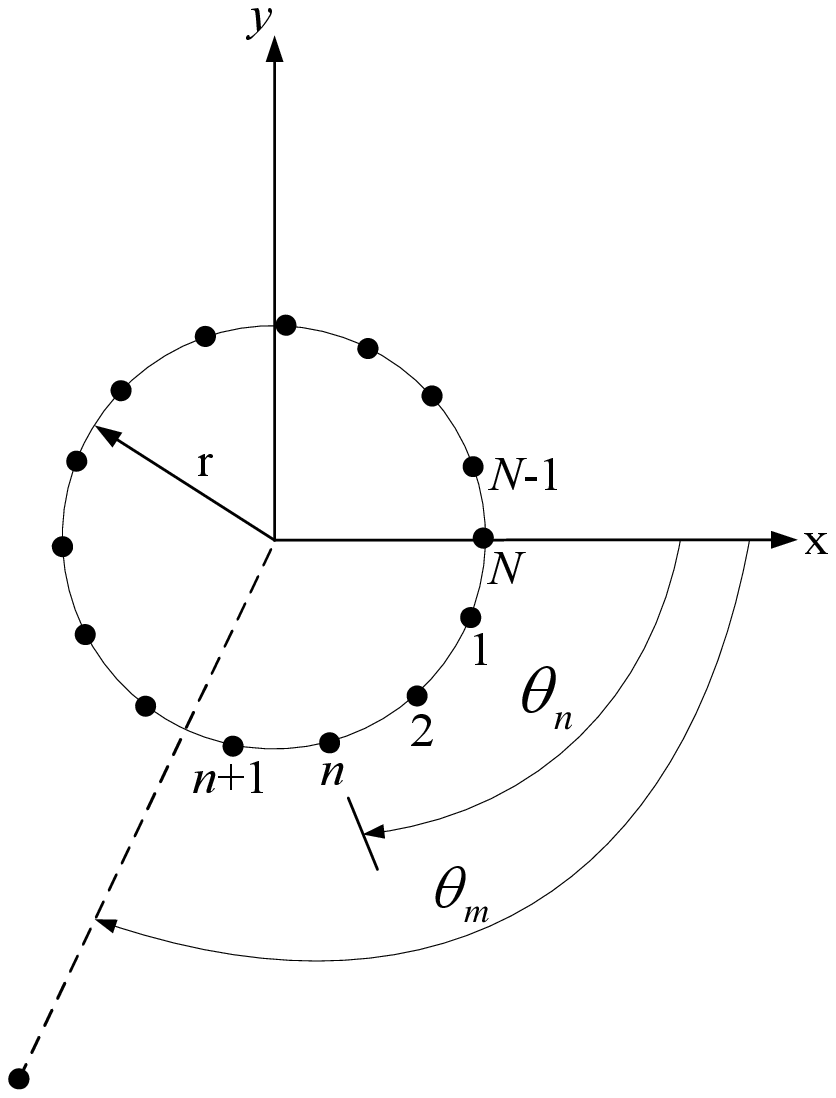}
        \caption[Top view of the N-element circular array in x-y plane]{Top view of the N-element circular array in x-y plane}
		 \label{Fig:ciruclar array}
\end{figure}

The Uniform Circular Array (UCA) is another adopted configuration of DOA, shown in Figure \ref{Fig:ciruclar array}. In it, a set of sensors (receivers) are being scattered in two dimensions, $x$-$y$ plane. To perform this configuration, let us assume a set of $N$ numbers of sensors. Those sensors are placed in a circular shape, making a ring that has a radius of $r$. It is crucial to maintain a uniform and equal distances between the sensors, $d$, along the ring circumference. It was established that the $N$-dimensional received data vector \textbf{x} at any time $t$ is given by equation \ref{4.4} \cite{tran-xuan_calibration_2009,ravindra_time_2013}.

The steering vector of a circular array is expressed by the following:

\begin{equation}%
\boldsymbol{\mathrm{a}}_s(\theta_m)=\left[\begin{array}{l}e^{j(\frac{2\pi r}{\lambda})\cos(\theta_m-\theta_n)}\end{array}\right]^T;1\leq n\leq N
\label{4.4}
\end{equation}%

where $\theta_n$ is the angular location of each element and is calculated using:

\begin{equation}%
\theta_n=2\pi \left(\dfrac{n-1}{N}\right)
\label{4.5}
\end{equation}%

\section{Type of Signals}
In communication systems and signal processing, a signal is defined as the function that carries data or conveys information about the attribute or behavior of some phenomenon. \textit{IEEE Transactions on Signal Processing} stretches the definition of the word “signal” to include all video, audio, speech, image, radar, sonar, communication, musical, geophysical or even medical signals \cite{priemer_introductory_1991}. Signals can be classified based on many different criteria. For example, in the physics world, signals can be categorized based on their exhibited variations in time or space or both. In this project, since we are considering practical environment, the classification that is mostly relevant is based on the correlation of the signals. In other words, the signals are cataloged into correlated signals and uncorrelated signals. 

\subsection{Uncorrelated Signals}
The definition of the uncorrelated signals comes from probability theory and statistics. In statistics, two real-valued random variables “A” and “B” are said to be uncorrelated if and only if their covariance is equal to zero. i.e.

\begin{equation}%
\mathrm{E(AB)-E(A)E(B)=0}
\label{4.6}
\end{equation}%

Consequently, if the two variables are considered uncorrelated, there cannot be linear relation between them. For example, uncorrelated white noise refers to the face that no two points in the time domain of the noise is associated with each other. This also means that the noise value at time ($t$+1) cannot be predicted if noise at time ($t$) is known \cite{bain_introduction_2000}.
In the detection of DOA, when an uncorrelated signal is in use the method of detection is quite straightforward to perform. The standard MUSIC, Root-MUSIC, ESPRIT algorithms, or any other type of algorithm of detection can be performed and used. That is due to the fact that every signal is being treated as a separate individual entity. In other words, each signal has its unique patterns and trends. It will be discussed with more details in later chapters, but it is important to highlight that MUSIC algorithm, for example, can detect up to $N$-1 uncorrelated signals without problems.

\subsection{Correlated Signals}
\label{Correlated Signals}
The other type of signals that we are considering in this project is known as the correlated signal. Generally speaking, correlation is a mathematical operation that is very similar to convolution. It can be define as the degree of association between two random variables. For example, when speaking about two graphs of two set of data, the correlation between them refers to the degree of resemblance between them. Unlike the non-correlation, the covariance of two inputs must not necessarily equal to zero to be called correlated. i.e.

\begin{equation}%
\mathrm{E(AB)-E(A)E(B)\neq 0}
\label{4.7}
\end{equation}%

It is important to highlight that correlation is not the same as causality. Causality set the relation between to events by making one the cause of the other (i.e. cause and effect).  Cross correlation is the correlation between two different signals. It also can be between a signal and itself, in which it is called auto-correlation \cite{smith_scientist_1997}.”

In the detection of DOA, correlated signal(s) cannot be detected by standard methods. Standard usage of the methods, mentioned earlier, will result into a wrongful interpretation of the data. To avoid such interpretations, new methodology must be introduced to the standard detection methods such as spatial smoothing, phase mode excitation, and Toeplitz algorithm.  

\subsubsection{Phase Mode Excitation}
Practically, uniform circular array is of a great importance when it comes to estimate the DOA of incident signals. Compare to ULA, UCA provides a complete coverage for all azimuthal angles as well as a uniform performance in detecting them. However, UCA falls behind ULA in the way its steering vector is designed. ULA has a steering vector of Vandermonde structure. This structure is unique as it allows efficient computational algorithms like Root-MUSIC and de-coherent techniques to be implemented. For UCA, it does not possess Vandermonde structure due to the dependence of UCA steering vector with the sensor angular location.

To overcome such limitation, phase mode excitation is used. Phase mode excitation is a beamforming technique to convert the steering vector of UCA into a virtual steering vector characterized by a Vandermonde structure. In other word, the UCA is mapped and converted into a virtual ULA. Thus, all methods applicable to ULA can now be employed to UCA indirectly through the phase mode excitation \cite{belloni_beamspace_2006}.

To comprehend the concept behind the phase mode excitation, we first need to extend our UCA modeling from 2-D into 3-D, where an elevation angle $\theta_e$ is introduced as shown in the Figure 4.3.

\begin{figure} [h]
       \centering
        \includegraphics[width=0.6\textwidth,clip]{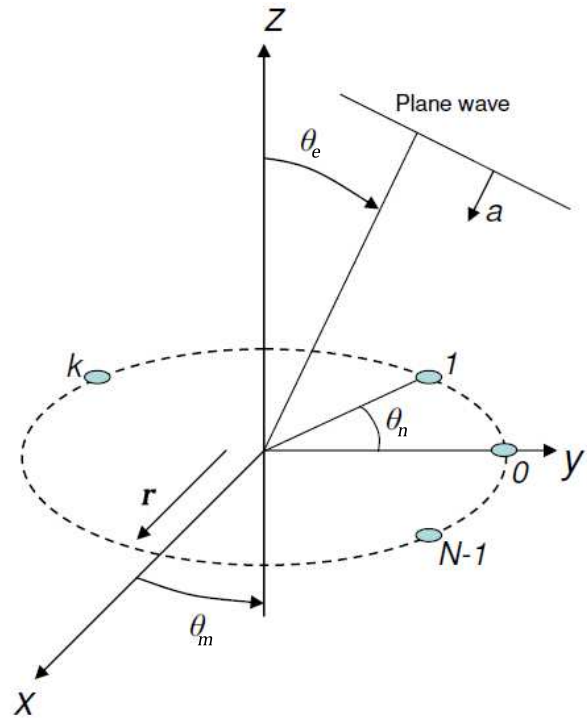}
        \caption[Array geometry for UCA including the elevation angle ]{Array geometry for UCA including the elevation angle }
		 \label{Fig:elevation angle}
\end{figure} 

In the previous UCA model discussed in section 4.1.2, the elevation angle was assumed to be fixed at $90^o$ for all incident signals. In the new UCA model, we will keep the value of elevation angle constant for any arbitrary incident signals but the actual angle can vary from $0^o$ up to $90^o$.

Consequently, the resulting steering vector for UCA having $N$ elements with radius $r$ and receiving a narrowband signal arriving from azimuth angle  $\theta_m\in[\begin{array}{l}-\pi~,~\pi\end{array}]$ and $\theta_e\in[\begin{array}{l}0~,~\frac{\pi}{2}\end{array}]$ is modeled as:  

\begin{equation}%
\boldsymbol{\mathrm{a}}_s(\theta_m)=\left[\begin{array}{l}e^{j(\frac{2\pi r}{\lambda})\sin(\theta_e)\cos(\theta_m-\theta_n)}\end{array}\right]^T,~1\leq n\leq N 
\label{4.8}
\end{equation}%

The elevation angle $\theta_e$ is measured from the incident signal down to the z-axis while the azimuth angle $\theta_m$ is measured from the projected vector of the incident signal to the x-axis in a clockwise direction. As $\theta_e$  is assumed to be fixed for all received signals, we can simplify equation \ref{4.8} by defining $\zeta=(\frac{2\pi r}{\lambda})\sin(\theta_e)$. Now, the elevation angle $\theta_e$ is dependent through $\zeta$ and the equation \ref{4.8} become: 

\begin{equation}%
\boldsymbol{\mathrm{a}}_s(\theta_m)=\left[\begin{array}{l}e^{j\zeta \cos(\theta_m-\theta_n)}\end{array}\right]^T,~1\leq n\leq N
\label{4.9}
\end{equation}%

\paragraph{Phase Mode Excitation Principle} \mbox{}\\
The basic idea of phase mode excitation is about applying a beamforming vector to transform the beam pattern of UCA into a beam pattern similar to that of ULA and, simultaneously, reserving the original steering direction of UCA. To carry out this beamforming, we first need to produce and visualize the array pattern of ULA. This procedure is done to give us an idea of how to synthesis the phase mode excitation beamformer.

Then, the phase mode excitation beamformer will be derived for continuous circular array (CCA) and extended to uniform circular array (UCA). 

In case of CCA, there is an infinite number of sensors arranged over ring, so the effect inter-element spacing in this scenario is ignored. Undoubtedly, CCA cannot be realized practically. However, from theoretical point of view, it considered as an ideal free-error scenario for employing the phase mode excitation beamformer. Then, the beamformer will be employed on the UCA which represents a practical model. At the latter model, we will understand why a quantization error arisein the UCA geometry \cite{belloni_beamspace_2006}.

\paragraph{Array Pattern of Uniform Linear Array (ULA)} \mbox{}\\
The Array pattern of an array can be obtained using beamforming technique. Beamforming is simply a spatial filtering operation where weights are assigned to the array output followed by a summing process to obtain a beam. Assume \textbf{x}($t$) to be the array output at time $t$ and $\boldsymbol{w}\mathrm{^H}=[w_0,w_1,\cdots,w_{(N-1)}]$  is the corresponding beamforming weight vectors assigned for each $N$ element. The equation of beam output \textbf{$\boldsymbol{y}$}($t$) is given as \cite{chandran_advances_2005}:

\begin{equation}%
\boldsymbol{y}(t)=\boldsymbol{w}^\mathrm{H}\boldsymbol{x}(t)
\label{4.10}
\end{equation}%

As we are interested in obtaining the beam pattern of an array, \textbf{$\boldsymbol{x}$}($t$) is replaced by $\boldsymbol{\mathrm{a}}_s(\theta)$. That is because the array response of an array to a source signal impinging from arbitrary angle $\theta$ is the steering vector to that angle. The elements in the steering vector $\boldsymbol{\mathrm{a}}_s(\theta)$ are separated by equally phase shifts. Thus, for $N$ elements array, its beam pattern equation is defined as

\begin{equation}%
\boldsymbol{\mathrm{f}}(\theta)=\boldsymbol{w}^\mathrm{H}\boldsymbol{a}_s(\theta)=\sum\limits^{N-1}_{n=0}w_n\boldsymbol{\mathrm{a}}_s(\theta)~~-90^o\leq\theta\leq90^o
\label{4.11}
\end{equation}%

The array pattern \textbf{f}($\theta$)  in equation \ref{4.11} is given in complex-value form. By taking the absolute value, the array pattern  $\left| \boldsymbol{\mathrm{f}}(\theta)\right|$ usually follow an oscillatory function with main-lobe and side-lobes. The main-lobe corresponds to the pass-band of the beamformer so signals received at main-lobe pass unattenuated. Similarly, signals received at side-lobes will be severely attenuated. Thus, beamformer weights can be selected to make the steered array ‘look’ in a specific direction.   

In cophasal beamforming, the weights $\boldsymbol{w}^H$ are assigned to steer the array to the desired direction $\theta_m$ [11]. This can be simply achieved by assigning $\boldsymbol{w}^\mathrm{H}=\dfrac{\boldsymbol{a}_s(\theta_m)^H}{N}$ so signal coming from $\theta_m$ will be unattenuated. The weights are divided by $N$ for normalized purposes. Thus, the array pattern under cophasal beamformer for a signal received at azimuth angle $\theta_m$ is given as: 

\begin{equation}%
\boldsymbol{\mathrm{f}}(\theta)=\frac{\boldsymbol{\mathrm{a}}_s(\theta_m)\boldsymbol{\mathrm{a}}_s(\theta)}{N},~-90^o\leq\theta\leq90^o
\label{4.12}
\end{equation}%

The array pattern of ULA under cophasal beamformer is determined by substituting $\boldsymbol{\mathrm{a}}_s(\theta)$ and $\boldsymbol{\mathrm{a}}_s(\theta_m)$  in equation \ref{4.12} with their equivalent quantity in ULA given in  equation \ref{4.2}. let $v=\frac{(2d)sin(\theta)}{\lambda}$, so $\boldsymbol{\mathrm{a}}_s(v)$ and $\boldsymbol{\mathrm{a}}_s(v_m)$ can be written as:

\begin{equation}%
\boldsymbol{\mathrm{a}}_s(v)=\left[\begin{array}{llll}1&e^{j\pi v}&...&e^{j(N-1)\pi v}\end{array}\right]^T
\label{4.13}
\end{equation}%

\begin{equation}%
\boldsymbol{\mathrm{a}}_s(v_m)=e^{j\pi v_m}
\label{4.14}
\end{equation}%

By substituting $\boldsymbol{\mathrm{a}}_s(v)$ and $\boldsymbol{\mathrm{a}}_s(v_m)$ in equation \ref{4.12}, the resulting array pattern for ULA becomes:

\begin{equation}%
 \mathrm{f}(v)=e^{j(N-1)\frac{\pi}{2}(v-v_m)}\frac{\sin(N\frac{\pi}{2}(v-v_m))}{N~\sin(\frac{\pi}{2}(v-v_m))}~,-1\leq \mathrm{v}\leq1
\label{4.15}
\end{equation}%

\begin{figure} [h]
       \centering
        \includegraphics[width=0.6\textwidth,clip]{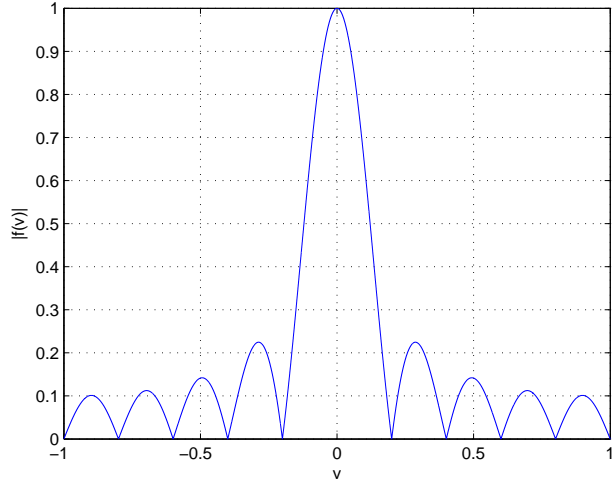}
        \caption[Array pattern of ULA with 10 elements receiving a signal from $\theta_m$=0]{Array pattern of ULA with 10 element receiving a signal from $\theta_m$=0}
		 \label{Fig:Array pattern of ULA}
\end{figure}

To visualize the beam pattern of ULA, let us consider a narrowband signal impinging on ULA from $\theta_m=0^o$   with $N$=10 and inter-element spacing of $d$=0.5$\lambda$. These values will result in $v_m=\frac{2d sin(\theta_m)}{\lambda}$. Thus, by substituting $v_m$ in equation \ref{4.15} and plotting the result for $-1\leq v\leq1$, we obtain the ULA array pattern as shown in Figure \ref{Fig:Array pattern of ULA}. 

Clearly from Figure \ref{Fig:Array pattern of ULA}, the cophasal beamformer has steered the ULA into direction $\theta_m=0^o$ as the main-lobe of the beam pattern is centered at $v$=0 which correspond to $\theta_m=0^o$.

\paragraph{Phase of Excitation of Continuous Circular Array (CCA)} \mbox{}\\
From the previous section, we noticed that a desired beam pattern of ULA is obtained by applying weights directly on the individual elements. Unfortunately, this methodology cannot be used with circular array due to the existence of angular position. However, there exists a power approach to overcome this problem. A beamformer weight called excitation function is applied with angular argument $\theta_n$. In the excitation function, the phase of the weight are assigned in a linear manner based on the angular location of an element to a reference element (array element 0). The linear phase increase of the excitation function weights, is done in a similar pattern compared to the cophasal excitation of a ULA. As $\theta_n$ is periodic with a period of $2\pi$ , then any excitation function $w(\theta_n)$ can be expressed using Fourier series with each of its harmonic is termed with the phase mode $m$. The general representation for the excitation function $w(\theta_n)$ is given as \cite{belloni_beamspace_2006}:

\begin{equation}%
w(\theta_n)=\sum\limits^{\infty}_{p=-\infty}c_{p}e^{jp\theta_n}
\label{4.16}
\end{equation}%
where $c_p$  is the Fourier series coefficient for the corresponding  m$^{th}$ phase mode obtained as:

\begin{equation}%
c_p=\frac{1}{2\pi}\int\limits^{2\pi}_{0}w(\theta_n)e^{-jp\theta_n}d\theta_n
\label{4.17}
\end{equation}%

The excitation function for the $p^{th}$ phase mode is modeled as $w_p=e^{(jp\theta_n)}$ which simply represents a spatial harmonic for the generalized excitation function $w(\theta_n)$. To obtain the normalized far-field of $p^{th}$ phase mode, $w_p$ is multiplied with circular array steering vector for each element followed by summation process. As we are dealing with CCA, then the summation approach integral taken from 0 to 2$\pi$. The resulting normalized far-field of phase mode $p$ is given as:
 
\begin{equation}%
f_p^c(\theta)=\frac{1}{2\pi}\int\limits^{2\pi}_{0}w(\theta_n)e^{-j\zeta \cos(\theta_m-\theta_n)}d\theta_n
\label{4.18}
\end{equation}%
where $\theta=(\zeta ,\theta_m)$ represents the angle of received signal in term of both azimuth and elevation angles while superscript $c$ donates continuous array. 

Using Bessel function property, $f_p^c(\theta)$ can be expressed in a compact form as:

\begin{equation}%
f_p^c(\theta)=j^pJ_p(\zeta)e^{jp\theta_m}
\label{4.19}
\end{equation}%

where $J_p(\zeta)$ is a first kind Bessel function of order $p$. 

By analyzing equation \ref{4.19}, the following observation is deduced:
\begin{itemize}
 \item	In the far-field pattern  $f_p^c(\theta)$, the resulting azimuthal variation given by  $e^{jp\theta_m}$ have the same form as the excitation function $e^{jp\theta_n}$.  This property of phase mode excitation is considered as the basis for pattern synthesis to be employed with UCA.
 \item 	The effect of amplitude and elevation in far-field pattern  $f_p^c(\theta)$ is determined through the Bessel function. This relationship imposes a limit on the number of modes to be synthesized. 
 As the visual region for $\theta_e$ is bounded by $0\leq\theta_e\leq\dfrac{\pi}{2}$ , its corresponding $\zeta$ will be bounded within $0\leq\zeta\leq \dfrac{2\pi r}{\lambda}$. For a given Bessel function $J_p(\zeta)$, its amplitude becomes small as the order $p$ exceed the argument $\zeta$. Thus, the highest order $h$ to excite a mode with a reasonable strength is given as
\begin{equation}%
h\approx\dfrac{2\pi r}{\lambda}
\label{4.20}
\end{equation}%

This implies that the total excitation modes for CCA is $p\in[-h,h]$.
The following example will demonstrate the validity of equation \ref{4.20}. Considering a CCA with $r=\lambda$, then equation \ref{4.20} suggests that the maximum order to be excited is $h=6$. This is clear from Figure \ref{Fig:Bessal} where  $J_7 (\zeta)$ corresponds to mode 7 is very small in the visible region $0 \leq \zeta \leq 2\pi$  and hence can be safely ignored. Therefore, we conclude that the phase modes in the range $ p \in [-6,6]$ considered enough to be excited by CCA with a reasonable strength.

\begin{figure} [H]
       \centering
        \includegraphics[width=1\textwidth,clip]{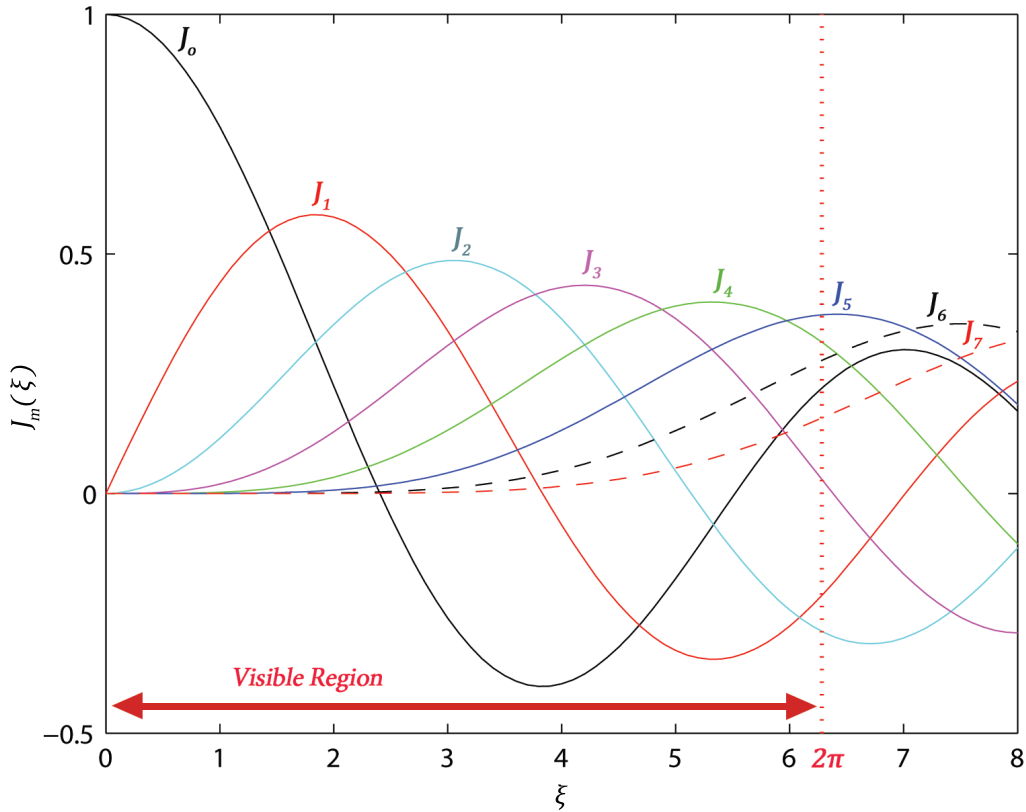}
        \caption[Bessel functions for  $J_0(\zeta)$  up to $J_7(\zeta)$  where $0 \leq \zeta \leq 2\pi$]{Bessel functions for  $J_0(\zeta)$  up to $J_7(\zeta)$  where $0 \leq \zeta \leq 2\pi$}
		 \label{Fig:Bessal}
\end{figure} 

 \end{itemize} 
\paragraph{Phase of Excitation of Uniform Circular Array (UCA)} \mbox{}\\
As UCA is a sampled version of CCA, its beamforming vector $w_m$  is also a sampled version of that for the CCA. For a UCA with $N$ elements, its normalized beamforming vector to excite an array with phase mode $p$, where $p\in[-h,h]$, is modeled as:

\begin{equation}%
\boldsymbol{\mathrm{w}}_p=\frac{1}{N}\left[\begin{array}{llll}1&e^{-\frac{j2\pi p}{N}}&...&e^{-\frac{j2\pi p}{N}}\end{array}\right]
\label{4.21}
\end{equation}%

The beam pattern of UCA is obtained by applying the normalized beamforming vectors on the UCA steering vector given as \cite{belloni_beamspace_2006}: 

\begin{equation}%
f^s_p(\theta)=\boldsymbol{w}_p^H\boldsymbol{a}_s(\theta)=\sum\limits^{N-1}_{n=0}e^{jp\theta_n}e^{-j\zeta \cos(\theta_m-\theta_n)}
\label{4.22}
\end{equation}%
where superscript $s$ donates a sampled array

After mathematical simplification explained in \cite{mathews_signal_1994}, equation \ref{4.22} is written as:

\begin{equation}%
f^s_p(\theta)=j^pJ_p(\zeta)e^{jp\theta_m}+\sum\limits^{\infty}_{q=1}(j^{g_1}J_g(\zeta)e^{-jg2\theta_m}+j^hJ_h(\zeta)e^{jh\theta_m})=j^mJ_m(\zeta)e^{jm\theta_m}+\epsilon_m
\label{4.23}
\end{equation}%
where variable $\epsilon_m$ represents the induced error which is the summation term, while indexes $g_1$ and $g_2$ corresponds to $g_1=N_q-p$ and $g_2=N_q-p$. 

By analyzing equation \ref{4.23}, the following observation is deduced:
\begin{itemize}
\item 	There are two main terms in equation (4.23), the first term is called the principle term which is identical to far-field pattern obtained from CCA. The second term is the residual term $\epsilon_p$ generated due to sampling CCA with $N$ elements to form the UCA.

\item 	The exact value of the error $\epsilon_p$ is not constant but it varies depending on the DOA of the received signal. We conclude that $\epsilon_p$  follows an exponential decrease as $\theta_e$ is lowered from $90^o$ to $0^o$. However, $\epsilon_p$ follows a sinusoidal function as with the changes in $\theta_m$.  



\item In order to make UCA far-field pattern close the ideal CCA case, $\epsilon_p$  must be minimized .This can be achieved by setting $N$ to be far larger than the any excitation mode which is given by the relationship  $N>2\left|p \right|$ [11]. As $h$ is the largest possible phase mode to be excited, then N  must meet the criteria $N>2h$ to make
\begin{equation}%
f^s_p(\theta)=j^pJ_p(\zeta)e^{jp\theta_m}
\label{4.24}
\end{equation}%

\end{itemize}

\subparagraph{Transformation matrix based on Phase mode excitation on UCA} \mbox{}\\
From our analysis on phase mode excitation, the steering vector of virtual array $\boldsymbol{\mathrm{a}}_{sv}(\theta)$ can be constructed to contain all individual elements $e^{(jp\theta_m)}$ specified by $f_p^s(\theta)$ given in equation \ref{4.24} for $p\in[-h,h]$. Mathematically, the virtual steering vector will have the following form:

\begin{equation}%
\boldsymbol{a}_{sv}(\theta)=\left[\begin{array}{llll}e^{-jm\theta_m}&...&1&e^{jm\theta_m}\end{array}\right]^T
\label{4.25}
\end{equation}%
where superscript $v$ donates a virtual array

To obtain $\boldsymbol{\mathrm{a}}_{sv}(\theta)$, all the beamformer weights $\boldsymbol{\mathrm{w}}_{p}^H$ are collected to form a matrix called \textbf{F} of size $(2h+1)\times N$ which given as:

\begin{equation}%
\boldsymbol{F}=\frac{1}{N}=\left[\begin{array}{ccccc}1&w^{-h}&w^{-2h}&\cdots&w^{-(N-1)h}\\\vdots&\vdots&\vdots&\cdots&\vdots\\1&w^{-1}&w^{-2}&\cdots&w^{-(N-1)}\\1&1&1&\cdots&1\\1&w^1&w^2&\cdots&w^{(N-1)}\\\vdots&\vdots&\vdots&\cdots&\vdots\\1&w^h&w^{2h}&\cdots&w^{(N-1)h}\end{array}\right]
\label{4.26}
\end{equation}%
where $w=e^{j2\pi/N)}$. The $\boldsymbol{\mathrm{F}}$ matrix will provide us with all possible $f_p^s(\theta)$ for $p\in[-h,h]$. However, as we are only interested with the azimuth angle $\theta_m$ provided in  $f_p^s (\theta)$, we need to get rid of the term  $j^pJ_p(\zeta)$   by introducing a $(2h+1) \times (2h+1)$ diagonal matrix $\boldsymbol{\mathrm{J}}$ defined as:

\begin{equation}%
\boldsymbol{\mathrm{J}}=\mathrm{diag}({\dfrac{1}{j^pJ_p(\zeta)}}), where -h \leq p \leq h
\label{4.27}
\end{equation}%

By multiplying \textbf{F} matrix with \textbf{J}, we obtain a transformation matrix $\boldsymbol{T}_v$  to map the element of UCA into the elements of the virtual uniform linear array (VULA). The transformation matrix $\boldsymbol{T}_v$ will have the size of $(2h+1) \times N$ and it is defined as:

\begin{equation}%
\boldsymbol{\mathrm{T}}_v=\boldsymbol{\mathrm{FJ}}
\label{4.28}
\end{equation}%

By applying $\boldsymbol{\mathrm{T}}_v$ on the steering matrix of UCA where $N$ $\geq 2h$ , the steering matrix of the VULA will become:

\begin{equation}%
\boldsymbol{\mathrm{A}}_{sv}=\boldsymbol{\mathrm{T}}_v\boldsymbol{\mathrm{A}}_s=\left[\begin{array}{ccc}e^{-jh\theta_1}&\ldots&e^{jh\theta_m}\\\vdots&\ddots&\vdots\\e^{jh\theta_1}&\cdots&e^{jh\theta_m}\end{array}\right] 
\label{4.29}
\end{equation}%

It is clear that $\boldsymbol{\mathrm{A}}_{sv}$ has the same vandermonde structure as in the ULA. When implementing techniques like UCA-Root-MUSIC and UCA-ESPRIT, we will need to transform the whole data received by UCA into that of the VULA. In a similar manner to $\boldsymbol{\mathrm{A}}_{sv}$, the received data of VULA is given as:

\begin{equation}%
\boldsymbol{\mathrm{x}}_v(t)=\boldsymbol{\mathrm{T}}_v\boldsymbol{\mathrm{x}}(t)
\label{4.30}
\end{equation}%

\subsubsection{Spatial Smoothing Techniques}
As explained earlier, signals can be either correlated or uncorrelated. In practical environment, however, incident signals are mostly correlated on the sensor array (i.e. signals that have similar pattern or trend during the time of observation). As it was established earlier, the signal covariance matrix ($\boldsymbol{\mathrm{R}}_{s}$=$s(t)s(t)^H$), influences the performance of the DOA algorithms. That is because the correlation matrix loses its non-singularity. A method proposed to overcome the effect of correlation on incident signals is known as “Spatial Smoothing”. This technique proposes decomposing the sensor array into smaller subarrays. The methodology of Spatial Smoothing is derived for ULA but it can be extended to UCA using phase mode excitation. Specifically, this technique is categorized into two types: Forward Spatial Smoothing (FSS), Forward/Backward Spatial Smoothing (FBSS) \cite{pillai_forward/backward_1989,shan_spatial_1985}. 

\begin{figure} [h]
       \centering
        \includegraphics[width=0.6\textwidth,clip]{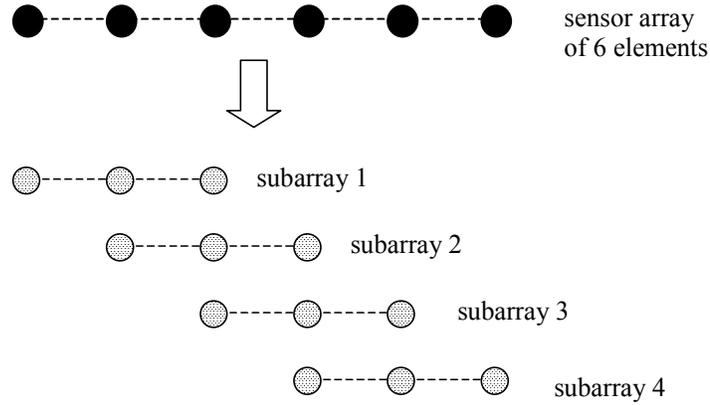}
        \caption[FSS spatial smoothing]{FSS spatial smoothing}
		 \label{Fig:FSS spatial smoothing}
\end{figure}

\paragraph{Forward Spatial Smoothing (FSS)} \mbox{}\\
This technique proposes dividing the sensor array into overlapping sub-arrays. This shall introduce phase shifts between them and so resolves the problem of the correlated incident signals. For the FSS, let us consider an array sensor of 6 elements as shown in Figure \ref{Fig:FSS spatial smoothing}. Those elements are to be divided into 4 overlapping sub-arrays of length 3 each ( $L_{ss}$=4, $p_{ss}$=3). The incident signal, thus, modeled as the following \cite{pillai_forward/backward_1989}.

\begin{equation}%
x^F_k=\boldsymbol{\mathrm{AD}}^{(k-1)}s(t)+n_k(t)
\label{4.31}
\end{equation}%
where ($k$-1) denotes the $k^{th}$ power of the diagonal matrix \textbf{D}:

\begin{equation}%
\boldsymbol{\mathrm{D}}=diag\left[\begin{array}{cccc}e^{-j\frac{2\pi}{\lambda}\sin(\theta_1)}&\ldots&\ldots&e^{-j\frac{2\pi}{\lambda}\sin(\theta_M)}\end{array}\right]
\label{4.32}
\end{equation}%

\begin{figure} [h]
       \centering
        \includegraphics[width=0.6\textwidth,clip]{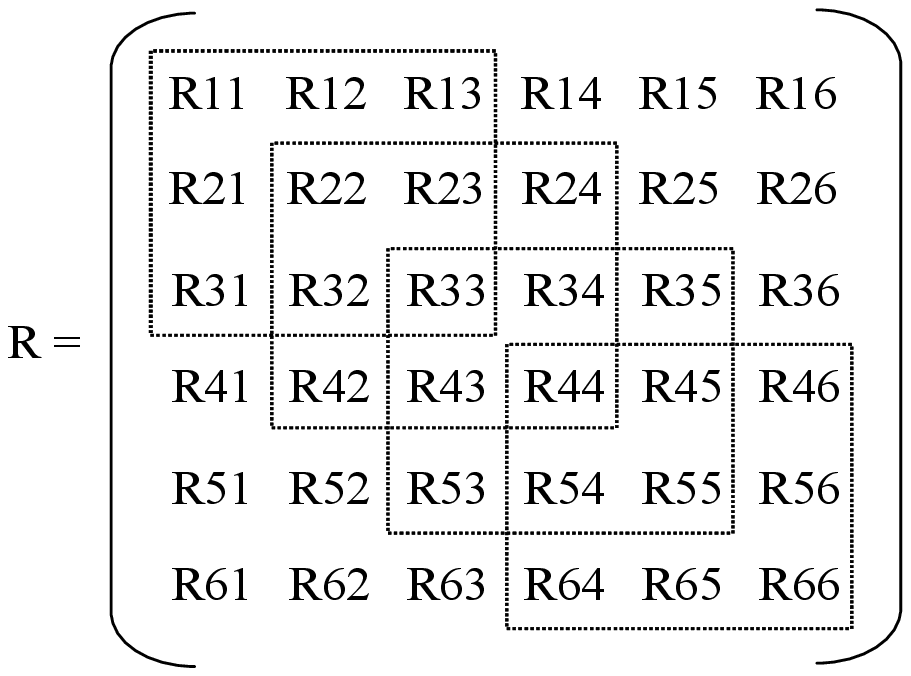}
        \caption[Applying FSS on Matrix R]{Applying FSS on Matrix R}
		 \label{Fig:R_matrix}
\end{figure}

The spatial covariance matrix \textbf{R} can be modeled as the covariance matrices of the forward sub-arrays as the following, Figure \ref{Fig:R_matrix}:

\begin{equation}%
\boldsymbol{\mathrm{R}}=\frac{1}{L_{ss}}\sum\limits^{L_{ss}-1}_{K=0}\boldsymbol{\mathrm{R^F}}_k
\label{4.33}
\end{equation}%

The division of the sensor arrays cannot be done randomly. There are some rules to be followed to obtain the optimum results of the FSS:
\begin{itemize}
\item The first rule implies that the number of the sub-arrays must be greater than the number of elements of the diagonal matrix \textbf{D}. 
\begin{equation}%
L>M\triangleq(N-p_{ss}+1)>M
\label{4.34}
\end{equation}%
\item The second rule implies that the number of elements in each sub-array must be greater than the number of elements of the diagonal matrix \textbf{D} yet less than $N$.
\begin{equation}%
N>p_{ss}>M
\label{4.35}
\end{equation}%
\end{itemize}

Combining the rules results in the following conclusion: the minimum value of “$p_{ss}$” can be obtained at $M_{max}$+1.

Substituting that in the equations leads to the following conclusion: the maximum number of correlated signals that may be detected by FSS method is equivalent to $N/2$. The number of uncorrelated signals that can be detected by conventional MUSIC algorithm, which will be discussed later, is $(N-1)$. It can be observed that this number of correlated signals that can be detected is less than the number of uncorrelated signals that can be detected by conventional MUSIC algorithm.

\begin{figure} [h]
       \centering
        \includegraphics[width=0.6\textwidth,clip]{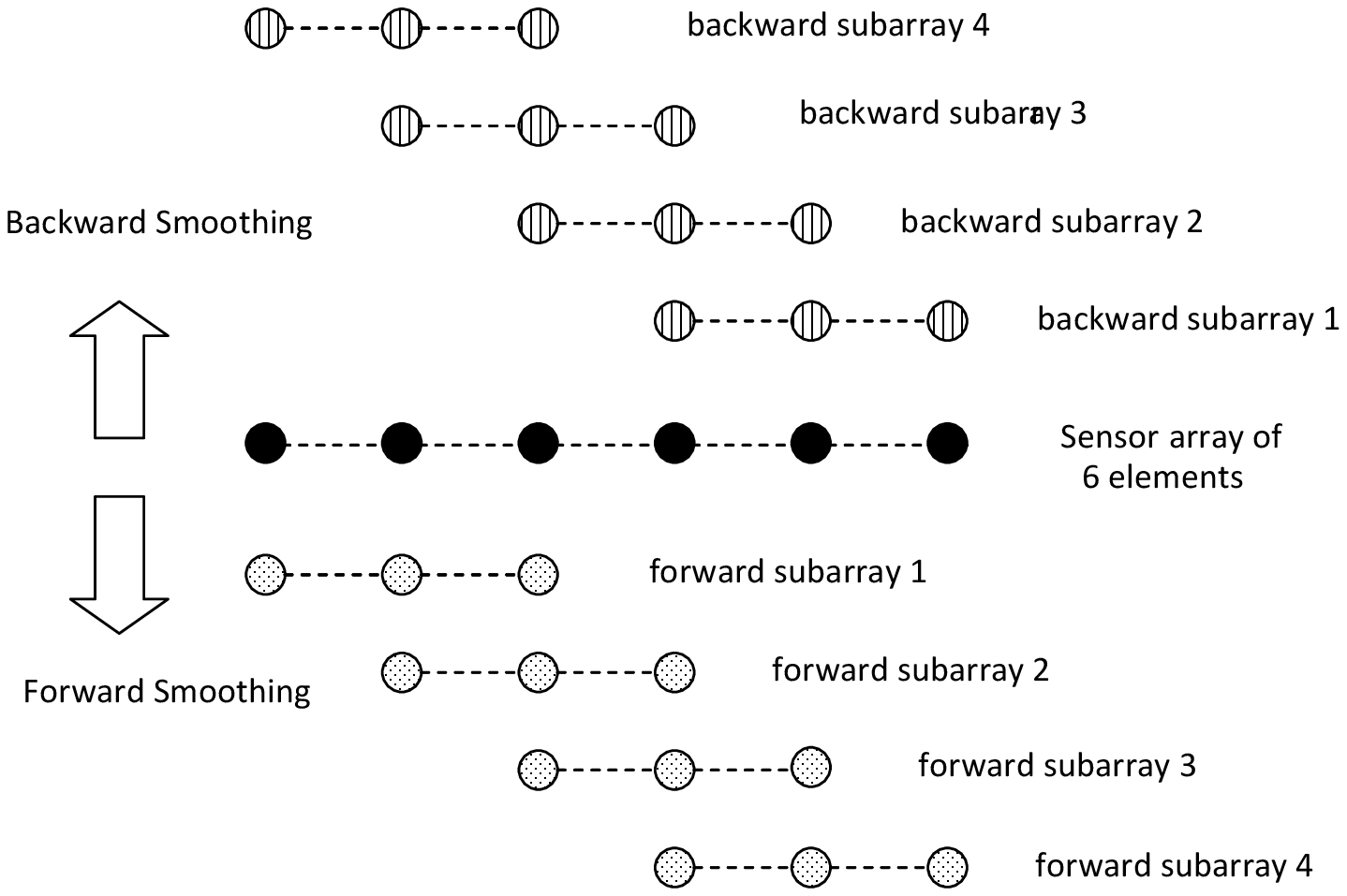}
        \caption[Forward/Backward spatial smoothing]{Forward/Backward spatial smoothing}
		 \label{Fig:FBSS}
\end{figure}

\paragraph{Forward/Backward Spatial Smoothing (FBSS)} \mbox{}\\
This technique proposes dividing the sensor array into overlapping sub-arrays. This shall introduces phase shifts between them. This will solve the problem of the correlated incident signals; it aims to increase the number of detectable correlated signal significantly from $N/2$ to $2N/3$. The principle of this technique is quite simple. It works by utilizing the principle of conjugate of the forward spatial smoothing (FSS). In other words, using a set of forward sub-arrays and their conjugate,as shown in Figure \ref{Fig:FBSS} \cite{pillai_forward/backward_1989}.

Figure \ref{Fig:FBSS} shows the principle of FBSS applied on a sensor array of 6 elements. The 6-element array is divided into 4 overlapping forward sub-arrays and 4 overlapping backward sub-arrays (i.e. $L_F=4$ and $L_B=4$). Each sub-array is of size 3 (i.e. $p_{ss}=3$). The received signals vector in the case of the FBSS at the k$^{th}$ backward sub-array is given as the following \cite{pillai_forward/backward_1989,_wiley_yes}:

\begin{equation}%
x^B_k=\boldsymbol{\mathrm{AD}}^{(k-1)}[\boldsymbol{\mathrm{D}}^{(N-1)}s(t)]^*+n_k(t)
\label{4.36}
\end{equation}%

And the spatial covariance matrix \textbf{R} in this case can be calculated by the following equation:

\begin{equation}%
\boldsymbol{\mathrm{R}}=\frac{\boldsymbol{\mathrm{R^F}}+\boldsymbol{\mathrm{R^B}}}{2}
\label{4.37}
\end{equation}%
where $\boldsymbol{\mathrm{R^F}}$ represents the average covariance matrix in the case of the forward sub-array vectors, and $\boldsymbol{\mathrm{R^B}}$ resembles the average covariance matrix of the backward sub-array vectors. 
The rules of dividing the sensor array into sub-array in the FSS case are also applicable in the case of the FBSS. That is 
\begin{enumerate}
\item The number of the sub-arrays must be greater than the number of elements of the diagonal matrix \textbf{D}.
\item The number of elements in each sub-array must be greater than the number of elements of the diagonal matrix \textbf{D} yet less than $N$.
\end{enumerate}

It is important to highlight that these rules put restraints upon the size of $p_{ss}$ and the number of sub-array $L_{ss}$.  In other words, for a successful detection of correlated signals, these rules and constrains must be met or else the detection will be inaccurate. If we go back to the example of the sensor array of elements, the rules suggest that the system will fail to detect three signals if the antenna array is divided into $L_{ss}$=2 overlapping sub-array with each having 5 elements ($p_{ss}$=5). While, it will succeed if it is divided into three sub-array ($L_{ss}$=3) with each having 4 elements in ($p_{ss}$=4).

It is significant to mention that the FSS may detect up to  $N/2$ correlated signals whereas the FBSS may detect up to $2N/3$ correlated signals. This means that the FBSS will succeed in detecting signals that FSS will fail at if the range of the signals is above $N/2$ yet less than $2N/3$. None of them though is comparable to the MUSIC algorithm when detecting uncorrelated signal since it can detect up to $N-1$ uncorrelated signal \cite{kulaib_efficient_2014,pillai_forward/backward_1989}. 

\subsubsection{Toeplitz Algorithm}
When all received signals are correlated, the resulting covariance matrix become singular. That is because the rank of the covariance matrix is diminished from $N$ to 1. A method to overcome rank deficient is by applying spatial smoothing technique on the covariance matrix. This method as previous explained is just a rank reconstruction allowing $N/2$ and $2N/3$ of coherent signals to be detected using FSS and FBSS. However, the spatial smoothing method is done by dividing the main array into subarrays and the smoothed covariance is obtained by averaging the individual covariance matrix of each subarrays. This clearly implies that the de-correlation performance of spatial smoothing is done at the cost of reducing the size of the main array. 

Unlike spatial smoothing, the Toeplitz algorithm does not reduce the size of the main array. Precisely, the Toeplitz algorithm is implemented by constructing a Toeplitz matrix from the covariance matrix of the correlated source. The generated Toeplitz matrix is a diagonal matrix of rank $N$. Therefore, Toeplitz algorithm provides a full de-correlation for received coherent sources allowing $N$-1 coherent signals to be detected \cite{zhou_new_2011}.

To explain the modeling of Toeplitz algorithm, we will begin by the mathematical model of the singular covariance matrix $\boldsymbol{\mathrm{R}}_c$ of correlated signals. Then, we will show how the Toeplitz algorithm construct a non-signular matrix $\boldsymbol{\mathrm{R}}_T$ from  $\boldsymbol{\mathrm{R}}_c$.

\paragraph{Modeling of Singular Covariance Matrix} \mbox{}\\
In a coherent environment, all the signals impinging on ULA will have the same phase.
Assuming $M$ received correlated signal with amplitude $\rho_m$ where $m=\left[\begin{array}{l}1,2,\ldots,M\end{array}\right]$. The signal covariance matrix is realized as:

\begin{equation}%
\boldsymbol{\mathrm{R}}_{s}=\left[\begin{array}{cccc}\rho^2_1&\rho_1\rho_2&\ldots&\rho_1\rho_M\\\rho_2\rho_1&\rho^2_2&\cdots&\rho_2\rho_M\\\vdots&\vdots&\ddots&\vdots\\\rho_M\rho_1&\rho_M\rho_2&\cdots&\rho^2_M\end{array}\right]
\label{4.38}
\end{equation}%

Consequently, the covariance matrix $\boldsymbol{\mathrm{R}}_c$ of the ULA is given as  

\begin{equation}%
\boldsymbol{\mathrm{R}}_c=\boldsymbol{\mathrm{A}}_s^H\boldsymbol{\mathrm{R}}_s\boldsymbol{\mathrm{A}}_s=\left[\begin{array}{cccc}\boldsymbol{\mathrm{R}}_{c(1,1)}&\boldsymbol{\mathrm{R}}_{c(1,2)}&\ldots&\boldsymbol{\mathrm{R}}_{c(1,N)}\\\boldsymbol{\mathrm{R}}_{c(2,1)}&\boldsymbol{\mathrm{R}}_{c(2,2)}&\ldots&\boldsymbol{\mathrm{R}}_{c(2,N)}\\\vdots&\vdots&\ddots&\vdots\\\boldsymbol{\mathrm{R}}_{c(N,1)}&\boldsymbol{\mathrm{R}}_{c(N,2)}&\ldots&\boldsymbol{\mathrm{R}}_{c(N,N)}\end{array}\right]
\label{4.39}
\end{equation}%
Due to the multiplication of $\boldsymbol{\mathrm{R}}_s$ with \textbf{A} and its complex conjugate $\boldsymbol{\mathrm{A}}^H$, $\boldsymbol{\mathrm{R}}_{(c(i,j))}$ will represent the complex conjugate of $\boldsymbol{\mathrm{R}}_{(c(j,i))}$. Hence $\boldsymbol{\mathrm{R}}_c$ can be simplified as:

\begin{equation}%
\boldsymbol{\mathrm{R}}_c=\boldsymbol{\mathrm{A}}_s^H\boldsymbol{\mathrm{R}}_s\boldsymbol{\mathrm{A}}_s=\left[\begin{array}{cccc}\boldsymbol{\mathrm{R}}_{c(1,1)}&\boldsymbol{\mathrm{R}}_{c(1,2)}&\ldots&\boldsymbol{\mathrm{R}}_{c(1,N)}\\\boldsymbol{\mathrm{R}}^*_{c(2,1)}&\boldsymbol{\mathrm{R}}_{c(2,2)}&\ldots&\boldsymbol{\mathrm{R}}_{c(2,N)}\\\vdots&\vdots&\ddots&\vdots\\\boldsymbol{\mathrm{R}}^*_{c(N,1)}&\boldsymbol{\mathrm{R}}^*_{c(N,2)}&\ldots&\boldsymbol{\mathrm{R}}_{c(N,N)}\end{array}\right]
\label{4.40}
\end{equation}%

With the coherence between received signals,  $\boldsymbol{\mathrm{R}}_c$ becomes a singular matrix. In a singular matrix, any row can be written as linear combination of other rows and the same thing is applied to the column. In order to de-correlate $\boldsymbol{\mathrm{R}}_c$, we need to remove the linear combination relationship  among the rows and the columns. 

\paragraph{Realization of Toeplitz Algorithm} \mbox{}\\
By using the Toeplitz algorithm, the linear combination relation is overcome by taking the first row and column in matrix $\boldsymbol{\mathrm{R}}_c$, while omitting the rest of the elements. Then, the taken elements are used to construct a diagonal matrix that is non-singular.  Due to the fact,that the first row in $\boldsymbol{\mathrm{R}}_c$ is a complex conjugate for the first column, we only need to obtain the actual values of the first row. By using the steering vector $\boldsymbol{\mathrm{a}}_s(\theta_m)$ of ULA given in equation \ref{4.2} the first row elements of  $\boldsymbol{\mathrm{R}}_c$ are given through vector $\boldsymbol{\mathrm{V}}_c$ as \cite{chen_cumulants-based_2014}:

\begin{equation}%
\boldsymbol{\mathrm{V}}_C=\left[\begin{array}{c}\boldsymbol{\mathrm{R}}_{C(1,1)}\\\boldsymbol{\mathrm{R}}_{C(1,2)}\\\vdots\\\boldsymbol{\mathrm{R}}_{C(1,N)}\end{array}\right]=\left[\begin{array}{c}\rho_1\alpha+\cdots+\rho_M\alpha\\\rho_1\alpha e^{-j\Phi_1}+\cdots+\rho_M\alpha e^{-j\Phi_m}\\\vdots\\\rho_1\alpha e^{-j(N-1)\Phi_1}+\cdots+\rho_M\alpha e^{-j(N-1)\Phi_m}\end{array}\right]
\label{4.41}
\end{equation}%
where $\alpha$ is a constant defined as $\alpha$=$\rho_1$+$\cdots$+$\rho_M$  
 
Also, vector $\boldsymbol{\mathrm{R}}_c$ can be realized a multiplication of two matrix given as:
\begin{equation}%
\boldsymbol{\mathrm{V}}_C=\left[\begin{array}{c}\boldsymbol{\mathrm{R}}_{C(1,1)}\\\boldsymbol{\mathrm{R}}_{C(1,2)}\\\vdots\\\boldsymbol{\mathrm{R}}_{C(1,N)}\end{array}\right]=\left[\begin{array}{cccc}1&1&\ldots&1\\e^{-j\Phi_1}&e^{-j\Phi_2}&\cdots&e^{-j\Phi_m}\\\vdots&\vdots&\ddots&\vdots\\e^{-j(N-1)\Phi_1}&e^{-j(N-1)\Phi_2}&\cdots&e^{-j(N-1)\Phi_M}\end{array}\right]\left[\begin{array}{c}\rho_1\\\rho_2\\\vdots\\\rho_M\end{array}\right]\alpha
\label{4.42}
\end{equation}%

\begin{equation}%
\boldsymbol{\mathrm{V}}_C=\boldsymbol{\mathrm{A}}_s\boldsymbol{\mathrm{S}}_C
\label{4.43a}
\end{equation}%

Lastly, based on the values specified in vector $\boldsymbol{\mathrm{V}}_c$ , Toeplitz matrix $\boldsymbol{\mathrm{R}}_T$ is constructed to replace the original  covariance matrix $\boldsymbol{\mathrm{R}}_c$ and its value is modeled as \cite{zhou_new_2011,chen_cumulants-based_2014}:

\begin{equation}%
\boldsymbol{\mathrm{R}}_T=\left[\begin{array}{cccc}\boldsymbol{\mathrm{R}}_{C(1,1)}&\boldsymbol{\mathrm{R}}_{C(1,2)}&\cdots&\boldsymbol{\mathrm{R}}_{C(1,N)}\\ \boldsymbol{\mathrm{R}}^*_{C(1,2)}& \boldsymbol{\mathrm{R}}_{C(1,1)}&\ldots& \boldsymbol{\mathrm{R}}_{C(1,N-1)}\\\vdots&\vdots&\ddots&\vdots\\ \boldsymbol{\mathrm{R}}^*_{C(1,N)}& \boldsymbol{\mathrm{R}}^*_{C(1,N-1)}&\ldots& \boldsymbol{\mathrm{R}}_{C(1,1)}\end{array}\right]=\boldsymbol{\mathrm{A}}_s^H\widetilde{\boldsymbol{\mathrm{R}}_S}\boldsymbol{\mathrm{A}}_s
\label{4.43b}
\end{equation}%
where $\widetilde{\boldsymbol{\mathrm{R}}_s}$  is a diagonal matrix containing the values of $\boldsymbol{\mathrm{R}}_{(c(1,1))}$. Hence, 

\begin{equation}%
\widetilde{\boldsymbol{\mathrm{R}}_S}=\mathrm{diag}\{\alpha\rho_1,\ldots,\alpha\rho_M\}
\label{4.43c}
\end{equation}%
By analyzing $\boldsymbol{\mathrm{R}}_T$ , the following observation is deduced:

\begin{itemize}
 \item	Toeplitz algorithm can completely resolve correlated signal as $\boldsymbol{\mathrm{R}}_T$ is a diagonal matrix of rank $N$. In other words, Toeplitz algorithm allows $N$-1 coherent signal to be detected compared to $N/2$ and $2N/3$ detected coherent signals using FSS and FBSS. 
 \item 	Toeplitz algorithm enhances the power of received signals from  $\rho_m^2$ to $\alpha\rho_m=(\rho_1+\rho_2+\ldots+\rho_M)\rho_m$. This will lead to a more sharp peaks on MUSIC angular spectrum compared to spatial smoothing indicating that Toeplitz algorithm has more robust performance \cite{zhou_new_2011}. Due to this superior performance, the Toeplitz algorithm can even detect a mixture of correlated and uncorrelated signals up to $N$-1.
  \item The computation load of Toeplitz algorithm is less compared to spatial smoothing technique as it has a diagonal structure
 \end{itemize} 

\section{DOA Algorithms}
DOA algorithm are used to provide an estimation of the DOA for incident signals impinging on a sensor array. A type of DOA algorithm which provides a high-resolution and accurate DOA estimation are known as subspace-based algorithms. Such techniques operate on the input covariance matrix which can be decomposed into eigenvalues and eigenvectors belong to the signal subspace while the reset belong to the noise subspace. The reason that subspace-based algorithm adopted this name is due to the existence of both noise and signal subspace surrounding the sensor array. In our project, we will focus on three most efficient subspace-based techniques for DOA estimation which are MSUC, Root-MSUIC and ESPRIT. The latter two technique are only applicable to ULA so phase mode excitation is used to obtain their equivalent version in UCA namely UCA-root-MUSIC and UCA-ESPRIT.

\subsection{MUSIC}
The MUSIC algorithm is one of the earlier technique proposed by Schmidt \cite{schmidt_multiple_1986} to offer a high-resolution detection for incoming signal DOA impinging on the sensor array. The main principle of MUSIC algorithm is based on exploiting the eigenstructure of the covariance matrix \textbf{R}, which is practically calculated by taking $K$ snapshots of the incident signal x($t$)  followed by avenging process over $K$. Thus, the covariance matrix, is expressed as: 

\begin{equation}%
\boldsymbol{\mathrm{R}}=\frac{1}{K}\sum\limits^{K^{}}_{t=1}\boldsymbol{\mathrm{x}}(t)\boldsymbol{\mathrm{x}}(t)^H
\label{4.44}
\end{equation}%
To obtain the full equation of covariance matrix \textbf{R}, the signal model \textbf{x}($t$) in equation \ref{4.1} is substituted into \ref{4.44}, which yields:  

\begin{equation}%
\boldsymbol{\mathrm{R}}=\boldsymbol{\mathrm{A}}_s\boldsymbol{\mathrm{R}}_{s}\boldsymbol{\mathrm{A}}_s^H+\sigma^2_n\boldsymbol{\mathrm{I}}
\label{4.45}
\end{equation}%
where $\boldsymbol{\mathrm{R}}_{s}$ is the signal covariance matrix, $\sigma^2_n$ is the noise variance and \textbf{I} is an identity matrix of dimension $N \times N$ \cite{AlArdi_Smart_2005}.

The eigenvalues of \textbf{R} can be expressed as $\{\gamma_1,~\gamma_2,~\gamma_3,~\ldots ~,\gamma_N\}$ which are obtained using the equation:

\begin{equation}%
\left| \boldsymbol{\mathrm{R}}-\gamma_i\boldsymbol{\mathrm{I}}\right|=0
\label{4.46}
\end{equation}%

Similarly, the eigenvectors of \textbf{R} is expressed as $\{q_1~q_2~q_3~\ldots ~q_N\}$ and they must satisfy the following condition:

\begin{equation}%
(\boldsymbol{\mathrm{R}}-\gamma_i\boldsymbol{\mathrm{I}})\boldsymbol{\mathrm{q}}_i=0
\label{4.47}
\end{equation}%

Using equation \ref{4.45}, equation \ref{4.46} can be extended as:

\begin{equation}%
\left| \boldsymbol{\mathrm{A}}_s \boldsymbol{\mathrm{R}}_{s} \boldsymbol{\mathrm{A}}_s^H+\sigma^2_n\boldsymbol{\mathrm{I}}-\gamma_i \boldsymbol{\mathrm{I}}\right|=\boldsymbol{\mathrm{A}}_s\boldsymbol{\mathrm{R}}_{s}\boldsymbol{\mathrm{A}}_s^H-(\gamma_i-\sigma^2_n)\boldsymbol{\mathrm{I}}
\label{4.48}
\end{equation}%

Based on the definition of eigenvalues, we conclude that the terms $\boldsymbol{\mathrm{A}}_s\boldsymbol{\mathrm{R}}_s \boldsymbol{\mathrm{A}}_s^H$ has $\gamma_i-\sigma^2_n$ eigenvalues.

The matrix $\boldsymbol{\mathrm{A}}_s$ contains a linearly independent steering vector signals with dimension $N \times M$. Thus, to perform eigen decomposition on \textbf{R}, signal covariance matrix  $\boldsymbol{\mathrm{R}}_{s}$ must be non-singular which is guaranteed as long as the incident signals are uncorrelated and the incoming signals $M$ is less than the elements $N$ in the sensor array. Based on this condition, it is concluded that the term $\boldsymbol{\mathrm{A}}_s\boldsymbol{\mathrm{R}}_s\boldsymbol{\mathrm{A}}_s^H$ has zero eigenvalues for $(N-M)$. Thus, $N$ eigenvalues of \textbf{R} can be sorted into signal eigenvalues with largest \textbf{M} eigenvalues while the remaining $(N-M)$ eigenvalues corresponds to the noise variance $\sigma^2_n$ \cite{oumar_comparison_2012}.

The key secret behind MUSIC algorithm is that the noise subspace eigenvectors $\boldsymbol{\mathrm{V}}_n$ associated with the noise eigenvalues are orthogonal to the steering vectors making up the matrix $\boldsymbol{\mathrm{A}}_s$. This can be proved by modeling the noise eigenvectors, which corresponds to $(N-M)$ eigenvalues as:  

\begin{equation}%
(\boldsymbol{\mathrm{R}}-\sigma^2_n\boldsymbol{\mathrm{I}})\boldsymbol{\mathrm{q}}_i=\boldsymbol{\mathrm{A}}_s\boldsymbol{\mathrm{R}}_{s}\boldsymbol{\mathrm{A}}_s^H\boldsymbol{\mathrm{q}}_i+\sigma^2_n\boldsymbol{\mathrm{Iq}}_i-\sigma^2_n\boldsymbol{\mathrm{Iq}}_i=0
\label{4.49}
\end{equation}%

 \begin{equation}%
\boldsymbol{\mathrm{A}}_s\boldsymbol{\mathrm{R}}_s\boldsymbol{\mathrm{A}}_s^H\boldsymbol{\mathrm{q}}_i=0
\label{4.50}
\end{equation}%

As $\boldsymbol{\mathrm{R}}_{s}$ is non-singular, the only way for $\boldsymbol{\mathrm{A}}_s\boldsymbol{\mathrm{R}}_s\boldsymbol{\mathrm{A}}_s^H\boldsymbol{\mathrm{q}}_i=0$ to be zero is by setting $\boldsymbol{\mathrm{A}}_s^H\boldsymbol{\mathrm{q}}_i$=0 which proves the orthogonality between the noise eigenvectors and matrix $\boldsymbol{\mathrm{A}}_s$. 

This allows the MUSIC angular spectrum to be expressed as:
 \begin{equation}%
P(\theta)=\frac{\boldsymbol{\mathrm{A}}_s(\theta)^H\boldsymbol{\mathrm{A}}_s(\theta)}{\boldsymbol{\mathrm{A}}_s^H(\theta)\boldsymbol{\mathrm{V}}_n\boldsymbol{\mathrm{V}}^H_n\boldsymbol{\mathrm{A}}_s(\theta)}
\label{4.51}
\end{equation}%

The orthogonality between $\boldsymbol{\mathrm{A}}_s$ and $\boldsymbol{\mathrm{V}}_n$ will minimize the denominator which gives peaks in the MUSIC angular spectrum. The locations of the peaks will correspond to the correct angle of arrival of the incoming incident signals. To determine the right DOA, MUSIC require a search through all possible steering vectors until the correct steering vector embodied on the covariance matrix \textbf{R} is found.  Given a sensor array of $N$ elements, the MUSIC can detect up to ($N$-1) uncorrelated signals in both ULA and UCA geometry \cite{schmidt_multiple_1986}.

\subsection{Root-MUSIC}
Root-MUSIC was developed to reduce the computational load of MUSIC algorithm. In MSUIC algorithm, the DOA is estimated by an exhaustive search through all possible steering vectors that are orthogonal to the noise eigenvectors. In Root-MUSIC, the DOA is estimated via the zeros of a polynomial so the exhaustive spectral search employed by MSUIC is avoided.  Root-MUSIC is only applicable to ULA as the steering vector of ULA has a vandermonde structure allowing the denominator of MUSIC spectrum equation to be modelled as a polynomial \cite{mathews_signal_1994,jing_improved_2012}. To formulate the polynomial, the notation of the steering vector in ULA is slightly modified to have the below form:   

\begin{equation}%
\boldsymbol{\mathrm{a}}_s(\theta)=\left[\begin{array}{c}1\\e^{2\pi(\frac{d}{\lambda})sin(\theta)}\\\vdots\\e^{2\pi(\frac{d}{\lambda})(N-1)sin(\theta)}\end{array}\right]=\left[\begin{array}{c}1\\z\\\vdots\\z^{N-1}\end{array}\right]=\boldsymbol{\mathrm{a}}_s(z)
\label{4.52}
\end{equation}%

By substituting $\boldsymbol{\mathrm{a_s}}$(z) in \ref{4.51}, the denominator of MUSIC algorithm can be modelled by the following polynomial:

\begin{equation}%
Q_{Root-MUSIC}(z)=\boldsymbol{\mathrm{a}}_s^T(1/z)\boldsymbol{\mathrm{V}}_n\boldsymbol{\mathrm{V}}^H_n\boldsymbol{\mathrm{a}}_s(z)
\label{4.53}
\end{equation}%

The polynomial  $Q_{Root-MUSIC}(z)$ become zero when z=$z_m$=$e^{j2\pi(\frac{d}{\lambda})\sin(\theta_m)}$ which correspond the actual DOA. Therefore, the true DOA can be identified by finding the roots of $Q_{Root-MUSIC}(z)$=0 and identifies the one close to the unit circle. Also to mention that the polynomial $Q_{Root-MUSIC}(z)$ has a 2($N$-1) roots that comes as reciprocal conjugate pair. In other words, if $z$ is a root of  $Q_{Root-MUSIC}(z)$ , then $\dfrac{1}{z^*}$  must also be one of its roots. After the closet roots $z_m$  to the unit circle are identified, the true DOA are given as:  

\begin{equation}%
\theta_m=\arcsin\left((\frac{\lambda}{2\pi d}\right)\arg(z_m))
\label{4.54}
\end{equation}%

\subsection{UCA-Root-MUSIC}
This algorithm was mainly developed to make Root-MUSIC applicable to UCA. As we know, the standard Root-MUSIC is implemented by expressing the denominator of MUSIC spectrum as a polynomial. However, constructing a polynomial is done on the assumption that the steering vector has a vandermonde structure which is true for ULA. As we proceed to UCA, this assumption is no longer true due to the dependence of the sensor angular location. To overcome this problem, phase mode excitation is used where the UCA beam pattern is transformed into a beamspace in which the steering vector follows a vandermonde structure. A transformation matrix $\boldsymbol{\mathrm{T}}_v$ , in equation \ref{4.26}, is derived based on the phase mode excitation to directly map UCA into desired VULA. 

The implementation of UCA-Root-MUSIC is summarized in the following step:
\begin{itemize}
\item 	Obtain the transformed received data of UCA given as $\boldsymbol{\mathrm{x}}_v$($t$)=$\boldsymbol{\mathrm{T}}_w\boldsymbol{\mathrm{x}}$($t$) where $\boldsymbol{\mathrm{T}}_w$=
$\boldsymbol{\mathrm{(T}}_v^H\boldsymbol{\mathrm{T}}_v)^{(-1/2)}\boldsymbol{\mathrm{T}}_v$. Here, we can not directly apply $\boldsymbol{\mathrm{T}}_v$ to the UCA received data as $\boldsymbol{\mathrm{T}}_v^H\boldsymbol{\mathrm{T}}_v$ $\neq$ $I$ which, in terms, leads to a VULA having a color noise.  This will impose a problem as Root-MUSIC algorithm assume a while background noise. To overcome this problem, $\boldsymbol{\mathrm{T}}_w$ is used instead which obeys a unitary transformation($\boldsymbol{\mathrm{T}}_w^H\boldsymbol{\mathrm{T}}_w$=$\boldsymbol{\mathrm{I}}$). The unitary prosperity of  $\boldsymbol{\mathrm{T}}_w$ is obtained by using a prewhitening scheme embodied in the term $\boldsymbol{\mathrm{(T}}_v^H\boldsymbol{\mathrm{T}}_v)^{(-1/2)}$ to convert the color noise back into a white noise \cite{zoltowski_beamspace_1993,li_effective_2014}.
\item Calculate the VULA covariance matrix as $\boldsymbol{\mathrm{R}}_v=\dfrac{1}{K}\sum\limits^{K}_{t=1}\boldsymbol{\mathrm{x}}_v(t)\boldsymbol{\mathrm{x}}_v(t)^H$ for $K$ snapshots
\item 	Perform Eigen-value decomposition on $\boldsymbol{\mathrm{R}}_v$  to obtain $\boldsymbol{\mathrm{V}}_n$ noise eigenvectors which correspond to the smallest $h-M$ eigenvalues
\item To construct a polynomial, the VULA  steering vector in \ref{4.23} is realized as   
\begin{equation}%
\boldsymbol{\mathrm{a}}_{sv}(\theta)=\left[\begin{array}{c}e^{-jh\theta}\\\vdots\\1\\\vdots\\e^{jh\theta}\end{array}\right]=\left[\begin{array}{c}z^{-h}\\\vdots\\1\\\vdots\\z^h\end{array}\right]=\boldsymbol{\mathrm{a}}_{sv}(z)
\label{4.55}
\end{equation}%
where z=$e^{j\theta}$
\item 	Substituting  $\boldsymbol{\mathrm{a}}_{sv}(z)$ in \ref{4.55}, a prewhitened polynomial is constructed as \cite{tuncer_classical_2009}:
\begin{equation}%
Q_{UCA-Root-MUSIC}(z)=\boldsymbol{\mathrm{a}}_{sv}^T(1/z)(\boldsymbol{\mathrm{T}}^H_v\boldsymbol{\mathrm{T}}_v)^{-1/2}\boldsymbol{\mathrm{V}}_n\boldsymbol{\mathrm{V}}_n^H(\boldsymbol{\mathrm{T}}^H_v\boldsymbol{\mathrm{T}}_v)^{-1/2}\boldsymbol{\mathrm{a}}_{sv}(z)^H
\label{4.56}
\end{equation}%

\item Just like in Root-MUSIC, the DOAs of UCA are estimated from the largest-magnitude roots $z_m$ of $Q_{UCA-Root-MUSIC}(z)$ and their estimated values are given as:
\begin{equation}%
\theta_m=\arg(z_m)
\label{4.57}
\end{equation}%
 
\item Due to the approximation in \ref{4.22}, there will be a bias in the estimated $\theta_m$ . To obtain $\theta_m$ with relatively less bias, $N$ must be selected based on the criteria $N \geq 2h$ and the larger the $N$ is, the lower the bias. 
\end{itemize}

\subsection{ESPRIT}
\label{ESPRIT}
\begin{figure} [h]
       \centering
        \includegraphics[width=0.6\textwidth,clip]{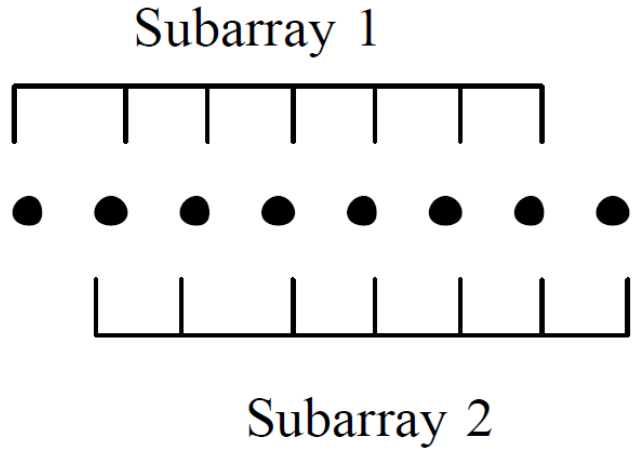}
        \caption[Illustration of Sensor Array Using ESPRIT Algorithm]{Illustration of Sensor Array Using ESPRIT Algorithm}
		 \label{Fig:ESPRIT}
\end{figure}

A different subspace-based method for estimating DOA was introduced by Roy and Kailath \cite{roy_esprit-estimation_1989} which called Estimation of signal parameters via rotation invariance techniques (ESPRIT). Such technique has an advantage over MUSIC that the DOA is estimated directly from the incident signal eigenvalues without going through the exhaustive search of all possible steering vectors.  This, in terms, reduces the overall computation and storage requirements as in Root-MSUIC.  ESPRIT operates by exploiting the property that a sensor array structure can be decomposed into two identical subarrays having the same size called doublets \cite{lavate_performance_2010}. These doublets are displaced from each other by a fixed distance $\Delta x$ as showing in Figure \ref{Fig:ESPRIT}. Therefore, ESPRIT can be applied to ULA as its structure can simply be divided into overlapping subarray

Assuming ULA receiving $M$  incident signal, then each subarrays, as shown in Figure \ref{Fig:ESPRIT}, will received a data vector $\boldsymbol{\mathrm{x}}_1(t)$  and $\boldsymbol{\mathrm{x}}_2(t)$ for subarray-1 and subarray-2, respectively. The combined output $\boldsymbol{\mathrm{x}}(t)$ of both subarraies is modelled as:

\begin{equation}%
\boldsymbol{\mathrm{x}}(t)=\left[\begin{array}{c}\boldsymbol{\mathrm{x}}_1(t)\\ \boldsymbol{\mathrm{x}}_2(t)\end{array}\right]=\left[\begin{array}{c}\boldsymbol{\mathrm{A}}_s\\ \boldsymbol{\mathrm{A}}_s\Phi\end{array}\right]s(t)+\left[\begin{array}{c}\boldsymbol{\mathrm{n}}_1(t)\\ \boldsymbol{\mathrm{n}}_2(t)\end{array}\right]
\label{4.57}
\end{equation}%
where $\boldsymbol{\mathrm{\Phi}}$ represents a diagonal matrix containing the phase shifts between the doublets for $M$ incident signals and it is given as: 

\begin{equation}%
\boldsymbol{\Phi}=\mathrm{diag}\left[\begin{array}{cccc}e^{jk\Delta xsin(\theta_1)}&e^{jk\Delta xsin(\theta_2)}&\cdots&e^{jk\Delta xsin(\theta_M)}\end{array}\right]
\label{4.58}
\end{equation}%

The $\boldsymbol{\mathrm{\Phi}}$ represents a scaling operator to relates the first subarray measurements $\boldsymbol{\mathrm{x_1}}(t)$ to the second subarray measurements $\boldsymbol{\mathrm{x_1}}(t)$. Also, $\boldsymbol{\mathrm{\Phi}}$ operation can be analysed as two dimension rotation and hence it is called a rotational operator [20]. From the received signal $\boldsymbol{\mathrm{x}}(t)$, the correlation matrix \textbf{R} can be formed where the largest $M$ eigenvalues corresponds to the signal eigenvectors $\boldsymbol{\mathrm{V}}_s$. Due to the existence of $\boldsymbol{\mathrm{\Phi}}$ between the subarrays, the signal eigenvectors $\boldsymbol{\mathrm{V}}_s$ is linked to the steering vector matrix 
$\boldsymbol{\mathrm{A}}_s$through non-singular matrix called \textbf{T} which defined as:

\begin{equation}%
\boldsymbol{\mathrm{V}}_s=\left[\begin{array}{c}\boldsymbol{\mathrm{A}}_s\\\boldsymbol{\mathrm{A}}_s\Phi\end{array}\right]\boldsymbol{\mathrm{T}}
\label{4.59}
\end{equation}%

Lastly, matrix $\boldsymbol{\mathrm{\Psi}}$ is defined as $\boldsymbol{\mathrm{\Psi}}=\boldsymbol{\mathrm{T^{-1}\Phi T}}$ where the eigenvalues of $\boldsymbol{\mathrm{\Psi}}$ corresponds to the diagonal elements of $\boldsymbol{\mathrm{\Phi}}$ while eigenvector of $\boldsymbol{\mathrm{\Psi}}$ corresponds to the column elements of $\boldsymbol{\mathrm{T}}$.
From equation \ref{4.59}, $\boldsymbol{\mathrm{V}}_s$ can be decomposed into $\boldsymbol{\mathrm{V}}_1=\boldsymbol{\mathrm{A}}_s\boldsymbol{\mathrm{T}}$ and $\boldsymbol{\mathrm{V}}_2= \boldsymbol{\mathrm{A}}_s\Psi \boldsymbol{\mathrm{T}}$ and hence:

\begin{equation}%
\boldsymbol{\mathrm{V}}_2=\boldsymbol{\mathrm{\Phi V}}_1
\label{4.60}
\end{equation}%

This shows that the key secret behind ESPRIT is the ability of  $\boldsymbol{\mathrm{\Psi}}$ eigenvalues which are $\boldsymbol{\mathrm{\Phi}}$ to map the signal subspace of $\boldsymbol{\mathrm{V}}_1$ that span signal subspace of $\boldsymbol{\mathrm{V}}_2$. Practically, it is impossible to achieve the relationship in equation \ref{4.60} due to the introduction of noise in the measurements and the system is over determinant, hence, $\boldsymbol{\mathrm{\Phi}}$ can be estimated using statistic techniques like Least Square (LS). After $\boldsymbol{\mathrm{\Phi}}$ is obtained, the DOA ($\theta_m$) is expressed as:

\begin{equation}%
\theta_m=\arcsin(\frac{\arg(\phi_m)}{k\Delta x})
\label{4.61}
\end{equation}%
where $\boldsymbol{\Phi}$ are the eigenvalues of $\boldsymbol{\mathrm{\Psi}}$, $k=\dfrac{2\pi}{\lambda}$ and $\Delta x$  is the displacement between the two subarrays.

\subsection{UCA-ESPRIT}
This algorithm was developed to make ESPRIT applicable to UCA. As discussed in section 4.3.4, The ESPRIT technique is designed to work with array geometry that can be decomposed into two identical overlayings. In other words, ESPRIT can be on array geometry that has a vandermonde structure just like the ULA. However, the UCA does not have a vandermonde structure. This problem is resolved by transforming the UCA into VULA using the transformation matrix $\boldsymbol{\mathrm{T}}_v$ give in \ref{4.26}. Then, the azimuth angles are estimated from VULA using the standard ESPRIT. 

The implementation of UCA-ESPRIT is summarized in the following steps:

\begin{itemize}
\item 	Obtain the observation vector of VULA as $\boldsymbol{\mathrm{x}}_v(t)=\boldsymbol{\mathrm{T}}_v\boldsymbol{\mathrm{x}}(t)$. Unlike UCA-Root-MUSIC, the noise-prewhitening technique cannot be employed here as it will destroy the shift-invariance between the two decomposed VULAs. By representing $\boldsymbol{\mathrm{T}}_v$ in its EVD form ($\boldsymbol{\mathrm{T}}_v=\boldsymbol{\mathrm{V \Lambda V}}^H$) , it is deduced that both VULAs will have their noise given as $W_1=\sigma^2 \boldsymbol{\mathrm{I}}$ and $W_2=\sigma_n^2\boldsymbol{\mathrm{\Lambda}}^2$. Even though the noise variance of the second VULA is proportional to the square of  $\boldsymbol{\mathrm{T}}_v$ eigenvalues, both noises are still spatially white and so we can proceed with the ESPRIT Algorithm \cite{buhren_new_2003}.
\item 	Calculate the VULA covariance matrix as  $\boldsymbol{\mathrm{R}}_v=\dfrac{1}{K}\sum\limits^{K}_{t=1}\boldsymbol{\mathrm{Q}}_c\boldsymbol{\mathrm{x}}_v(t)\boldsymbol{\mathrm{x}}_v(t)^H\boldsymbol{\mathrm{Q}}_c^H$ where $\boldsymbol{\mathrm{Q}}_c$ is centro-hermitian matrix that obeys $\boldsymbol{\mathrm{Q}}_c\boldsymbol{\mathrm{Q}}_c^H$=$\boldsymbol{\mathrm{I}}$.  The purpose of $\boldsymbol{\mathrm{Q}}_c$ is to reduce the computational load of EVD operation. 
\item 	Perform EVD on $\boldsymbol{\mathrm{R}}_v$  to obtain $\boldsymbol{\mathrm{E}}_s= \boldsymbol{\mathrm{Q}}_c\boldsymbol{\mathrm{V}}_s$ which corresponds to the largest $M$ eigenvalues.  As we are interested $\boldsymbol{\mathrm{V}}_s$, the equation is rearranged as $\boldsymbol{\mathrm{V}}_s=\boldsymbol{\mathrm{Q}}_c^H\boldsymbol{\mathrm{E}}_s$.
\item 	By Decomposing the main VULA into two identical subarrays with one inter-element spacing between them, as in Figure \ref{ESPRIT}, then, $\boldsymbol{\mathrm{V}}_s$ is also decomposed into $\boldsymbol{\mathrm{V}}_1$ and $\boldsymbol{\mathrm{V}}_2$ where $\boldsymbol{\mathrm{V}}_1$ has the first $M$-1 elements of $\boldsymbol{\mathrm{V}}_s$ while $\boldsymbol{\mathrm{V}}_2$ has the last $M$-1 element of $\boldsymbol{\mathrm{V}}_s$. 
\item Estimate $\Phi$ from the relationship $\boldsymbol{\mathrm{V}}_2 =\Phi \boldsymbol{\mathrm{V}}_1$ using statistic techniques like Least Square.
\item Due to phase mode excitation, $\boldsymbol{\mathrm{\Phi}}$ of the VULA will be in the theoretical form given as:

\begin{equation}%
\boldsymbol{\Phi}=\mathrm{diag}\left[\begin{array}{cccc}e^{j\theta_1}&e^{j\theta_2}&\ldots&e^{j\theta_N}\end{array}\right]
\label{4.62}
\end{equation}%
Hence, the estimated DOA ($\theta_m$) is estimated as:
\begin{equation}%
\theta_m=\arg(\Phi_m)
\label{4.63}
\end{equation}%
\item 	Due to the approximation in equation \ref{4.24}, $\boldsymbol{\Phi}$ will be close but not equal to its theoretical form given in equation \ref{4.62} and hence a bias will occur in the estimated $\theta_m$. To obtain $\theta_m$  with relatively less bias, $N$ must be selected based on the criteria $N\geq2h$ and the larger is $N$ , the lower the bias.
\end{itemize}

\subsection{Comparison in the Performance of ULA and UCA}
In this section, we compare the performances of ULA and UCA using MUSIC algorithm. MUSIC is selected here because it offers a spectrum which can be used to extract information about the characteristics of both geometries.  For a fair comparison, both the ULA and UCA will have parameters (SNR = 15dB, $N$ = 8 and $K$ = 100) in uncorrelated environment. The spacing between the array elements in both geometry is chosen to be 0.5$\lambda$ which is required as minimum distance for MUSIC algorithm to works. Both geometries will receive three signals with DOA ($85^o$, $0^o$ , $-85^o$) transmitted over AWGN channel. The results for ULA and UCA are plotted on Figure \ref{Fig:ULAvsUCA}.
Figure \ref{Fig:ULAvsUCA} shows a comparison of the tests of the ULA and the UCA, it can be seen that ULA is not very capable of interpreting correctly when the steering vectors are being emitted from wider angles, close to the endfire direction ($\theta =\pm 90^o$). This is because the power distribution gets weaker at the edges of the linear array geometry. Whereas in the case of the UCA, the geometry makes power distributed equally in all directions. Another drawback of the ULA is that it produces a symmetric angular spectrum that was seen in both tests of the ULA. In other words, when an angle of $15^o$ in needed, a peak at the angle appears as well as a peak at an angle of $180^o - 15^o = 165^o$ appears that is undesired. This introduction of an undesired angle causes an ambiguity in the interpretation of the results. As seen, this was resolved when using UCA, which only produced a single peak at the desired angle. This comparison yields the following conclusion:

\begin{figure} [H]
       \centering
        \includegraphics[width=1\textwidth,clip]{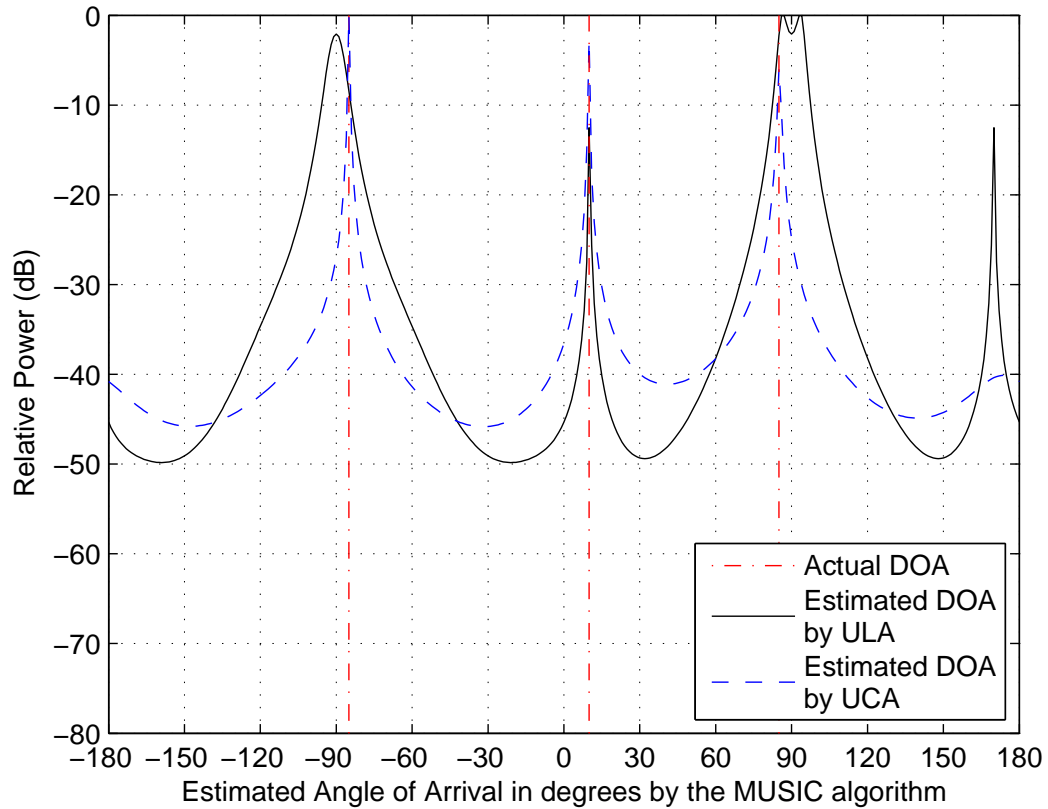}
        \caption[DOA estimation using ULA and UCA]{DOA estimation using ULA and UCA}
		 \label{Fig:ULAvsUCA}
\end{figure}

\subsection{DOA Algorithms Performance}
In this section, we will analyse thoroughly the performance of all DOA algorithms which explained in the report namely MUSIC, Root-MUSIC and ESPRIT. To have a well-established performance analysis of DOA algorithms, all the parameters involved in DOA algorithms to detect the DOA of the incident signals will be investigated separately. By this methodology, we will be able to realize the accuracy and the capacity of each algorithm. Also, we will be able to realize how to effectively implement them in real world with less possible requirements.

The parameters can be classified under two categories. The first category is related to the sensor array geometry used which represented by the number of sensor elements been arranged on the sensor array. The second category is related to the signal environment impact which includes the number of incoming signals, the angular separation between these them. Also, it includes the number of samples taken for the received signal, their SNR ratio and their types whether they are correlated or uncorrelated. 
In simulation, the noise is modelled as AWGN. It is crucial to note that that some algorithms, Root-MSUIC and ESPRIT, in UCA require the use of phase mode excitation (PME) to be implemented. However, the use of PME will result in a small quantization error ($\epsilon_q$). To consider this issue in our analysis, we will study DOA algorithms in respect to ULA and UCA. The true accuracy of DOA algorithms can be deduced from ULA simulation and its results is almost stable. In UCA, the results is not stable due to $\epsilon_q$. To reflect the impact of $\epsilon_q$ in the angle measurements, the maximum possible error will be measured under the variable ‘margin error’. $\epsilon_q$ can be reduced in UCA by increasing $N$ to allow accurate detection. Also, as PME is function of elevation angle it is assumed to be $20^o$ fixed for all signals received by UCA. 

In the simulation, the results of MUSIC algorithm represent angular spectrum graphs while the results of Root-MUSIC and ESPRIT are numerically listed in a table.

\subsubsection{DOA Algorithms Performance in ULA}
\begin{figure} [H]
       \centering
        \includegraphics[width=1\textwidth,clip]{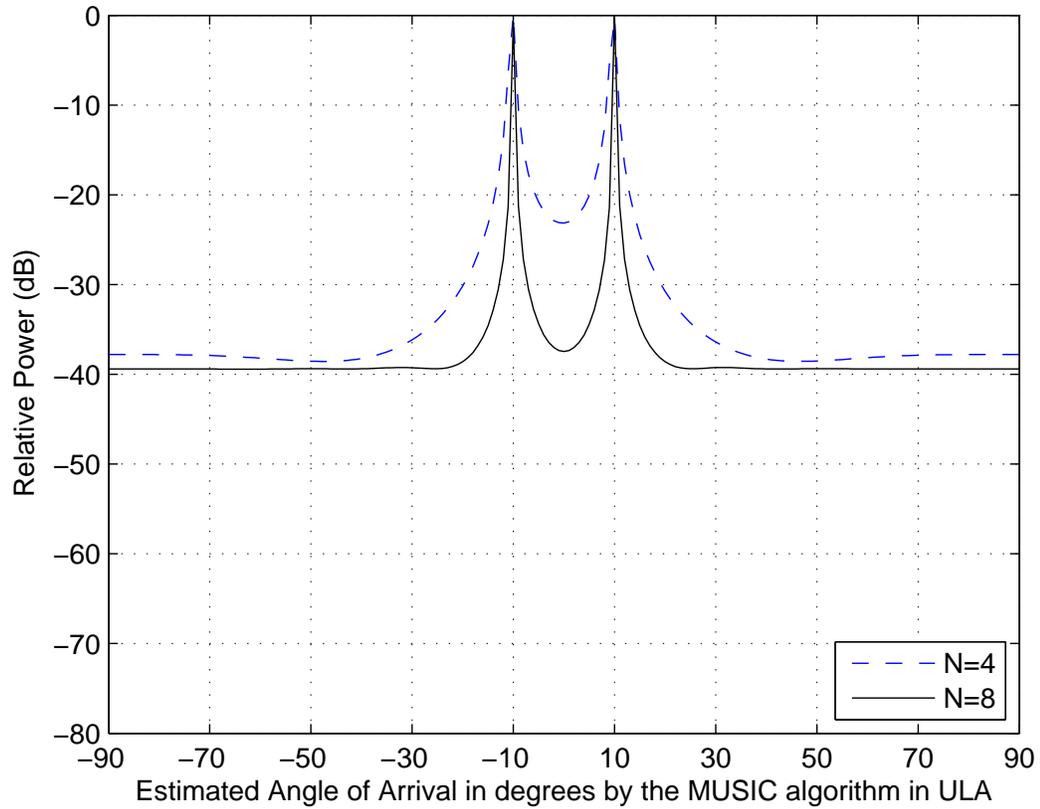}
        \caption[Impact of changing the number of elements in UCA on the performance of MUSIC algorithm with settings (M=2, $\theta=10^o$  and $-10^o$, $d=0.5\lambda$) , SNR=10dB and $K$=100)]{Impact of changing the number of elements in UCA on the performance of MUSIC algorithm with settings (M=2, $\theta=10^o$  and $-10^o$, $d=0.5\lambda$ , SNR=10dB and $K$=100}
		 \label{Fig:Elements_number}
\end{figure}

\paragraph{Number of Sensor Elements} \mbox{}\\
In this test, the impact of ULA number of sensors is investigated on DOA algorithms by applying two incoming signals ($-10^o$, $10^o$) with $N$=4 and $N$=8 on ULA. The results show that increasing the number of elements in ULA will improve the DOA estimation in all DOA algorithms. The improvement in MUSIC spectrum is evident by sharper peaks at the directions of incident signals and lower noise floor in Figure \ref{Fig:Elements_number}. This is also clear in Root-MUISC where the roots become closer to the unit circle allowing accurate detection as given in Table \ref{Root1}. The enhancement in ESPRIT is in a form of more accurate estimation as shown in Table \ref{Esprit1}.

\begin{table}[H]
\centering
  \caption{Impact of changing the number of elements in ULA on the performance of Root-MUSIC algorithm with settings ($M$=2, $\theta=10^o$  and $-10^o$, d=0.5$\lambda$ , SNR=10dB and $K$=100)} 
  \label{Root1}
\begin{tabular}{c}
\includegraphics[width=1\textwidth,clip]{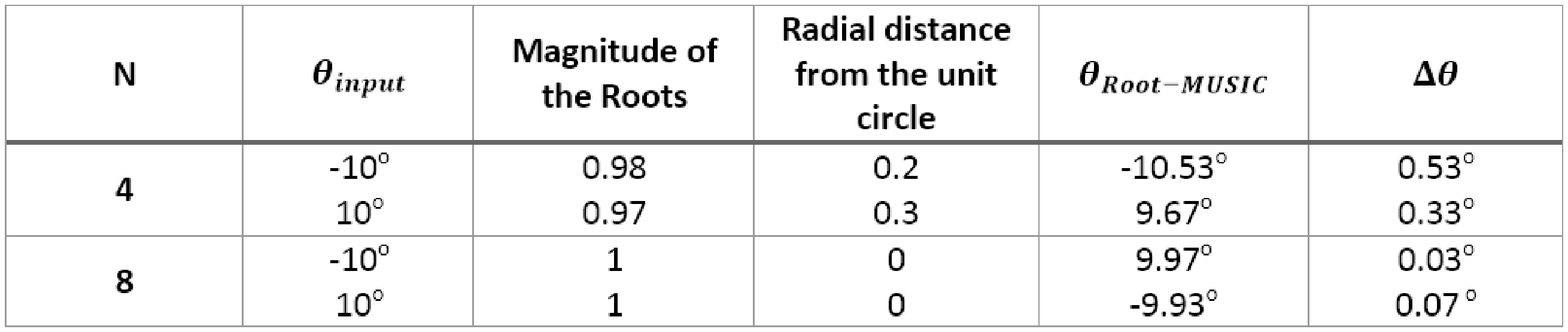}
\end{tabular}
\end{table}   

\begin{table}[H]
\centering
  \caption{Impact of changing the number of elements in ULA on the performance of ESPRIT algorithm with settings ($M$=2, $\theta=10^o$  and $-10^o$, d=0.5$\lambda$ , SNR=10dB and $K$=100)} 
  \label{Esprit1}
\begin{tabular}{c}
\includegraphics[width=1\textwidth,clip]{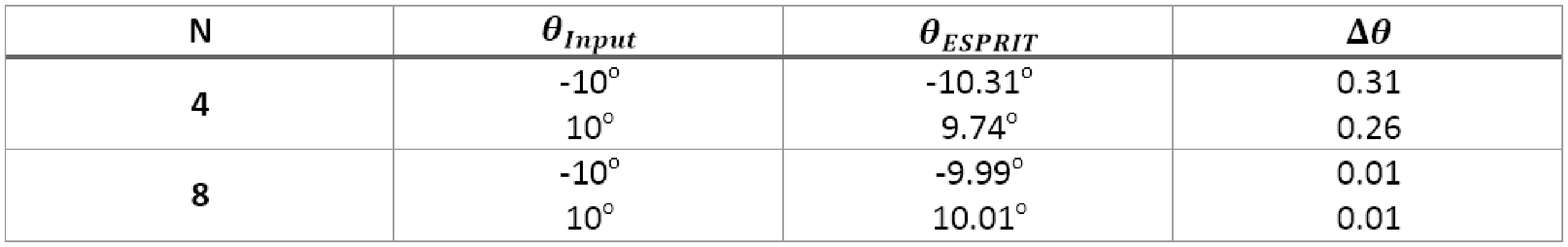}
\end{tabular}
\end{table}

\begin{figure} [H]
       \centering
        \includegraphics[width=1\textwidth,clip]{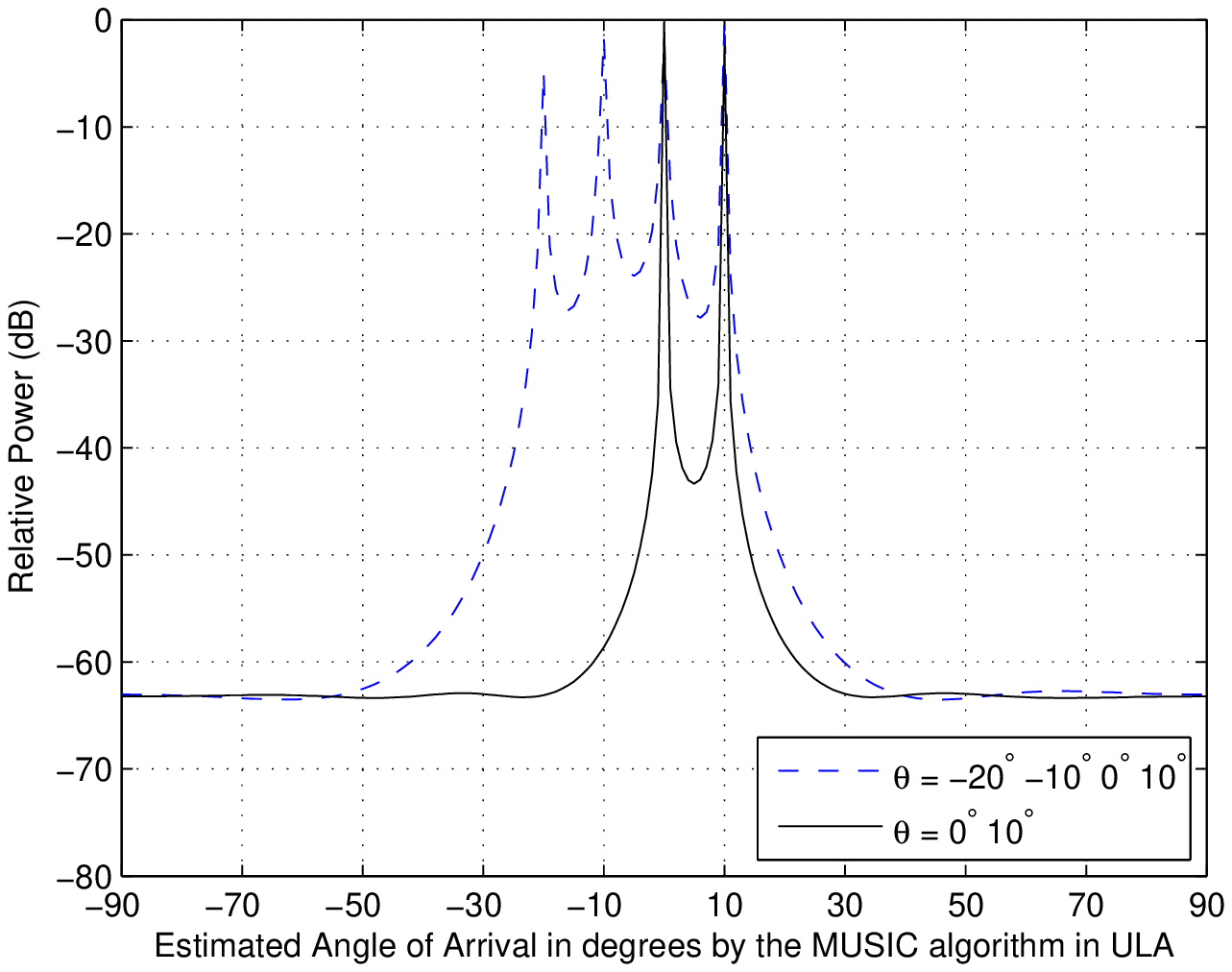}
        \caption[Impact of changing the number of incident signals impinging ULA on the performance of MUSIC algorithm with settings ($N$=6, $d$=0.5$\lambda$, SNR=20dB and $K$=100)]{Impact of changing the number of incident signals impinging ULA on the performance of MUSIC algorithm with settings ($N$=6, $d$=0.5$\lambda$, SNR=20dB and $K$=100)}
		 \label{Fig:Incident_Signals}
\end{figure}

\paragraph{Number of Incident Signal} \mbox{}\\
In the second test, the impact for the number of incoming signals on DOA algorithm is examined by considering two scenarios with 4 incoming signals ($-20^o$ , $-10^o$, $0^o$ , $10^o$) and 2 incoming signal ($0^o$, $10^o$) impinging on ULA. From Figure \ref{Fig:Incident_Signals}, we conclude that as the incident signals increases, the performance of MUSIC will starts to degrade leading to less sharp peaks. This effect is also clear from Table \ref{Root2} and \ref{Esprit2} where the detection accuracy of Root-MUSIC and ESPRIT is reduced. This problem can be resolved by increasing the number of sensor elements in an array. 

\begin{table}[H]
\centering
  \caption{Impact of changing the number of incident signals impinging ULA on the performance of  Root-MUSIC algorithm with settings ($N$=6, $d$=0.5$\lambda$, SNR=20dB and $K$=100)} 
  \label{Root2}
\begin{tabular}{c}
\includegraphics[width=1\textwidth,clip]{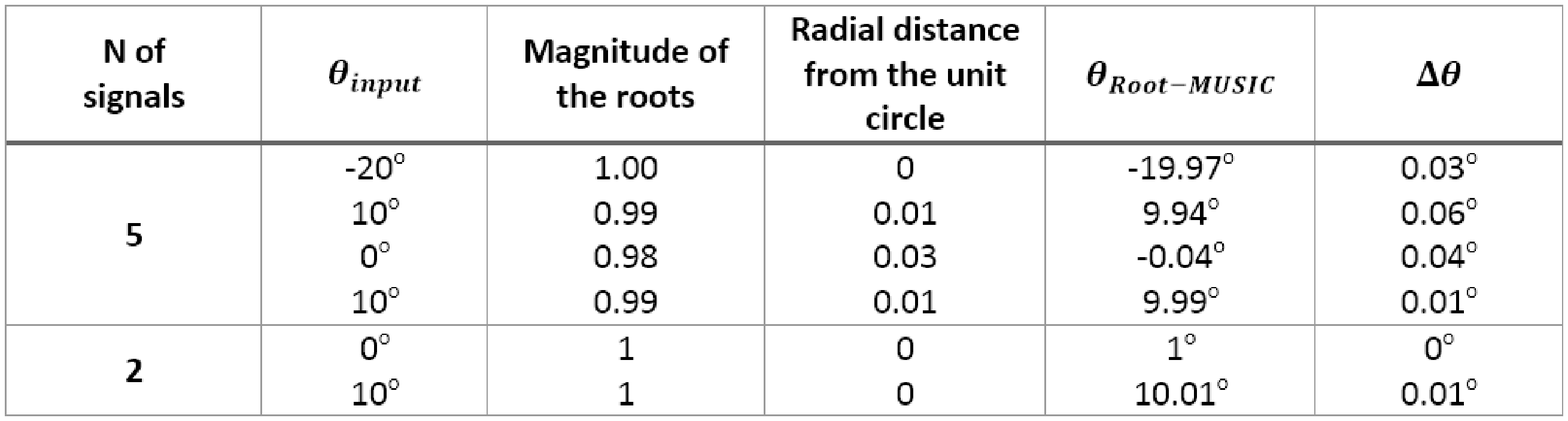}
\end{tabular}
\end{table}   

\begin{table}[H]
\centering
  \caption{Impact of changing the number of incident signals impinging ULA on the performance of  ESPRIT algorithm with settings ($N$=6, $d$=0.5$\lambda$, SNR=20dB and $K$=100)} 
  \label{Esprit2}
\begin{tabular}{c}
\includegraphics[width=1\textwidth,clip]{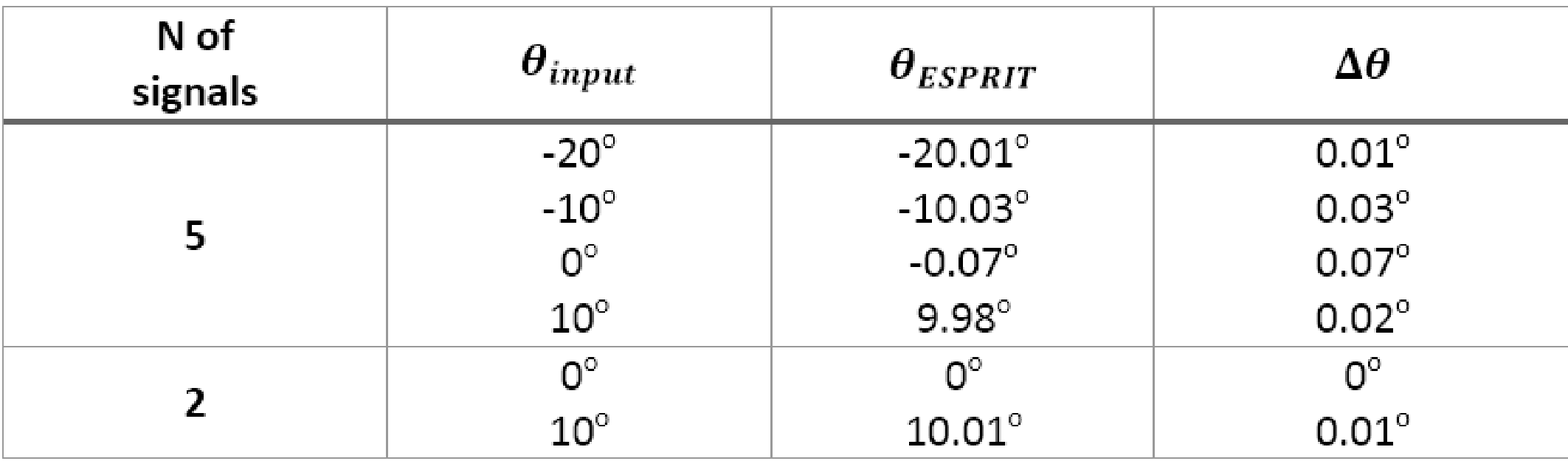}
\end{tabular}
\end{table} 

\begin{figure} [H]
       \centering
        \includegraphics[width=1\textwidth,clip]{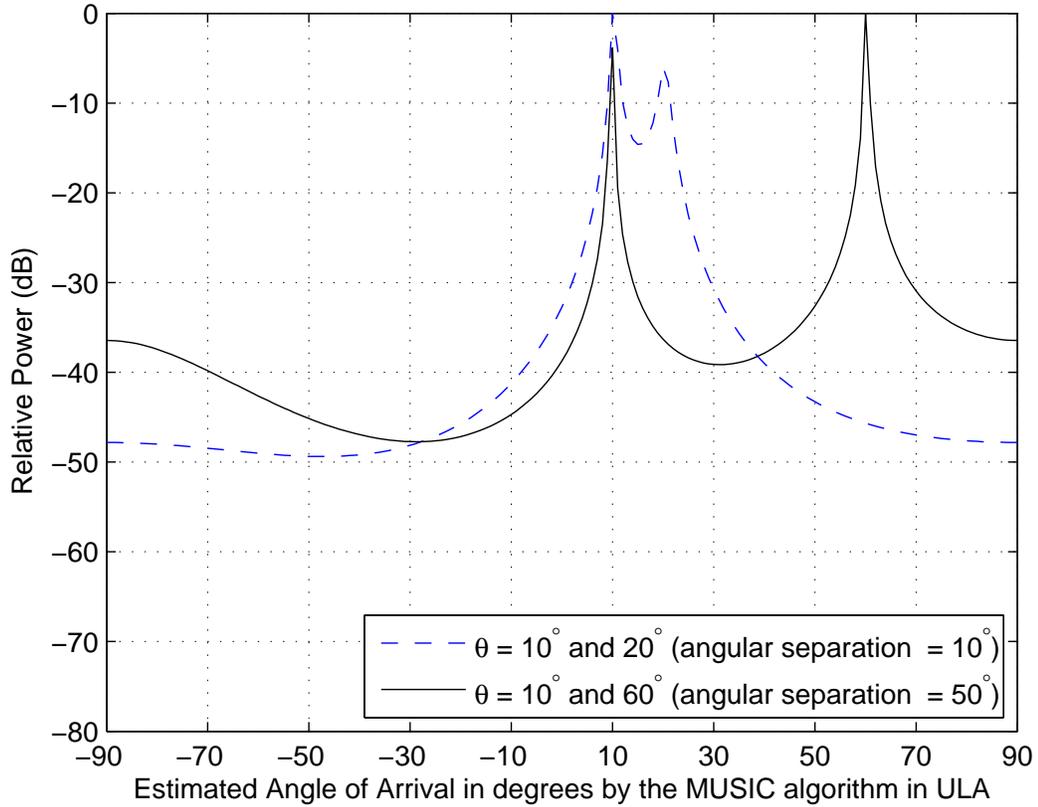}
        \caption[Impact of changing the angular separation between the incident signals impinging on ULA on the performance of MUSIC algorithm with settings ($N$=3, $d$=0.5$\lambda$ , SNR=10dB and $K$=100)]{Impact of changing the angular separation between the incident signals impinging on ULA on the performance of MUSIC algorithm with settings ($N$=3, $d$=0.5$\lambda$ , SNR=10dB and $K$=100)}
		 \label{Fig:Angular_Separation}
\end{figure}

\paragraph{Angular Separation between Incident Signals} \mbox{}\\
In this test, the impact of the angular separation between the incoming signals on DOA algorithm is examined by considering two scenarios. In the first scenario, ULA will receive two incident signals ($10^o$, $20^o$) having a small angular separation of $10^o$. In the second scenario, the angular separation will be increased to $50^o$ as the two incident signals will have a direction of arrival of $10^o$ and $60^o$. Clearly, From Figure \ref{Fig:Angular_Separation}, increasing the angular separation will improve the performance of MUSIC algorithm through producing sharp spectral peaks and reduce the noise floor. The same conclusion is deduced from Table \ref{RootEsp3} where the detection accuracy of Root-MUSIC and ESPRIT is increased as the angular separation is increased.

\begin{table}[H]
\centering
  \caption{Impact of changing the angular separation between the incident signals impinging on ULA on the performance of Root-MUSIC algorithm with settings ($N$=3, $d$=0.5$\lambda$, SNR=20dB and $K$=100)} 
  \label{RootEsp3}
\begin{tabular}{c}
\includegraphics[width=1\textwidth,clip]{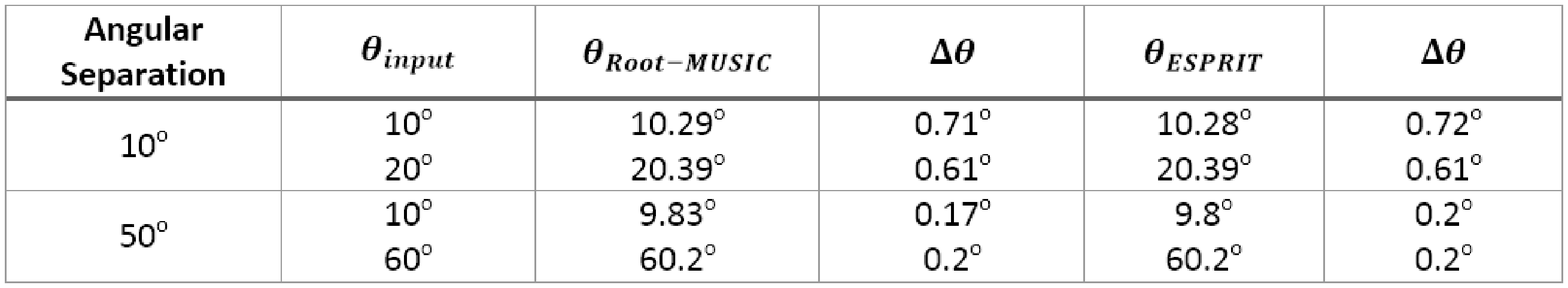}
\end{tabular}
\end{table}   

\begin{figure} [H]
       \centering
        \includegraphics[width=1\textwidth,clip]{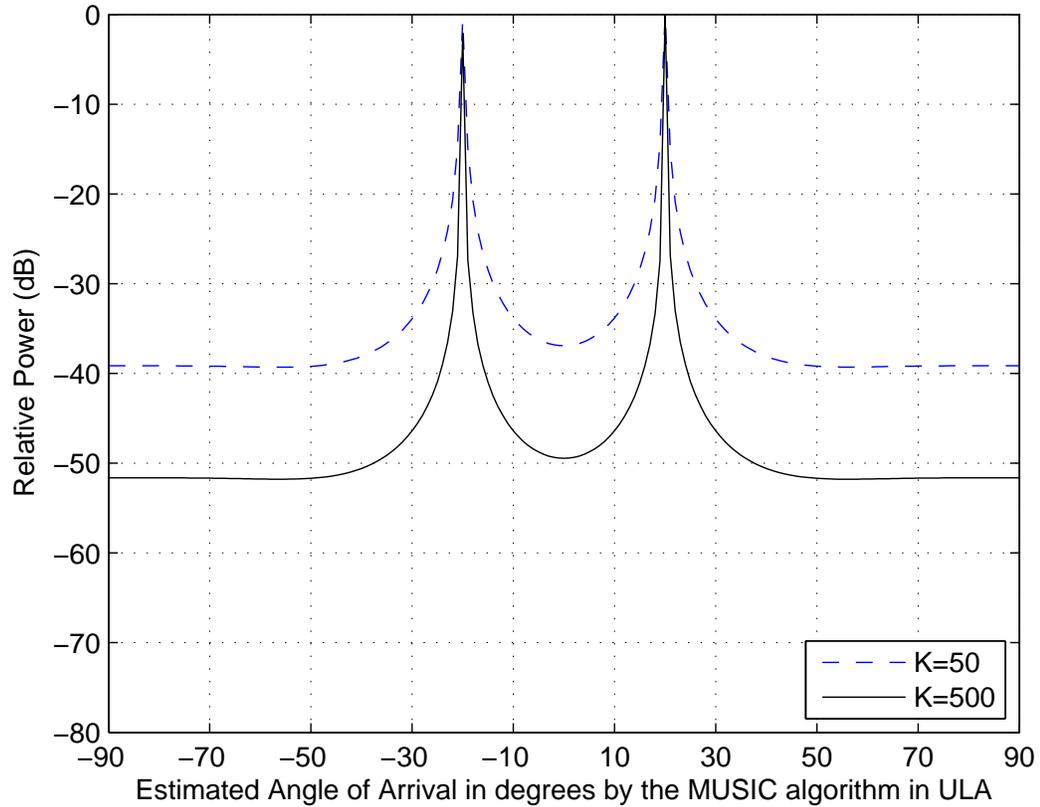}
        \caption[Impact of changing the number of samples of the incident signals impinging on ULA on the performance of MUSIC algorithm with settings ($N$=5, $\theta$=$20^o$  and $-20^o$, $d$=0.5$\lambda$ , SNR=10dB and $K$=100)]{Impact of changing the number of samples of the incident signals impinging on ULA on the performance of MUSIC algorithm with settings ($N$=5, $\theta$=$20^o$  and $-20^o$, $d$=0.5$\lambda$ , SNR=10dB and $K$=100)}
		 \label{Fig:Number_Samples}
\end{figure}

\paragraph{Number of Samples} \mbox{}\\
In the following test, the impact of samples’ number taken for the incoming signals on DOA algorithm is examined by applying two incoming signals ($-20^o$ , $20^o$) on with $K$=50 and $K$=500. From Figure \ref{Fig:Number_Samples}, we conclude that increasing the number of samples enhances the performance of MUSIC algorithm as the peak becomes sharper and the noise floor is lowered. This conclusion is also evident from Table \ref{Root4} and \ref{Esprit4} where the detection accuracy of root-MUSIC and ESPRIT has increased. The reason for this improvement is because increasing the number of samples will lead to a more accurate estimation of incident signals. Hence, a covariance matrix will be more accurate as well.

\begin{table}[H]
\centering
  \caption{Impact of changing the number of samples of the incident signals impinging on ULA on the performance of Root-MUSIC algorithm with settings ($N$=5, $\theta$=$20^o$  and $-20^o$, $d$=0.5$\lambda$ , SNR=10dB and $K$=100)} 
  \label{Root4}
\begin{tabular}{c}
\includegraphics[width=1\textwidth,clip]{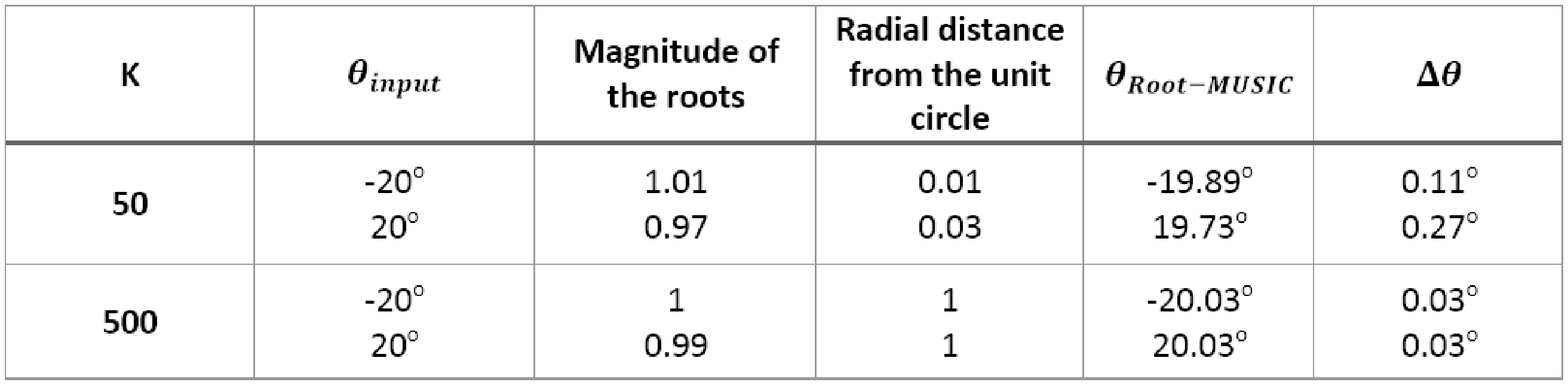}
\end{tabular}
\end{table}   

\begin{table}[H]
\centering
  \caption{Impact of changing the number of samples of the incident signals impinging on ULA on the performance of ESPRIT algorithm with settings ($N$=5, $\theta$=$20^o$  and $-20^o$, $d$=0.5$\lambda$ , SNR=10dB and $K$=100)} 
  \label{Esprit4}
\begin{tabular}{c}
\includegraphics[width=1\textwidth,clip]{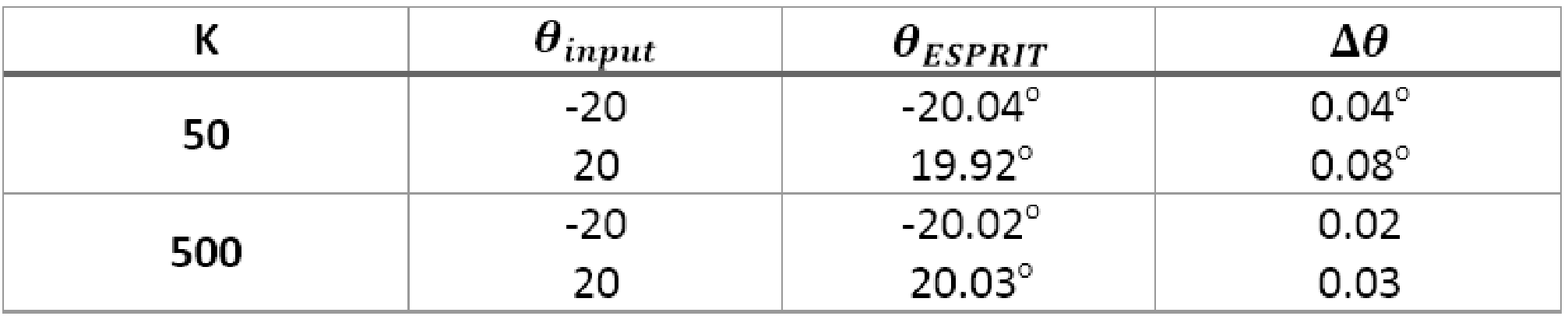}
\end{tabular}
\end{table} 

\begin{figure} [H]
       \centering
        \includegraphics[width=1\textwidth,clip]{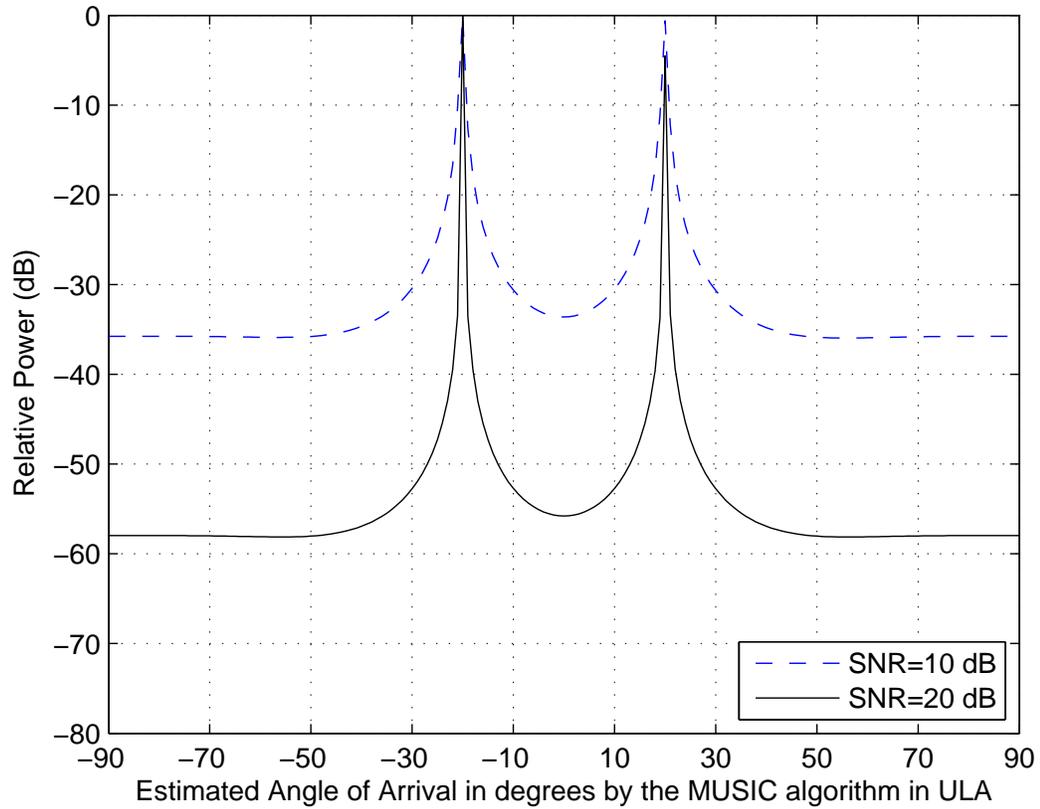}
        \caption[Impact of changing SNR for ULA on the performance of MUSIC algorithm with settings ($N$=5, $\theta=20^o$  and $-20^o$, d=0.5$\lambda$  and $K$=100)]{Impact of changing SNR for ULA on the performance of MUSIC algorithm with settings ($N$=5, $\theta=20^o$  and $-20^o$, d=0.5$\lambda$  and $K$=100)}
		 \label{Fig:SNR}
\end{figure}

\paragraph{Signal to Noise Ratio (SNR)} \mbox{}\\
In this test, the impact of SNR on DOA algorithm is investigated through limiting the noise power introduce by the channel to meet SNR=10 and SNR=20. Both condition of SNR is applied for detecting two incident signals ($-20^o$, $20^o$) on ULA. By analysing Figure \ref{Fig:SNR} with Table \ref{Root5} and \ref{Esprit5}, we conclude that increasing the SNR to higher values will improve the MUSIC algorithm as it will produce sharper peaks with reduced noise level. Also, the roots will come closer to unit circle allowing accurate estimation by Root-MUSIC whereas precise detection is evident in ESPRIT as SNR increases.

\begin{table}[H]
\centering
  \caption{Impact of changing SNR for ULA on the performance of Root-MUSIC algorithm with settings ($N$=5, $\theta=20^o$  and $-20^o$, d=0.5$\lambda$  and $K$=100)} 
  \label{Root5}
\begin{tabular}{c}
\includegraphics[width=1\textwidth,clip]{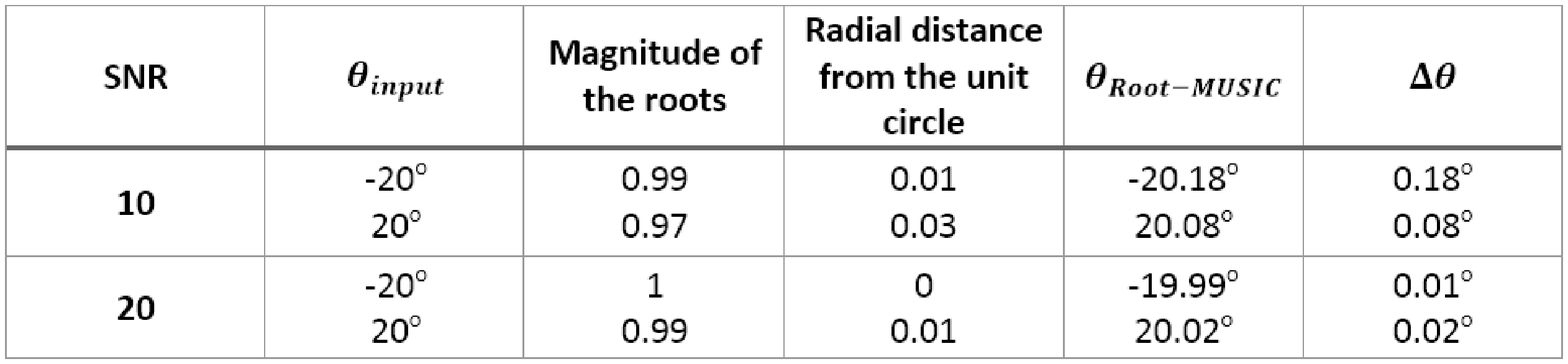}
\end{tabular}
\end{table}   

\begin{table}[H]
\centering
  \caption{Impact of changing SNR for ULA on the performance of ESPRIT algorithm with settings ($N$=5, $\theta=20^o$  and $-20^o$, d=0.5$\lambda$  and $K$=100)} 
  \label{Esprit5}
\begin{tabular}{c}
\includegraphics[width=1\textwidth,clip]{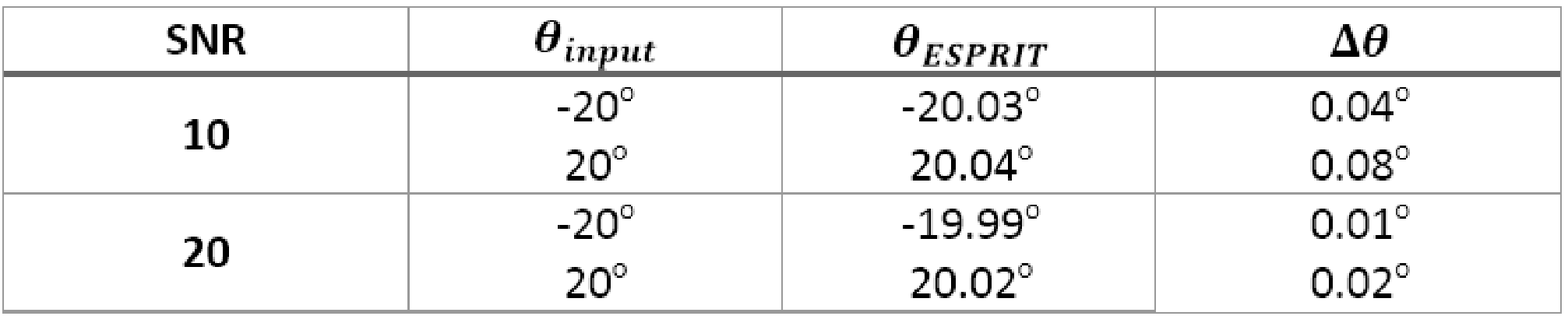}
\end{tabular}
\end{table} 

\paragraph{Signal Correlation} \mbox{}\\
Unlike the previous discussed parameters, this particular parameter which is “signal correlation” will make the input covariance matrix singular preventing the DOA algorithms from functioning. To resolve this issue, preprocessing techniques, explained in section \ref{Correlated Signals}, are needed to change the covariance matrix into non-singular. In another words, the preprocessing techniques will de-correlate the coherent signals so that the DOA algorithms can properly works. Generally, the preprocessing techniques can operate along all the DOA algorithms as they only modify the covariance matrix. In the following test, we will show the efficiency for each of the preprocessing techniques as they are employed with MUSIC Algorithm in ULA. The geometry of ULA allows the use of three preprocessing techniques namely Forward Spatial Smoothing (FSS), Forward-Backward Spatial Smoothing (FBSS) and Toeplitz Algorithm. 

\begin{figure} [H]
       \centering
        \includegraphics[width=1\textwidth,clip]{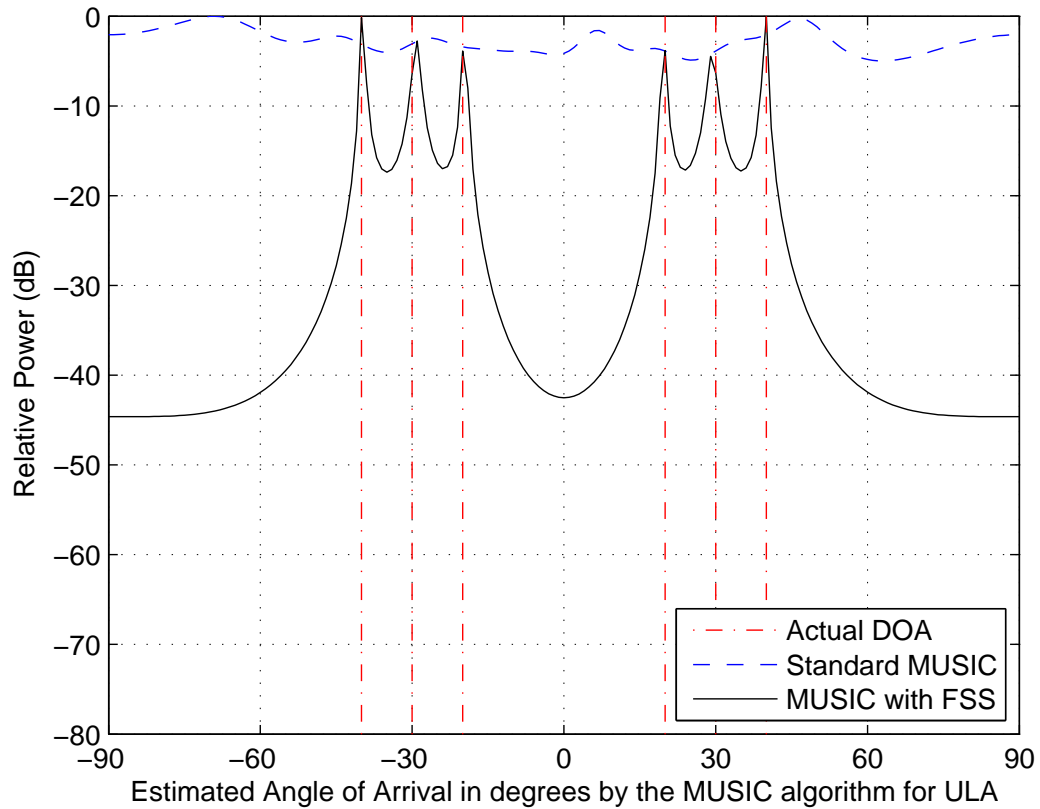}
        \caption[Implementation of Standard MUSIC and MUSIC with FSS for ULA in Correlated Environment with the settings ($N$=12, $\theta$=$-40^o,  -30^o, -20^o, 20^o, 30^o, 40^o$,  $d$=0.5$\lambda$ , SNR=20dB and $K$=100)]{Implementation of Standard MUSIC and MUSIC with FSS for ULA in Correlated Environment with the settings ($N$=12, $\theta$=$-40^o  -30^o, -20^o, 20^o, 30^o, 40^o$,  $d$=0.5$\lambda$ , SNR=20dB and $K$=100)}
		 \label{Fig:FSS_Corellated_12}
\end{figure}

\begin{figure} [H]
       \centering
        \includegraphics[width=1\textwidth,clip]{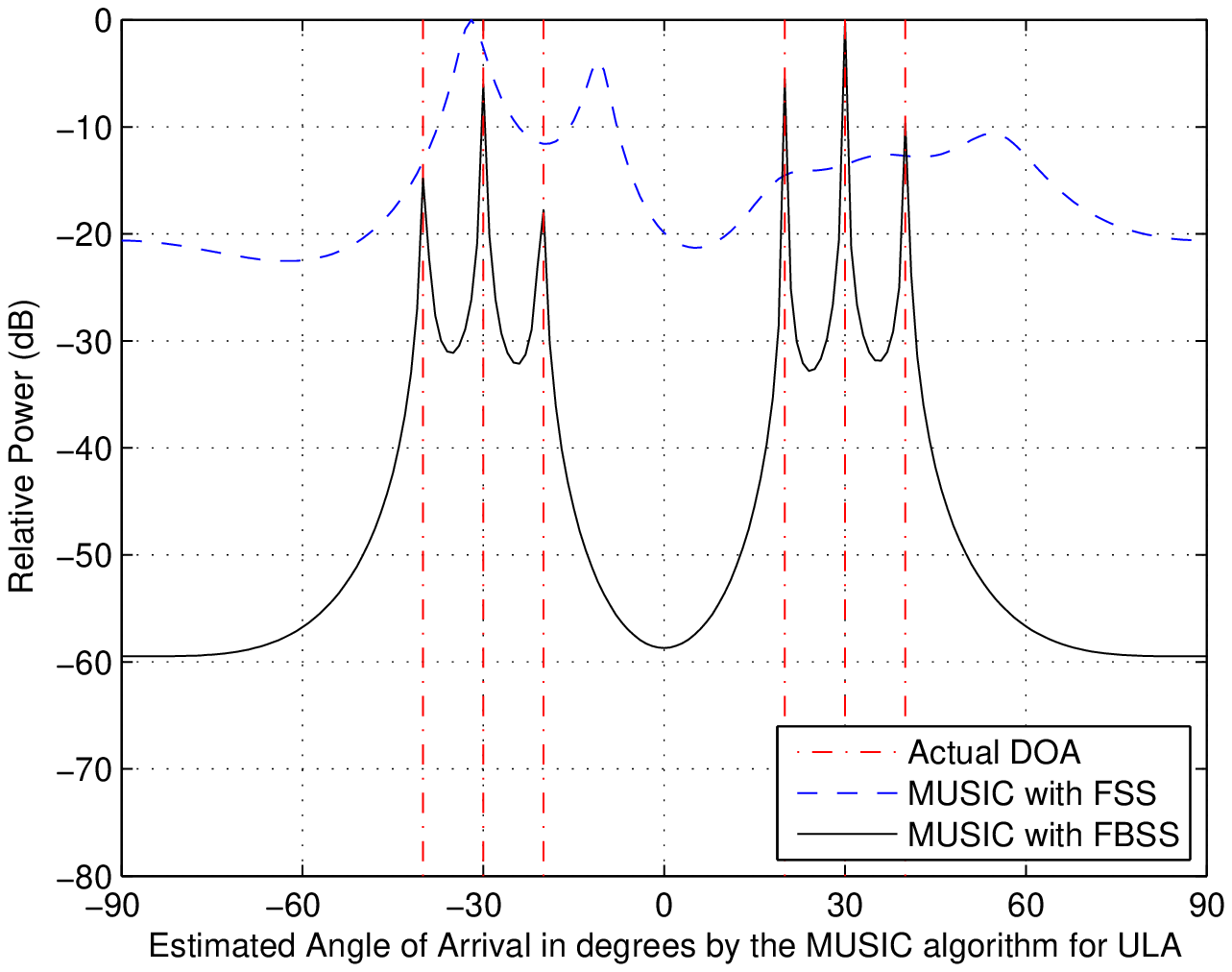}
        \caption[Implementation of FSS and FBSS using MUSIC algorithm for ULA in Correlated Environment with the settings ($N$=9, $\theta$=$-40^o,  -30^o, -20^o, 20^o, 30^o, 40^o$,  $d$=0.5$\lambda$ , SNR=20dB and $K$=100)]{Implementation of FSS and FBSS using MUSIC algorithm for ULA in Correlated Environment with the settings ($N$=9, $\theta$=$-40^o,  -30^o, -20^o, 20^o, 30^o, 40^o$,  $d$=0.5$\lambda$ , SNR=20dB and $K$=100)}
		 \label{Fig:FSS_Corellated_9}
\end{figure}

\begin{figure} [H]
       \centering
        \includegraphics[width=1\textwidth,clip]{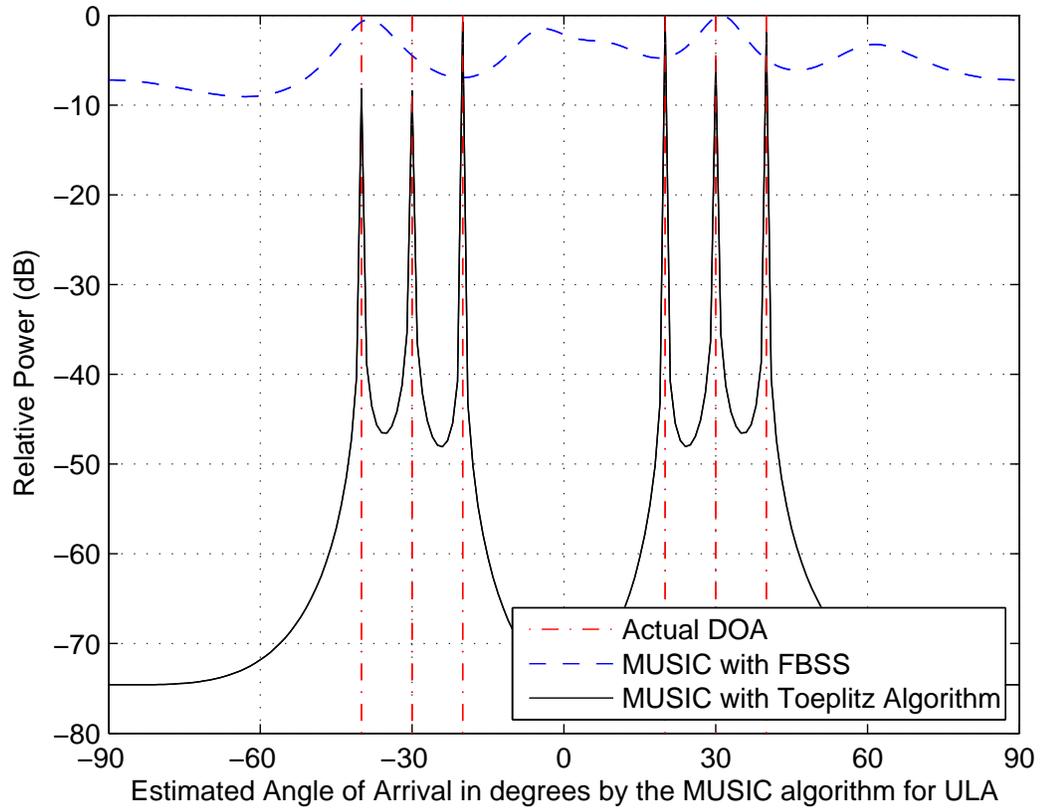}
        \caption[Implementation of FBSS and Toeplitz using MUSIC algorithm for ULA in Correlated Environment with the settings ($N$=7, $\theta$=$-40^o,  -30^o, -20^o, 20^o, 30^o, 40^o$,  $d$=0.5$\lambda$ , SNR=20dB and $K$=100)]{Implementation of FBSS and Toeplitz using MUSIC algorithm for ULA in Correlated Environment with the settings ($N$=7, $\theta$=$-40^o,  -30^o, -20^o, 20^o, 30^o, 40^o$,  $d$=0.5$\lambda$ , SNR=20dB and $K$=100)}
		 \label{Fig:FSS_Corellated_7}
\end{figure}

\begin{figure} [H]
       \centering
        \includegraphics[width=1\textwidth,clip]{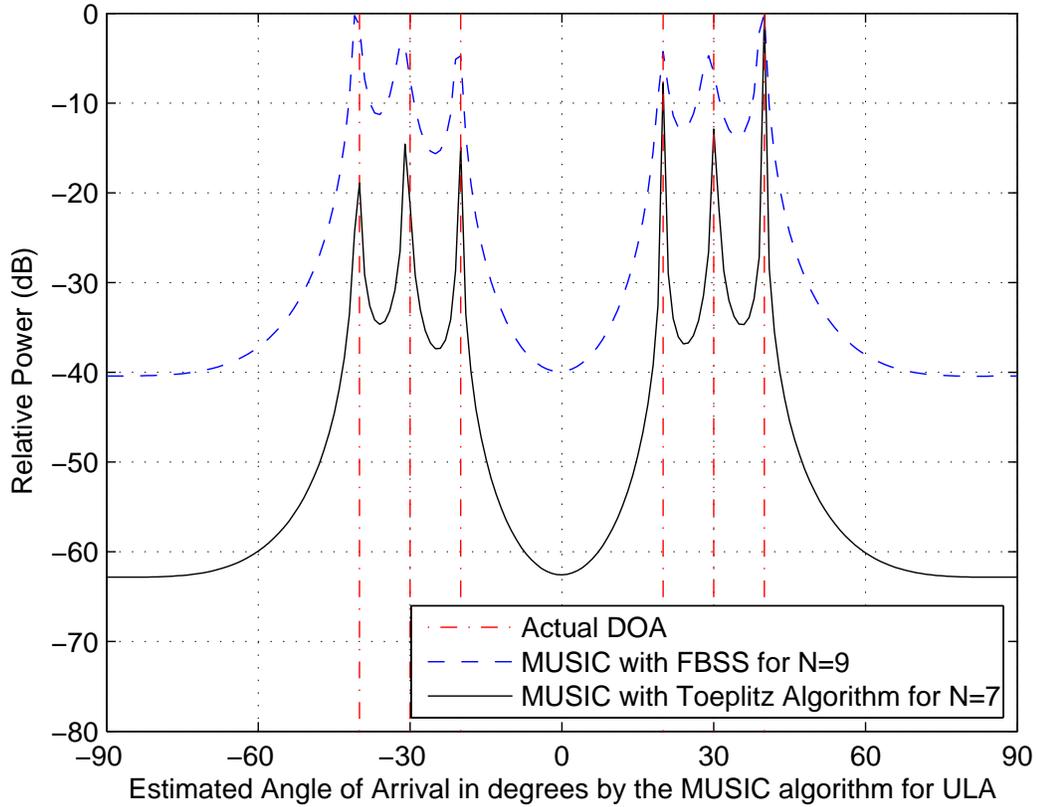}
        \caption[Performance Comparison between FBSS and Toeplitz using MUSIC algorithm for ULA in Correlated Environment with the settings ($\theta=-40^o,  -30^o, -20^o, 20^o, 30^o, 40^o$,  $d$=0.5$\lambda$ , SNR=10dB and $K$=100)]{Performance Comparison between FBSS and Toeplitz using MUSIC algorithm for ULA in Correlated Environment with the settings ($\theta=-40^o  -30^o, -20^o, 20^o, 30^o, 40^o$,  $d$=0.5$\lambda$ , SNR=10dB and $K$=100)}
		 \label{Fig:FSS_Corellated_All}
\end{figure}

In our test, we will consider a scenario where six correlated signals are impinging on ULA with angles ($-40^o$ ,$-30^o$ ,$-20^o$  ,$20^o$ ,$30^o$ ,$40^o$). Firstly, we will use the normal MUSIC and MUSIC with FSS to detect these correlated signals when they received by ULA having $N$=12. Clearly from Figure \ref{Fig:FSS_Corellated_12}, we conclude that MUSIC was unable to resolve the correlated signals alone but MUSIC succeeds when it is used with FSS. The FSS technique is theoretically capable of detecting $N/2$ correlated signals and this is proven from simulation results as $12/2$=6. By reducing to $N$=9, FSS will fail to detect the correlated signals but FBSS will succeed as shown in Figure \ref{Fig:FSS_Corellated_9}. That is because theoretically, FBSS can detect up to $2N/3$ signals which is in this case $  (2 \times 9)/3$=6. By further reducing the $N$ to 7, the FBSS will be unable to resolve the correlated signals but the Toeplitz algorithm will be as shown in Figure \ref{Fig:FSS_Corellated_7}. That is because Toeplitz algorithm can restore the whole rank of covariance matrix permitting the $N$-1 full range detection of MUSIC algorithm. As $N$=7, the whole six correlated signals can be perfectly detected with Toeplitz algorithm. Interestingly, Toeplitz Algorithm offer more robot performance with $N$=7 compared with FBSS with $N$=9 which is evident by the sharp peaks and lower noise floor as shown in Figure \ref{Fig:FSS_Corellated_All}. Therefore, we conclude that Toeplitz Algorithm is the best choice to be used practically with ULA in coherent environment due to its low physical requirement, strong performance and less computational complexity.    

\subsubsection{DOA Algorithms Performance in UCA}
\begin{figure} [H]
       \centering
        \includegraphics[width=1\textwidth,clip]{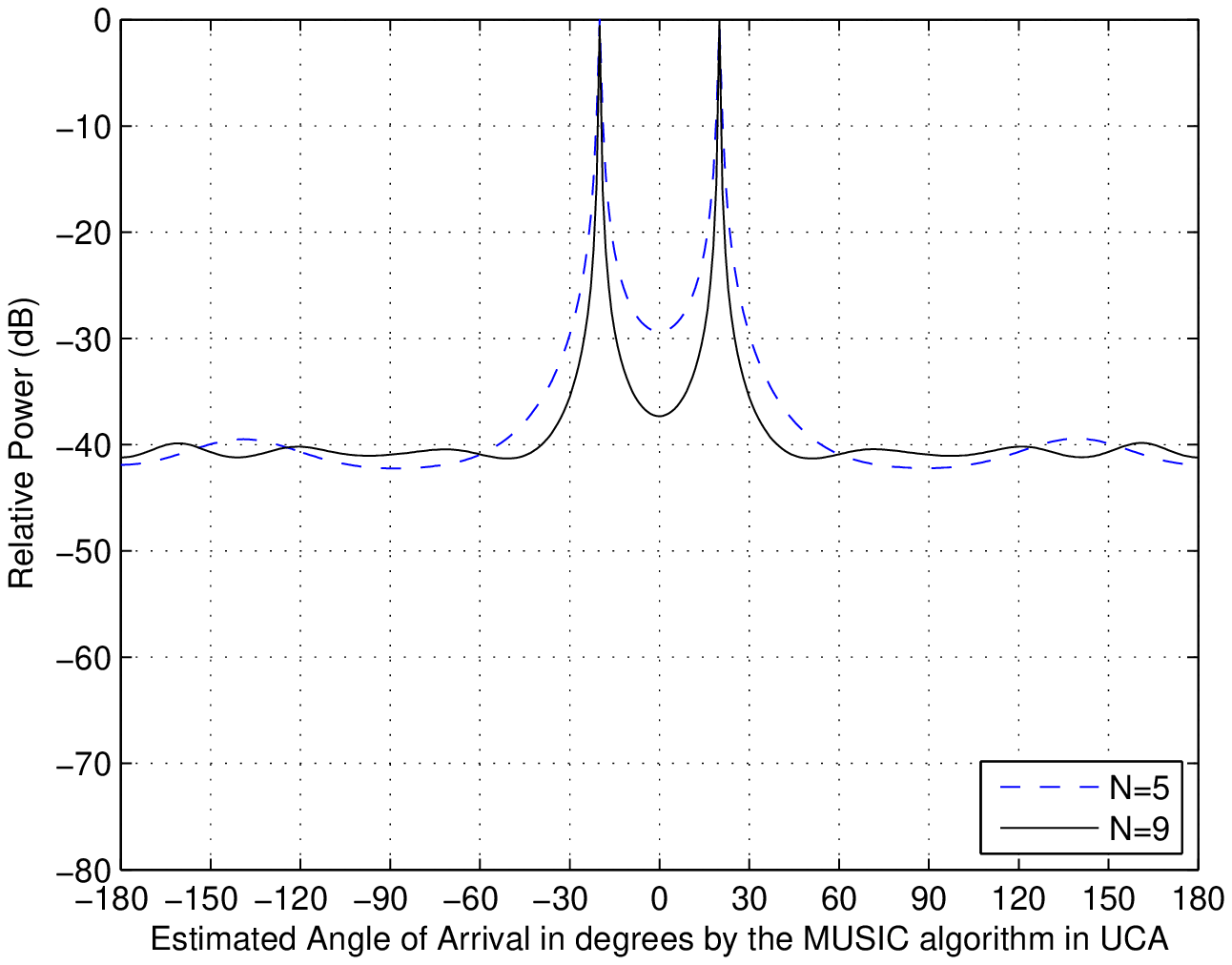}
        \caption[Impact of changing the number of elements in UCA on the performance of MUSIC algorithm with settings ($M$=2, $\theta$=$20^o$  and $-20^o$, $\theta_e$=$20^o$, $d$=0.5$\lambda$ , SNR=10dB and $K$=100)]{Impact of changing the number of elements in UCA on the performance of MUSIC algorithm with settings ($M$=2, $\theta$=$20^o$  and $-20^o$, $\theta_e$=$20^o$, $d$=0.5$\lambda$ , SNR=10dB and $K$=100)}
		 \label{Fig:Sensor_Elements_2}
\end{figure}

\paragraph{Number of Sensor Elements} \mbox{}\\
In this test, the impact of UCA number of sensors is investigated on DOA algorithms by applying two incoming signals ($-10^o$, $10^o$) with $N$=5 and $N$=9. The results from Figure \ref{Fig:Sensor_Elements_2}, show that increasing the number of elements in UCA will improve the DOA estimation by MUSIC algorithm as evident by the sharper peaks and lower noise floor in the spectrum. Table \ref{Root6} shows that the performance of Root-MUSIC will improve where the roots become closer to the unit circle as well as better detection.  Similarly, the accuracy of ESPRIT is improved as shown in Table \ref{Esprit6}.   By comparing all these results, it is noticeable to mention that MUSIC in UCA offer the most accurate results at low number of elements. That is because Root-MUSIC and ESPRIT has an error margin of almost $\pm1$ due to the quantization error introduced by PME. To reduce the quantization error,we need to rise $N$ to allow accurate transformation from UCA to VULA and this is clear from case $N$ = 9 were the margin error is reduced to almost $\pm0.3$ in both algorithms.

\begin{table}[H]
\centering
  \caption{Impact of changing the number of elements in UCA on the performance of Root-MUSIC algorithm with settings ($M$=2, $\theta$=$20^o$  and $-20^o$, $\theta_e$=$20^o$, $d$=0.5$\lambda$ , SNR=10dB and $K$=100)} 
  \label{Root6}
\begin{tabular}{c}
\includegraphics[width=1\textwidth,clip]{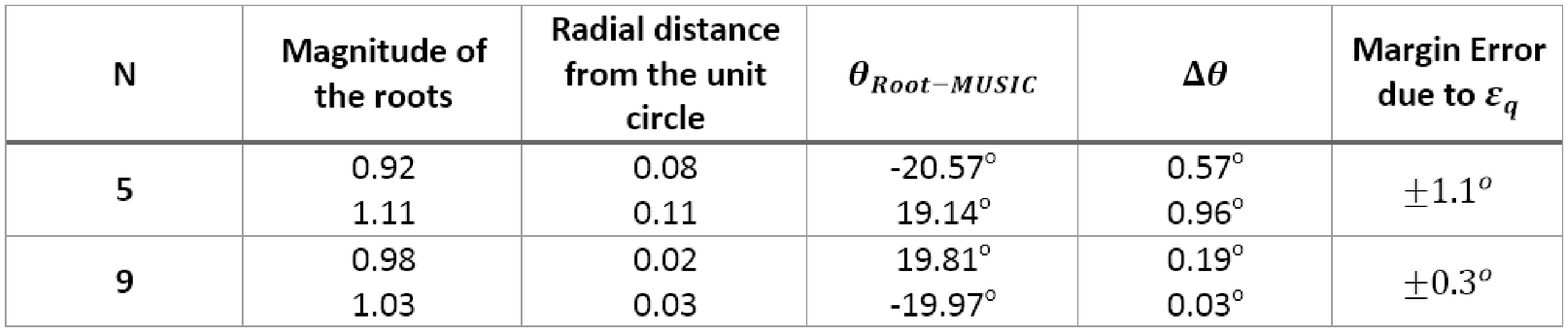}
\end{tabular}
\end{table}   

\begin{table}[H]
\centering
  \caption{Impact of changing the number of elements in UCA on the performance of ESPRIT algorithm with settings ($M$=2, $\theta$=$20^o$  and $-20^o$, $\theta_e$=$20^o$, $d$=0.5$\lambda$ , SNR=10dB and $K$=100)} 
  \label{Esprit6}
\begin{tabular}{c}
\includegraphics[width=1\textwidth,clip]{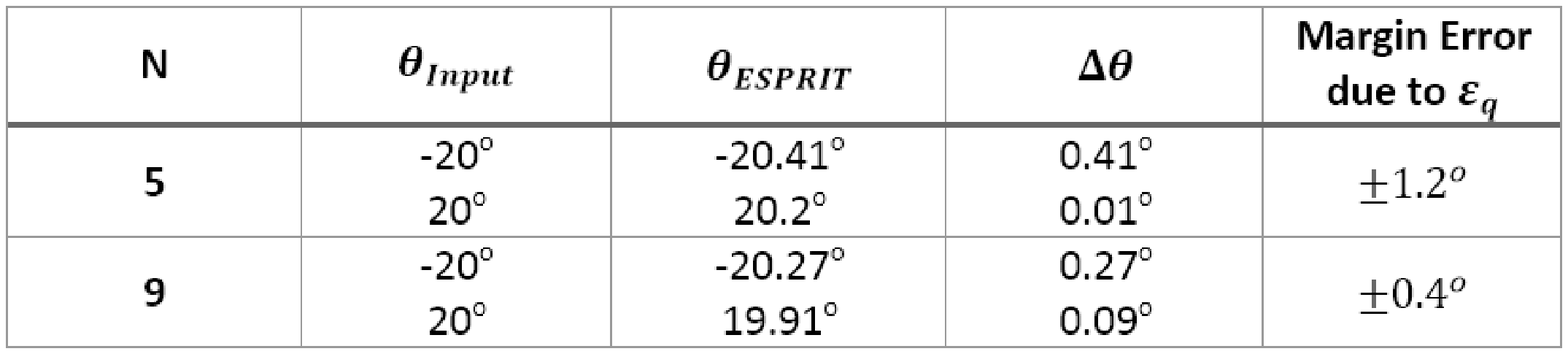}
\end{tabular}
\end{table} 

\begin{figure} [H]
       \centering
        \includegraphics[width=1\textwidth,clip]{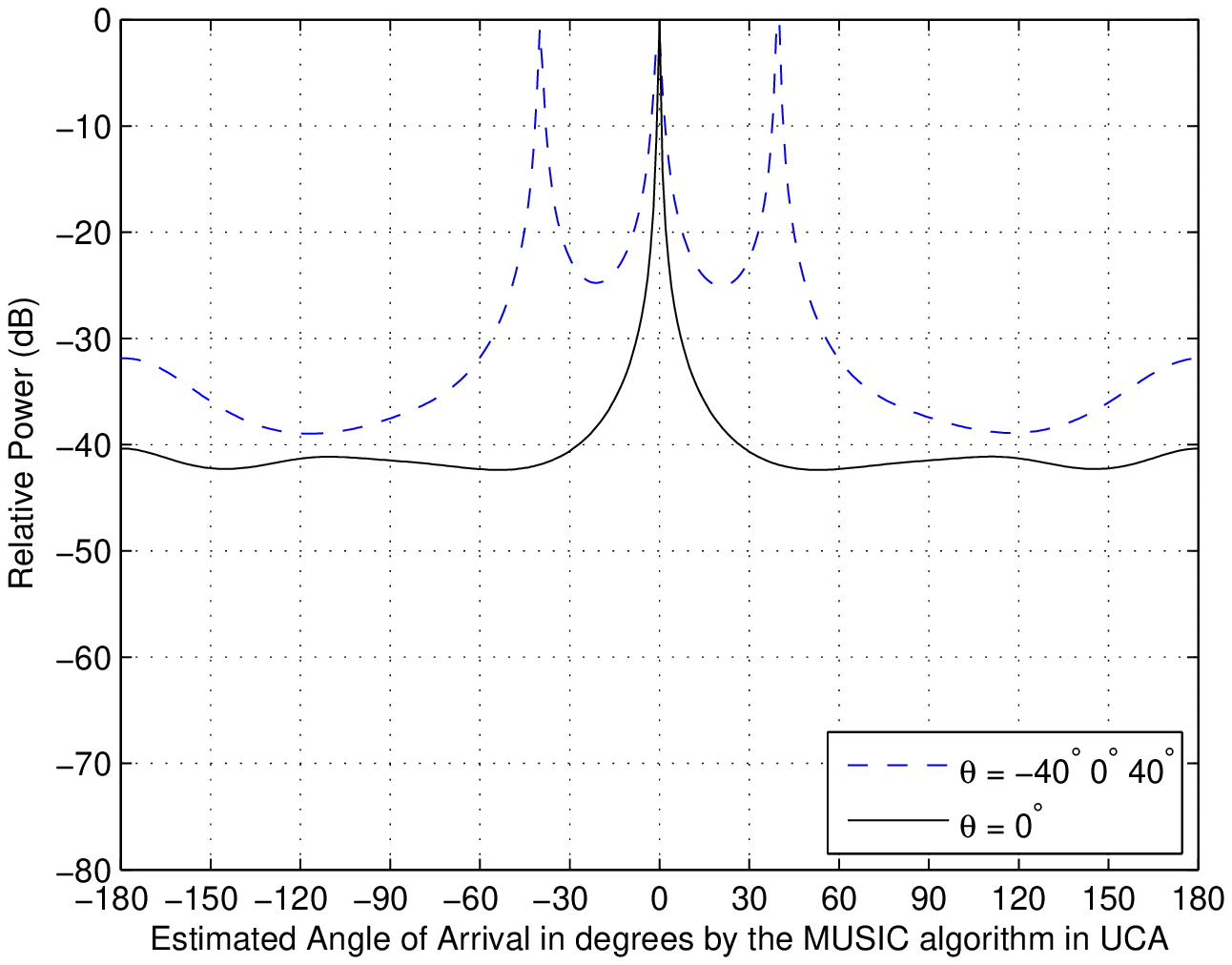}
        \caption[Impact of changing the number of incident signals impinging UCA on the performance of MUSIC algorithm with settings ($N$=5, $\theta_e$=$20^o$, $d$=0.5$\lambda$, SNR=10dB and $K$=100)]{Impact of changing the number of incident signals impinging UCA on the performance of MUSIC algorithm with settings ($N$=5, $\theta_e$=$20^o$, $d$=0.5$\lambda$, SNR=10dB and $K$=100)}
		 \label{Fig:Incident_Signals_2}
\end{figure}

\paragraph{Number of Incident Signal} \mbox{}\\
In the second test, the impact number of incoming signals on DOA algorithm is examined by considering two scenarios with 3 incoming signals ($-40^o$, $0^o$, $40^o$) against 1 incoming signal from ($0^o$) impinging on UCA. From Figure \ref{Fig:Incident_Signals_2}, we conclude that as the incident signals increases, the performance of MUSIC will starts to degrade leading to a less sharp peaks. This effect is also clear from Table \ref{RootEsp7} where the detection accuracy of Root-MUSIC and ESPRIT is reduced leading to a larger error margin. In fact the large margin error is due to the combined effect of increasing number incident signal as well as the quantization error due to PME. Thus, the margin error can be further reduced by increasing the number of elements in UCA.

\begin{table}[H]
\centering
  \caption{Impact of changing the number of incident signals impinging UCA on the performance of Root-MUSIC and ESPRIT algorithms with settings ($N$=5, $\theta_e$=$20^o$, $d$=0.5$\lambda$, SNR=10dB and $K$=100)} 
  \label{RootEsp7}
\begin{tabular}{c}
\includegraphics[width=1\textwidth,clip]{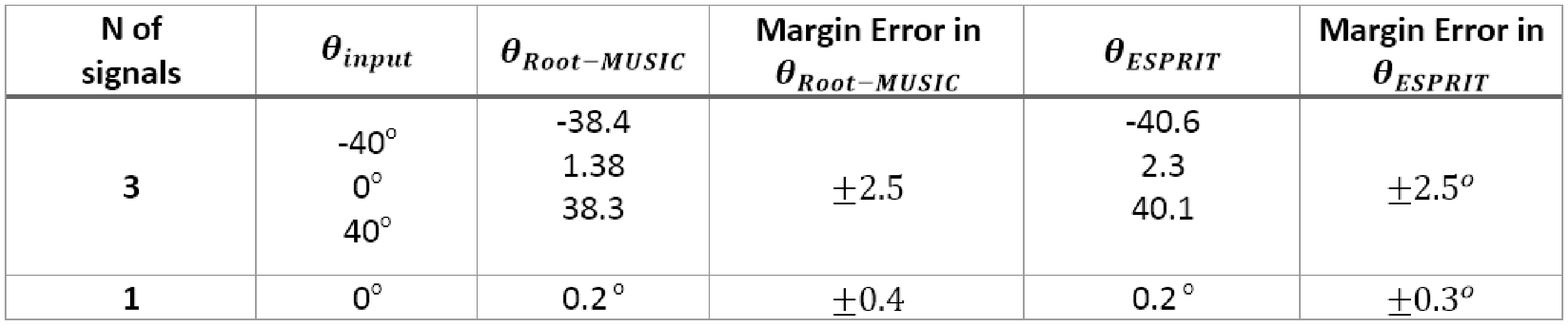}
\end{tabular}
\end{table}   

\begin{figure} [H]
       \centering
        \includegraphics[width=1\textwidth,clip]{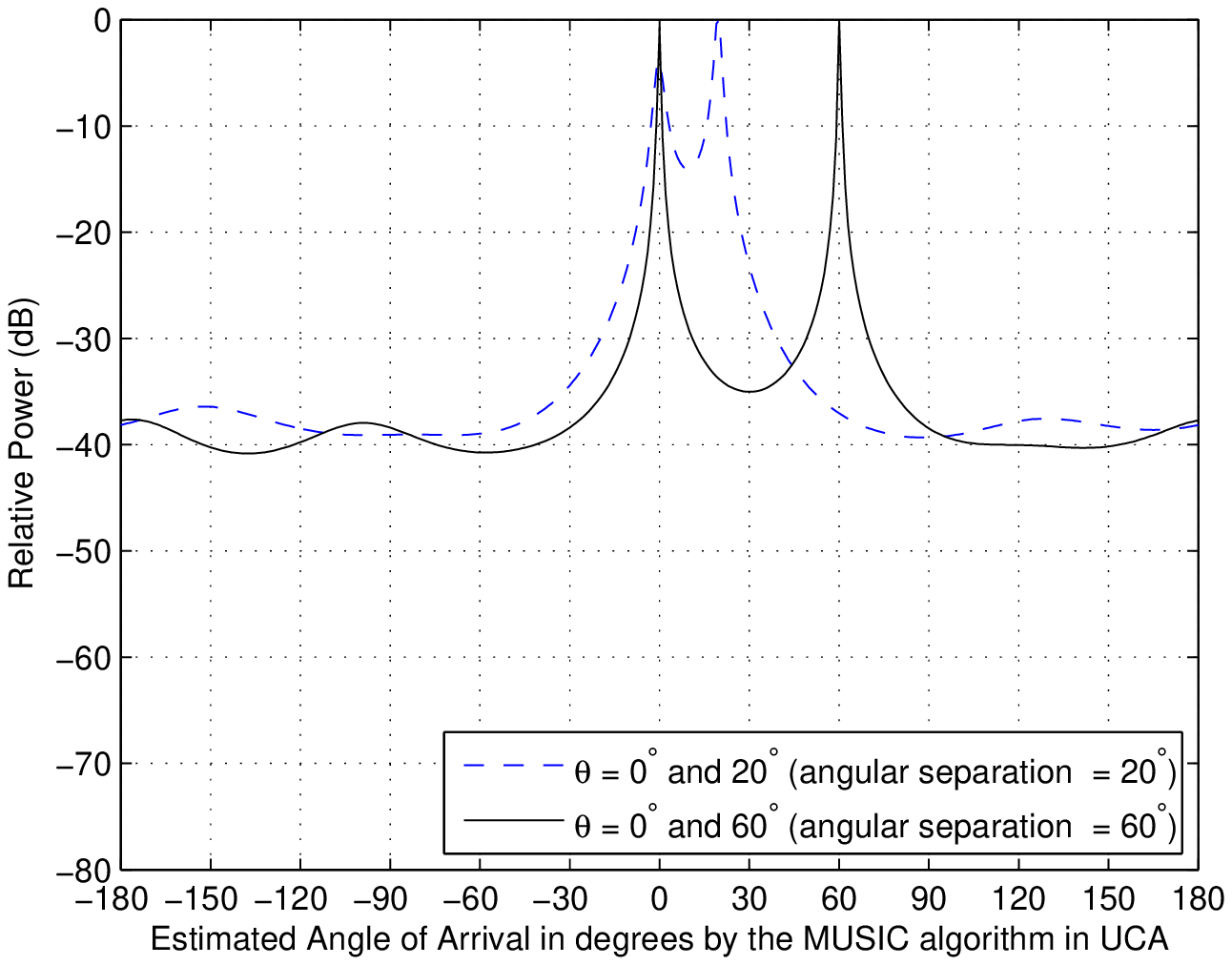}
        \caption[Impact of changing the angular separation between the incident signals impinging on UCA on the performance of MUSIC algorithm with settings ($N$=5, $\theta_e$=$20^o$, $d$=0.5$\lambda$ , SNR=10dB and $K$=100)]{Impact of changing the angular separation between the incident signals impinging on UCA on the performance of MUSIC algorithm with settings ($N$=5, $\theta_e$=$20^o$, $d$=0.5$\lambda$ , SNR=10dB and $K$=100)}
		 \label{Fig:Angular_Separation_2}
\end{figure}

\paragraph{Angular Separation between Incident Signals} \mbox{}\\
In this test, the impact of the angular separation between the incoming signals on DOA algorithm is examined by considering two scenarios. In the first scenario, UCA will receive two incident signals ($0^o$, $20^o$) having a small angular separation of $20^o$. In the second scenario, the angular separation will be increased to $50^o$ as the two incident signals will have a direction of arrival of $0^o$ and $60^o$. Clearly, From Figure \ref{Fig:Angular_Separation_2}, that increasing the angular separation will improve the performance of MUSIC algorithm through producing sharp spectral peaks and reduce the noise floor. The same conclusion is deduced from Table \ref{Root8} and \ref{Esprit8} where the detection accuracy of Root-MUSIC and ESPRIT is increased as the angular separation is increased. However, The MUSIC algorithm offers more results for smaller angular resolution. At such case, the resolution of Root-MUSIC and ESPRIT can be increased by improving the estimation of VULA which require increasing the number of elements.    

\begin{table}[H]
\centering
  \caption{Impact of changing the angular separation between the incident signals impinging on UCA on the performance of Root-MUSIC algorithm with settings ($N$=5, $\theta_e$=$20^o$, $d$=0.5$\lambda$ , SNR=10dB and $K$=100)} 
  \label{Root8}
\begin{tabular}{c}
\includegraphics[width=1\textwidth,clip]{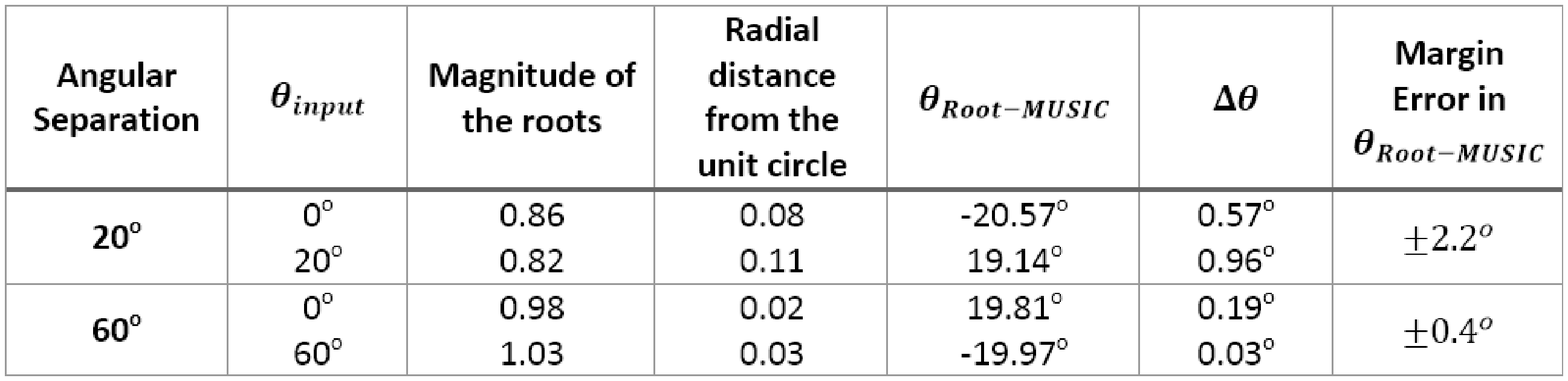}
\end{tabular}
\end{table}   

\begin{table}[H]
\centering
  \caption{Impact of changing the angular separation between the incident signals impinging on UCA on the performance of ESPRIT algorithm with settings ($N$=5, $\theta_e$=$20^o$, $d$=0.5$\lambda$ , SNR=10dB and $K$=100)} 
  \label{Esprit8}
\begin{tabular}{c}
\includegraphics[width=1\textwidth,clip]{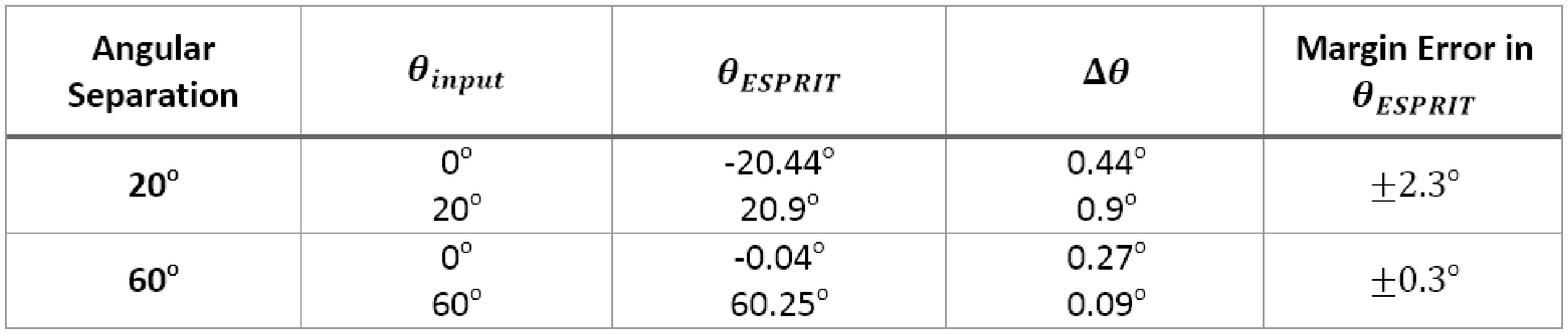}
\end{tabular}
\end{table}   

\begin{figure} [H]
       \centering
        \includegraphics[width=1\textwidth,clip]{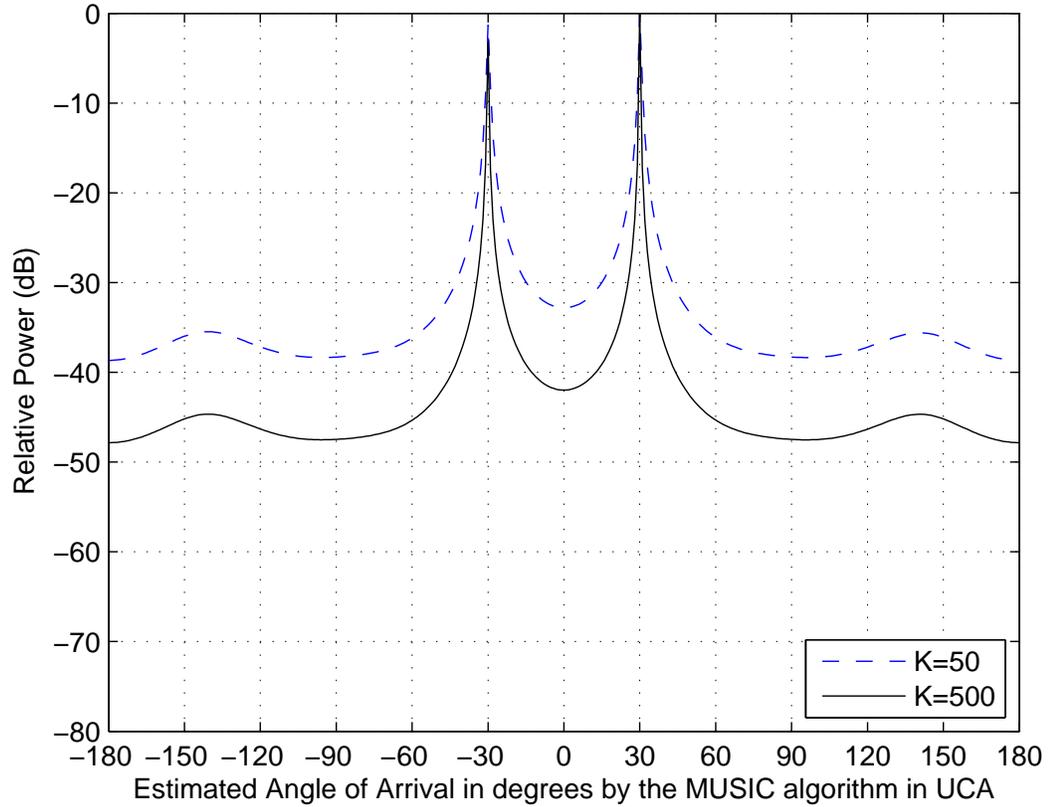}
        \caption[Impact of changing the number of samples of the incident signals impinging on UCA on the performance of MUSIC algorithm with settings ($N$=5, $\theta=30^o$  and $-30^o$, $\theta_e$=$20^o$, $d$=0.5$\lambda$ , SNR=10dB and $K$=100)]{Impact of changing the number of samples of the incident signals impinging on UCA on the performance of MUSIC algorithm with settings ($N$=5, $\theta=30^o$  and $-30^o$, $\theta_e$=$20^o$, $d$=0.5$\lambda$ , SNR=10dB and $K$=100)}
		 \label{Fig:Samples_2}
\end{figure}

\paragraph{Number of Samples} \mbox{}\\
In the next test, the impact of number of snapshots taken for the incoming signals on DOA algorithm is examined by applying two incoming signals ($-30^o$ , $30^o$) on with $K$=50 and $K$=500. From Figure \ref{Fig:Samples_2}, we conclude that increasing the number of snapshots improve the performance of MUSIC algorithm as the peak becomes sharper and the noise floor is lowered. This conclusion is also clear from Table \ref{Root9} and \ref{Esprit9} where the detection accuracy of Root-MUSIC and ESPRIT has increased leading to a lower noise margin. However, the does not approach zero due to present of 
quantization error from the PME. 

\begin{table}[H]
\centering
  \caption{Impact of changing the number of samples of the incident signals impinging on UCA on the performance of Root-MUSIC algorithm with settings ($N$=5, $\theta=30^o$  and $-30^o$, $\theta_e$=$20^o$, $d$=0.5$\lambda$ , SNR=10dB and $K$=100)} 
  \label{Root9}
\begin{tabular}{c}
\includegraphics[width=1\textwidth,clip]{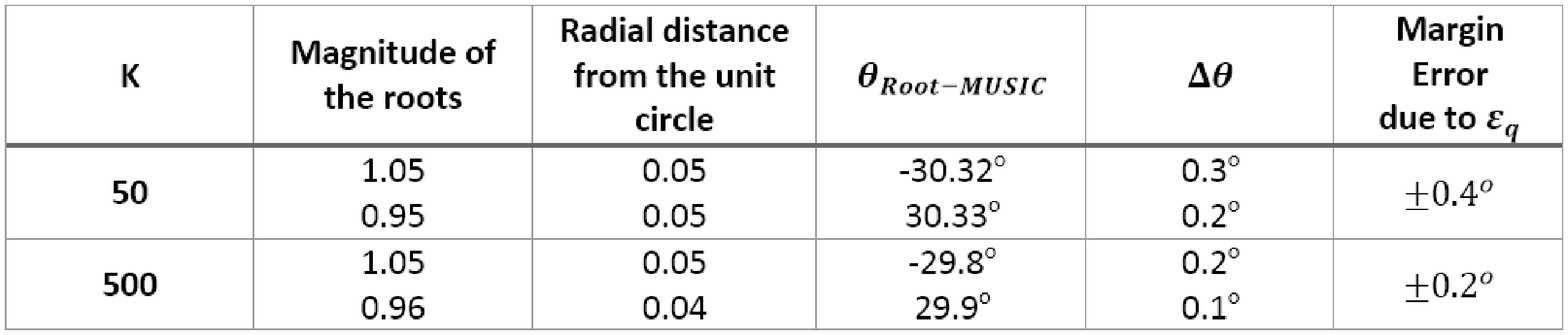}
\end{tabular}
\end{table}   

\begin{table}[H]
\centering
  \caption{Impact of changing the number of samples of the incident signals impinging on UCA on the performance of ESPRIT algorithm with settings ($N$=5, $\theta=30^o$  and $-30^o$, $\theta_e$=$20^o$, $d$=0.5$\lambda$ , SNR=10dB and $K$=100)} 
  \label{Esprit9}
\begin{tabular}{c}
\includegraphics[width=1\textwidth,clip]{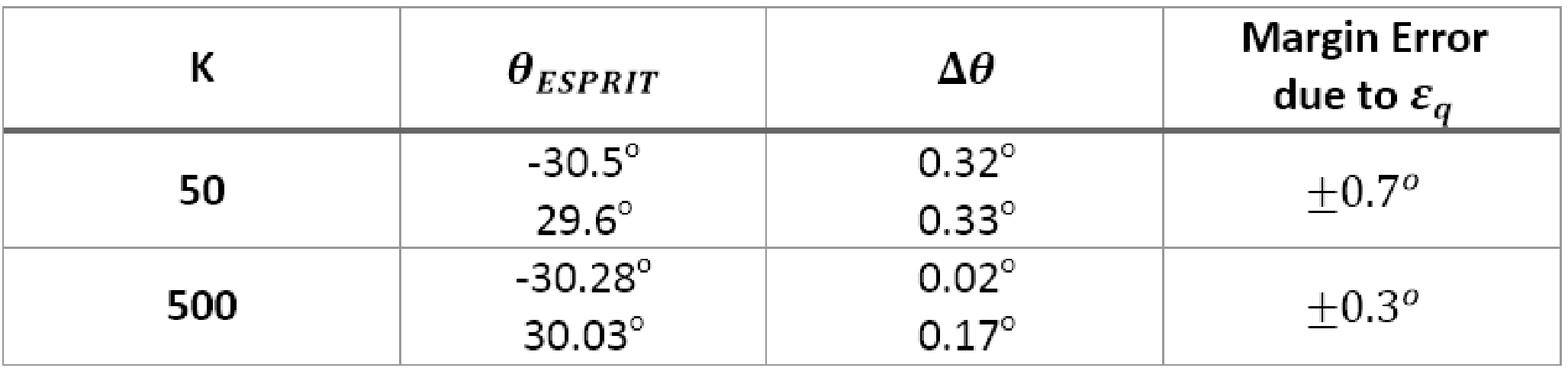}
\end{tabular}
\end{table}   

\begin{figure} [H]
       \centering
        \includegraphics[width=1\textwidth,clip]{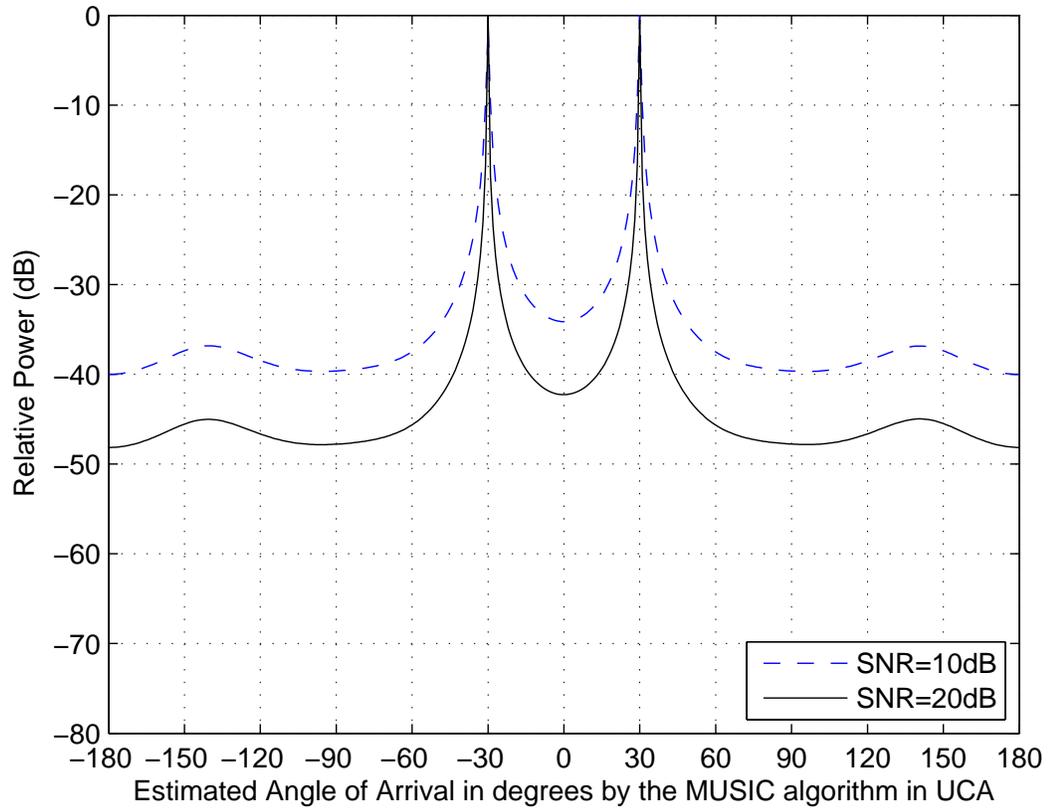}
        \caption[Impact of changing SNR for UCA on the performance of MUSIC algorithm with settings ($N$=5, $\theta$=$30^o$  and $-30^o$, $\theta_e$=$20^o$, $d$=0.5$\lambda$  and $K$=100)]{Impact of changing SNR for UCA on the performance of MUSIC algorithm with settings ($N$=5, $\theta$=$30^o$  and $-30^o$, $\theta_e$=$20^o$, $d$=0.5$\lambda$  and $K$=100)}
		 \label{Fig:SNR_2}
\end{figure}

\paragraph{Signal to Noise Ratio (SNR)} \mbox{}\\
In The last test for UCA, the impact of SNR on DOA algorithm is investigated through limiting the ratio between the signal power and noise to meet SNR=10 and SNR=20. Both condition of SNR is applied for detecting two incident signals ($-30^o$, $30^o$) on UCA. By examining Figure \ref{Fig:SNR_2} with Table 15 and 16, we conclude that increasing the SNR to higher values will improve the MUSIC algorithm as it will produce sharper peaks with reduced the noise level. Also, the detection of ESPRIT and Root-MSUIC is increased as evident from the reduction in quantization error. 

\begin{table}[H]
\centering
  \caption{Impact of changing SNR for UCA on the performance of Root-MUSIC algorithm with settings ($N$=5, $\theta$=$30^o$  and $-30^o$, $\theta_e$=$20^o$, $d$=0.5$\lambda$  and $K$=100)} 
  \label{Root10}
\begin{tabular}{c}
\includegraphics[width=1\textwidth,clip]{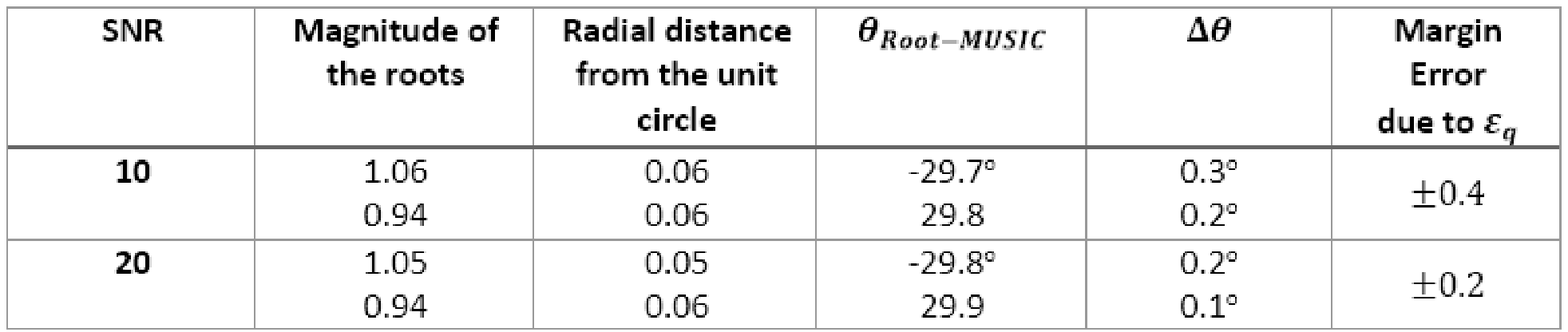}
\end{tabular}
\end{table}   

\begin{table}[H]
\centering
  \caption{Impact of changing SNR for UCA on the performance of ESPRIT algorithm with settings ($N$=5, $\theta$=$30^o$  and $-30^o$, $\theta_e$=$20^o$, $d$=0.5$\lambda$  and $K$=100)} 
  \label{Esprit10}
\begin{tabular}{c}
\includegraphics[width=1\textwidth,clip]{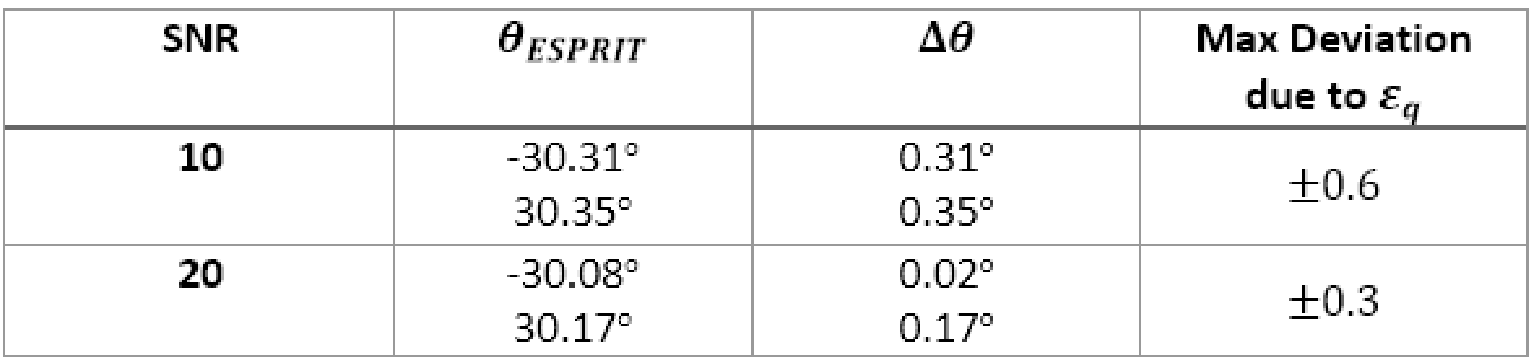}
\end{tabular}
\end{table}

\paragraph{Signal Correlation} \mbox{}\\
Unlike the previous discussed parameters, this particular parameter “signal correlation” will cause a problem for DOA estimation. Here, we will assume the worst scenario where all received signal are correlated. This in terms will reduce the rank of input covariance matrix to 1 preventing DOA algorithms from functioning. To resolve this issue, preprocessing techniques, explained in section \ref{Correlated Signals}, are need to restore the covariance matrix rank. Generally, the preprocessing techniques can operate along all the DOA algorithms as they only modify the covariance matrix. In the following test, we will show the efficiently each of the preprocessing techniques as they employed with MUSIC Algorithm in UCA. The geometry of UCA allows the use of two preprocessing techniques namely Forward Spatial Smoothing (FSS), Forward-Backward Spatial Smoothing (FBSS).These techniques are implemented through the PME.

In our test, we will consider a scenario where six correlated signals are impinging on UCA with angles ($-140^o$, $-80^o$ , $-20^o$ , $50^o$ , $80^o$ , $140^o$). Firstly, we will use the normal MUSIC and MUSIC with FSS to detect these correlated signals when they received by UCA having $N$=12. Clearly from Figure \ref{Fig:Uncorellated}, we conclude that MUSIC was unable to resolve the correlated signals alone but MUSIC succeeds when it used with FSS. The FSS technique is theoretically capable of detecting $N/2$ correlated signals and this is proven from simulation results as $12/2$=6. By reducing $N$=9, FSS will fail to detect the correlated signals but FBSS will succeed as shown in Figure \ref{Fig:Uncorellated_1}. That is because theoretically, FBSS can detect up to $2N/3$ signals which is in this case $  (2 \times 9)/3$=6. In term of computational complexity, FBSS requires more computation over FSS due to the fact that FBSS divides the main array into subarrays in both forward and backward direction while FSS operation is restricted to the forward direction. However, for expected high number of correlated signals, FBSS is the preferred choice. 

\begin{figure} [H]
       \centering
        \includegraphics[width=1\textwidth,clip]{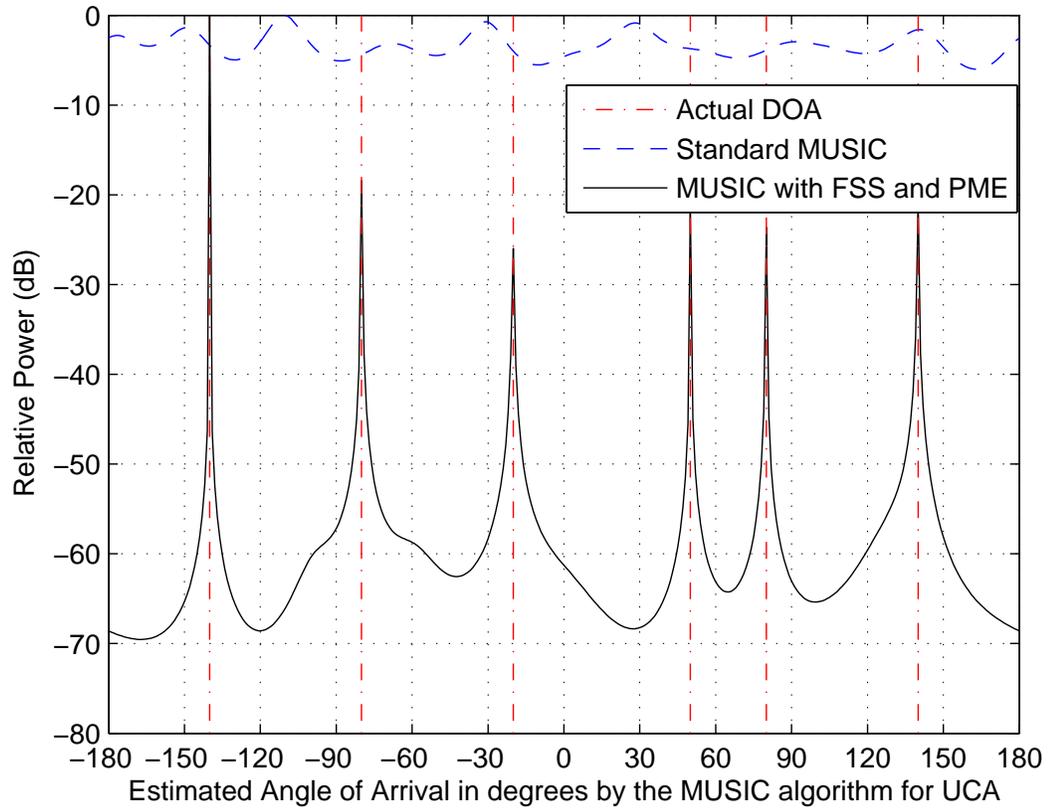}
        \caption[Implementation of standard MUSIC and MUSIC with FSS for UCA in Correlated Environment with the settings ($N$=12, $\theta=-140^o,  -80^o, -20^o, 50^o, 80^o, 140^o$, $\theta_e$=$20^o$,  $d$=0.5$\lambda$ , SNR=20dB and $K$=100)]{Implementation of standard MUSIC and MUSIC with FSS for UCA in Correlated Environment with the settings ($N$=12, $\theta=-140^o , -80^o, -20^o, 50^o, 80^o, 140^o$, $\theta_e$=$20^o$,  $d$=0.5$\lambda$ , SNR=20dB and $K$=100)}
		 \label{Fig:Uncorellated}
\end{figure}

\begin{figure} [H]
       \centering
        \includegraphics[width=1\textwidth,clip]{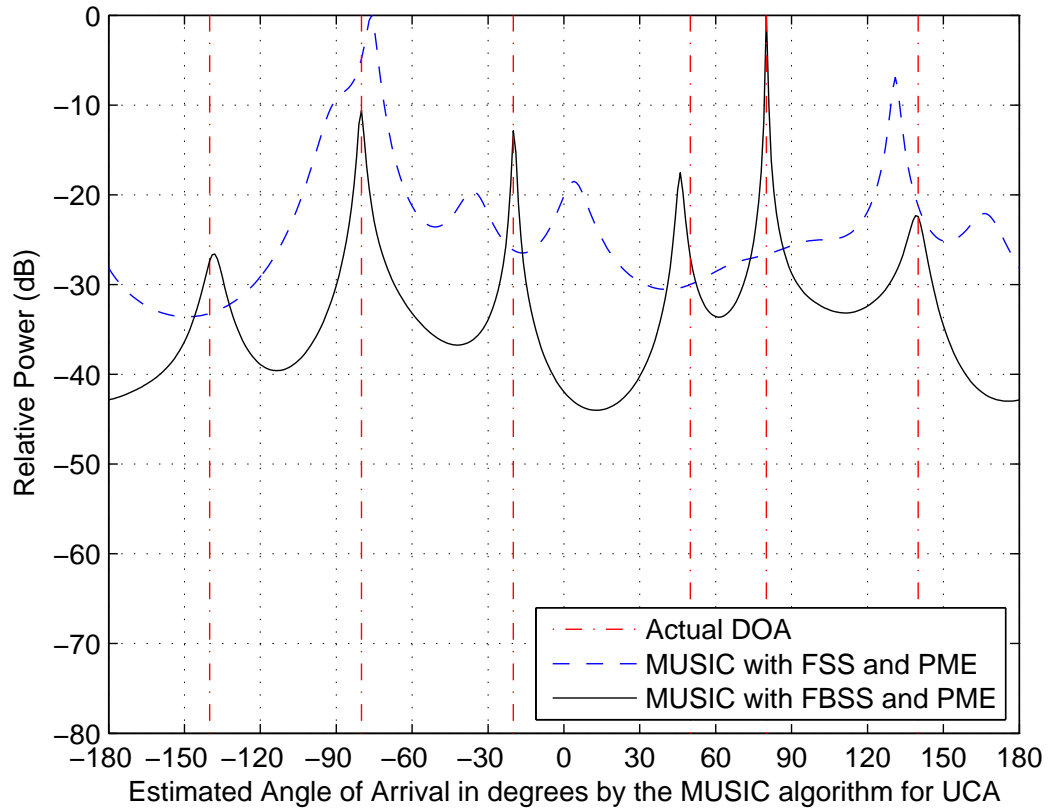}
        \caption[Implementation of FSS and FBSS using MUSIC algorithm for UCA in Correlated Environment with the settings ($N$=9, $\theta=-140^o,  -80^o, -20^o, 50^o, 80^o, 140^o$, $\theta_e$=$20^o$,  $d$=0.5$\lambda$ , SNR=20dB and $K$=100)]{Implementation of FSS and FBSS using MUSIC algorithm for UCA in Correlated Environment with the settings ($N$=9, $\theta=-140^o,  -80^o, -20^o, 50^o, 80^o, 140^o$, $\theta_e$=$20^o$,  $d$=0.5$\lambda$ , SNR=20dB and $K$=100)}
		 \label{Fig:Uncorellated_1}
\end{figure}

\subsubsection{System Modeling}  
In order to conform the usability, practicality, and accuracy of the work done, a virtual simulation using MATLAB was used. In the simulation, the conditions of a practical environment were mimicked with utmost precision possible. The following will introduce one of the simulated scenarios:

In this scenario, a land with an area of 10 $\times$ 10 m is being considered with frequency of interest equal to the 1 GHz and number of snapshots is 100.  This scenario tries to simulate the behavior of the MUSIC algorithm when using UCA-geometry sensor nodes with N=5. In the test carried out by this scenario, three nodes were chosen to have random locations and the location happened to be as in the following Figure:

\begin{figure} [H]
       \centering
        \includegraphics[width=1\textwidth,clip]{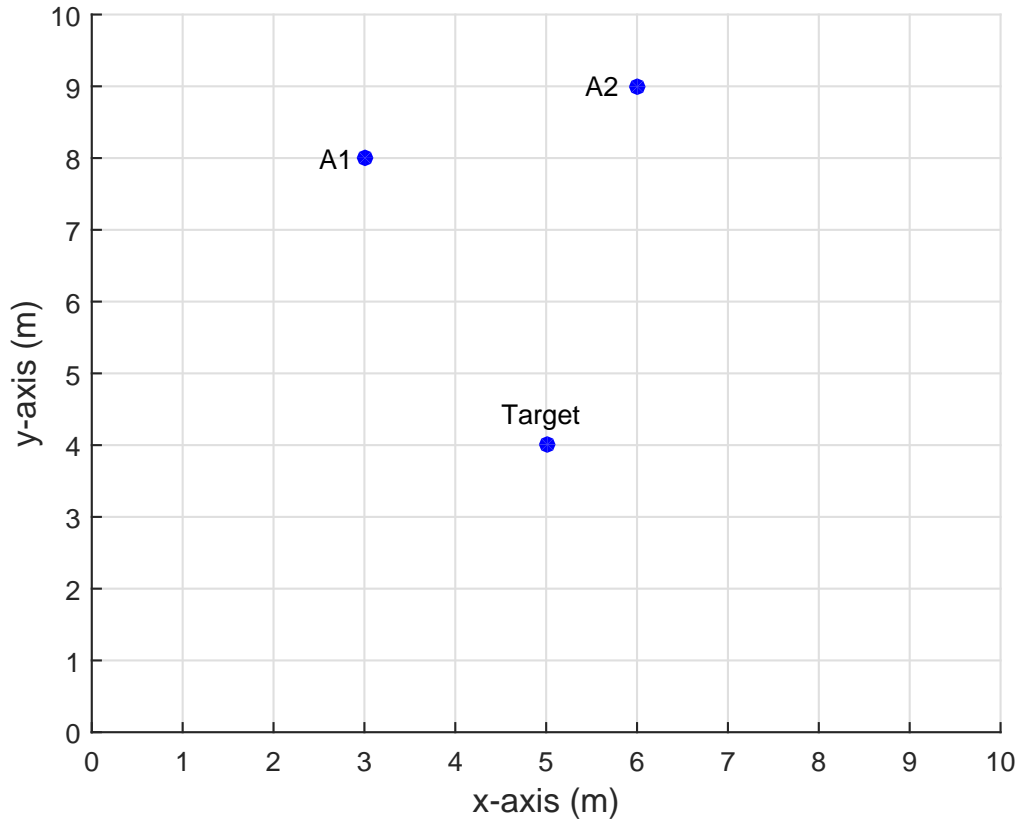}
        \caption[Location of the nodes in a practical environment]{Location of the nodes in a practical environment}
		 \label{Fig:location}
\end{figure}

\begin{table}[H]
\centering
  \caption{Nodes location on the environment} 
  \label{Esprit10}
\begin{tabular}{c}
\includegraphics[width=1\textwidth,clip]{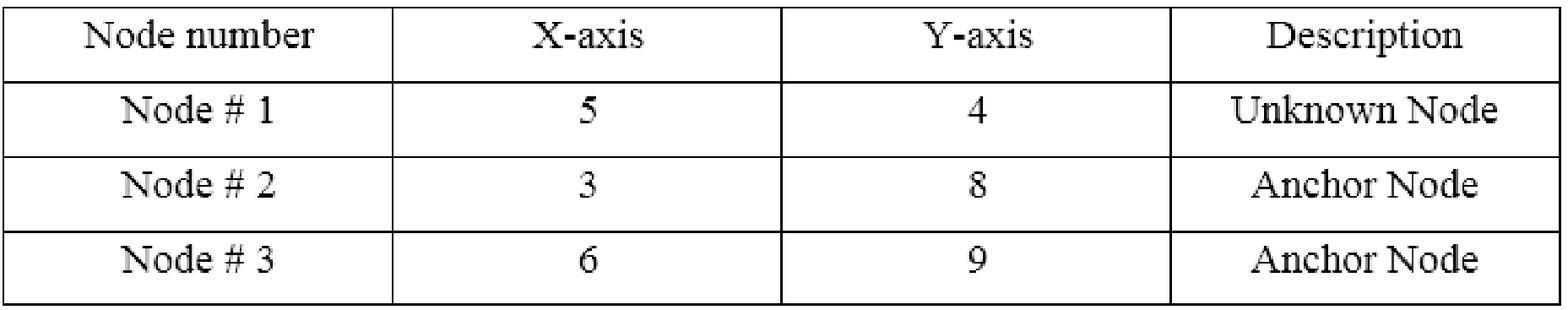}
\end{tabular}
\end{table}

The Figure above shows the location of the nodes in a practical environment. The scenario considers two anchor node, nodes with known location, and a single unknown node whose location will be determined using the anchor nodes using triangulation method. Another aim of this scenario is to see the how the SNR behave verse the RMSE of the algorithm. The algorithm used to determine the location is MUSIC algorithm since it results in accurate results in UCA whereas using UCA-ESPRIT or UCA Root-MUSIC algorithms require additional steps, such as phase mood excitation, in the case of the correlated signals, which introduce small error. The scenario is carried out with both correlated signals and uncorrelated signals.

\paragraph{Uncorrelated Signals} \mbox{}\\
The first part of this scenario deals with the uncorrelated signals. In the case of the uncorrelated signal, the implementation was straightforward. In this scenario, node \#2 has an angle of the $116.56^o$, and node \#3 has an angle of $78.69^o$ with the horizontal line. The behavior of the system is shown in the following Figure:

\begin{figure} [H]
       \centering
        \includegraphics[width=1\textwidth,clip]{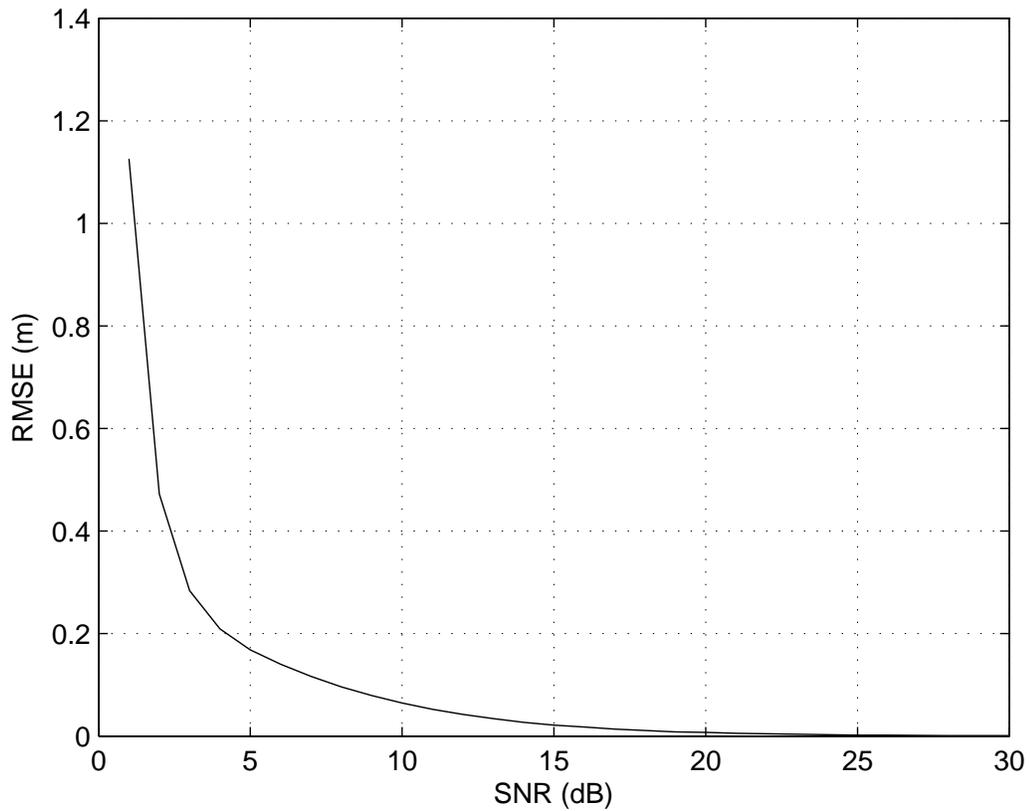}
        \caption[RMSE for different SNR for uncorrelated signals]{RMSE for different SNR for uncorrelated signals}
		 \label{Fig:Uncorellated}
\end{figure}

The Figure above shows the RMSE of different values of SNR in the practical environment chosen when uncorrelated signals are in use. It is important to highlight that the number of antennas in the UCA geometry is five. The general behavior that can be seen from the graph is that as the value of the SNR increases, the value of RMSE decreases. Moreover, as the value of the SNR grows greater, the RMSE shows an asymptotic behavior around 0. This indicates that the error as the SNR increases decreases drastically, which shows how accurate this method is in estimating the location of the nodes.

\paragraph{Correlated Signals} \mbox{}\\
The second part of this scenario deals with the correlated signals. as it was established earlier, the correlated signals defers from the uncorrelated signals in its need for more careful implementation in order not to obtain wrongful results. The same conditions were applied for this part of the scenario, same practical environment with same locations of anchor nodes and unknown node. The only additional step in this part is the need for the use of spatial smoothing technique, specifically Forward/Backward spatial smoothing FBSS. In this scenario, node \# 2 has an angle of the $116.56^o$, and node \# 3 has an angle of $78.69^o$ with the horizontal line. The behavior of this system is shown in the following Figure:

\begin{figure} [H]
       \centering
        \includegraphics[width=1\textwidth,clip]{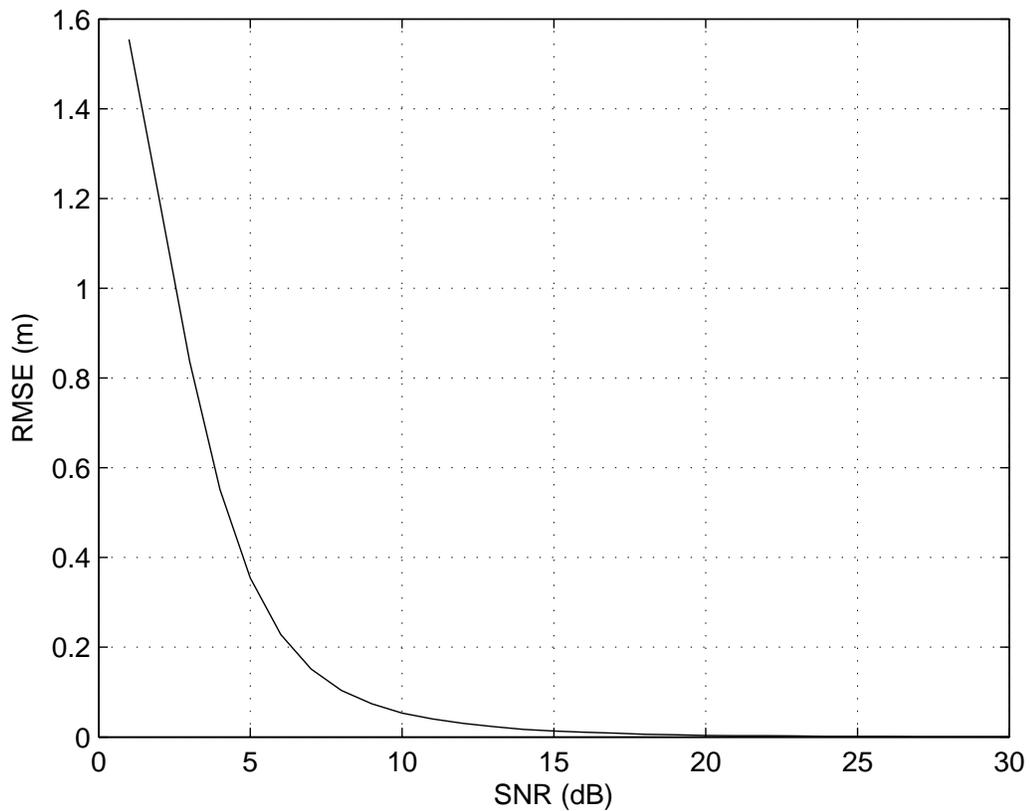}
        \caption[RMSE for different SNR for correlated signals]{RMSE for different SNR for correlated signals}
		 \label{Fig:Uncorellated}
\end{figure}

The Figure above shows the RMSE of different values of SNR in the practical environment chosen when correlated signals are in use. It is important to highlight that the number of antennas in the UCA geometry is five. The general behavior that can be seen from the graph is that as the value of the SNR increases, the value of RMSE decreases. Moreover, as the value of the SNR grows greater, the RMSE shows an asymptotic behavior around 0. This indicates that the error as the SNR increases decreases drastically, which shows how accurate this method is in estimating the location of the nodes.

\chapter{Hybrid Techniques}
\label{Hybrid Techniques}
In this chapter, different hybrid techniques is discussed to improve the accuracy of estimation of the location of the unknown node \cite{kulaib_efficient_2014,kulaib_investigation_2013,yick_wireless_2008}. This is because the measurement noise from various techniques comes from different sources. Consequently, the errors in the estimation of the position using different techniques are partially independent. The independence, in these measurements, allows creating an estimator with a better performance. On other contrary, there is high complexity in terms of time and computation \cite{al-ardi_computationally_2004,al-ardi_computationally_2005,alhajri_hybrid_2015,goian_fast_2015,alhajri_hybrid_2015-1}.

\section{Hybrid RSS and DOA using one Hybrid Node}
A well-known hybrid technique combines RSS and DOA. As previously stated, the RSS has low accuracy than DOA. The reason behind this is that it is difficult to perfectly model the signal propagation in the environment \cite{chan_hybrid_2014}. Therefore, DOA can compensate for this low accuracy and provide a hybrid system that has higher accuracy than the RSS alone \cite{kulaib_investigation_2013}.  In \cite{kulaib_investigation_2013}, it was proposed that one hybrid node can estimate the position of the unknown node. This hybrid node uses the antenna elements to find the target direction. This is equivalent to a line that originates from the array center to the target. Also, these elements measure RSS and each of these measurements is modelled by a circle which its center is the antenna element.  The line and each circle intersect at one point. These intersections are averaged to find the location of the unknown node. To model the line and circles mathematically, we will use the parametric of the line and implicit circle equations. Thus, the parametric equation of the line in 2D is:  
\begin{equation}%
\boldsymbol{\mathrm{Q_{parametric}}}=\boldsymbol{\mathrm{p}}+t\boldsymbol{\mathrm{d}}
\label{5.1}
\end{equation}%
where \textbf{Q} is the matric that contains $x$ , $y$ components and \textbf{p}=($x_0,y_0$) is a point on the line, $t$ is the parameter and $\boldsymbol{\mathrm{d_{dir}}}$ is the direction vector for line. 

The equation above is expressed in the $x$ and $y$ components form as the following:

\begin{equation}%
x=x_0+f_{dir}t
\label{5.2}
\end{equation}%
\begin{equation}%
y=y_0+g_{dir}t
\label{5.3}
\end{equation}%
where $f_{dir}$ and $g_{dir}$ are the components of the direction vector $\boldsymbol{\mathrm{d_{dir}}}$. In this hybrid approach this is evaluated by the product of arrow length and the $\cos$ and $\sin$ the angle from DOA for $f_{dir}$ and $g_{dir}$, respectively. 

The circle equation with center ($x_c,y_c$), other than origin is expressed as the following:
\begin{equation}%
(x-x_c)^2+(y-y_c)^2=r^2
\label{5.4}
\end{equation}%
where $r$ is the radius.

This line and each circle intersect at one point. Thus, substituting the equations \ref{5.2} and \ref{5.3} in equation \ref{5.4}, the result is the following: 
\begin{equation}%
(x_0+f_{dir}t-x_c)^2+(y_0+g_{dir}t-y_c)^2=r^2
\label{5.5}
\end{equation}%
By expanding the terms and eliminating the terms that are equivalent to $r$, the equation reduces to the following:
\begin{equation}%
(f_{dir}t)^2+(x_0-x_c)f_{dir}t=-(g_{dir}t)^2-(y_0-y_c)g_{dir}t
\label{5.6}
\end{equation}%
Dividing by $t$ and isolating it at one side results in the following:
\begin{equation}%
t=\dfrac{(x_c-x_0)f_{dir}+(y_c-y_0)g_{dir}}{(f_{dir})^2+(g_{dir})^2}
\label{5.7}
\end{equation}%

After evaluating this value, it is substituted in the equations \ref{5.2} and \ref{5.3} and $x_0$ and $y_0$ is the estimated position node with respect to the antenna element. Consequently, the equations \ref{5.2} and \ref{5.3} is $x$ and $y$ coordinates of the intersection point, respectively.  These intersections are averaged to get one point, which is the location of the unknown node.
\subsection{Simulation Results}
\begin{figure} [h]
       \centering
        \includegraphics[width=0.8\textwidth,clip]{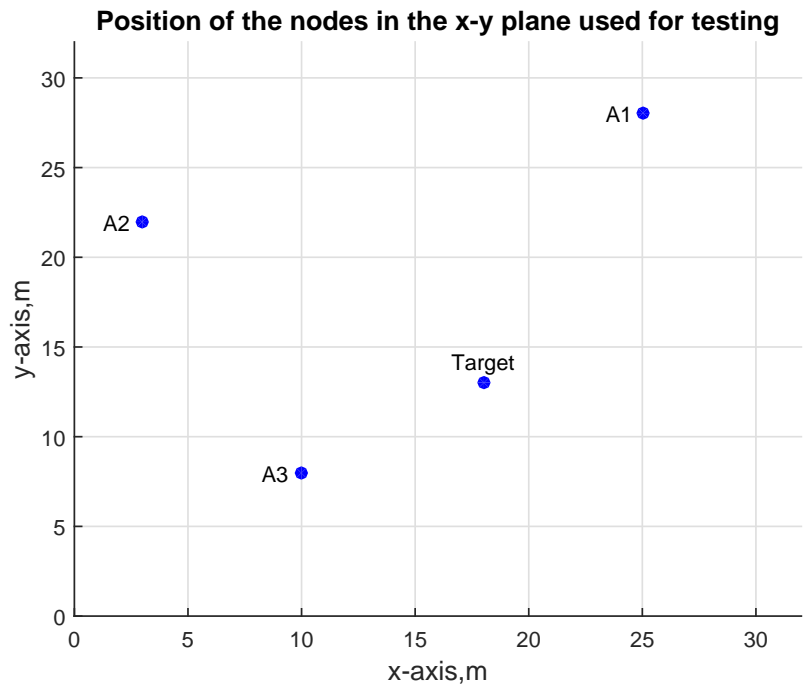}
        \caption[Position of the nodes in the x-y plane used for hybrid testing]{Position of the nodes in the x-y plane used for hybrid testing}
		 \label{Fig:Environment_Hybrid}
\end{figure}

To test this algorithm, 3 anchor nodes are placed on the $x$-$y$ plane with the size of (30m $\times$ 30m) as shown in Figure \ref{Fig:Environment_Hybrid}.  Each anchor node contains 4 antennas placed in circular configuration, the center frequency is 1 GHz and the RMSE is computed 150 times for every SNR value.     

Comparing the results with RSS alone, the hybrid shows better performance than the RSS technique. This is an expected result because the DOA has high accuracy which improves the hybrid technique.

\begin{figure} [h]
       \centering
        \includegraphics[width=1\textwidth,clip]{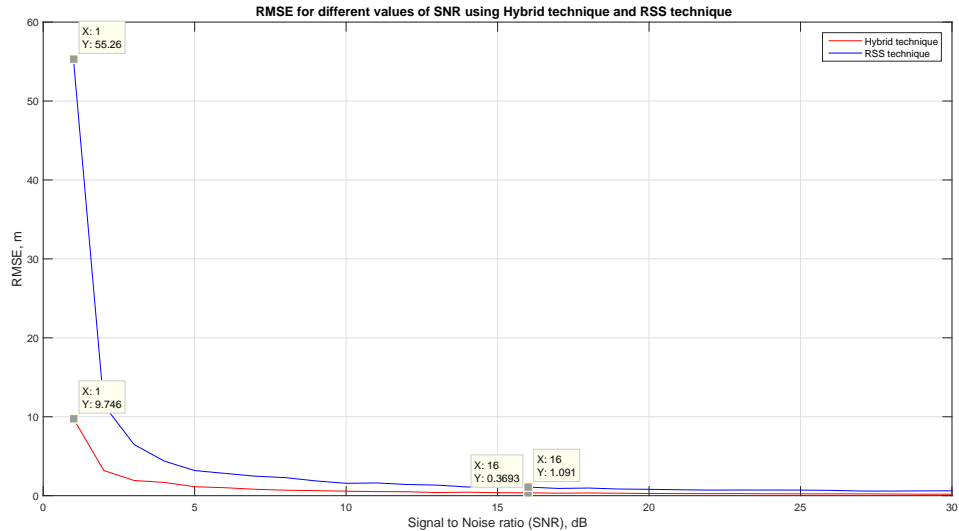}
        \caption[RMSE for different values of SNR using Hybrid technique and RSS technique]{RMSE for different values of SNR using Hybrid technique and RSS technique}
		 \label{Fig:RMSE_Hybrid_1}
\end{figure} 

Clearly, from the figure \ref{Fig:RMSE_Hybrid_1} the hybrid technique outperforms the RSS technique. The results in the hybrid are about 4 times better. One hybrid node is sufficient to produce these results. This matches the expected result because the presence of the DOA improves the accuracy, lower RMSE at each SNR value, hence, better estimation.However, DOA provides better accuracy compared to the hybrid technique \cite{kulaib_investigation_2013}.

However, this algorithm does not take into consideration an environment that has correlated signals. For this reason, the next algorithm will address this point by using the spatial smoothing in DOA technique.

\section{Hybrid RSS and DOA with Spatial Smoothing}
When the environment is correlated, the previous hybrid technique will no longer be able to locate the unknown node. That is because the correlation will interfere with the execution of DOA algorithms which is needed by the hybrid technique to operate. In fact, the DOA algorithms are designed to work under the assumption that the received signals is uncorrelated resulting in non-singular covariance matrix. However, the covariance matrix becomes singular when the received signals are correlated causing a violation in the principles of DOA algorithms. Considering a UCA-configuration for the used nodes, the correlation problem can be overcome by using spatial smoothing, particularly Forward Backward Spatial Smoothing FBSS. The spatial smoothing will divide the main array into overlaying subarrays and then average their covariance matrices to obtain a non-singular smoothed covariance matrix. However, FBSS is done linearly and hence the UCA must be converted into VULA using PME before FBSS can be used.
\subsection{Simulation Results}

\begin{figure} [h]
       \centering
        \includegraphics[width=1\textwidth,clip]{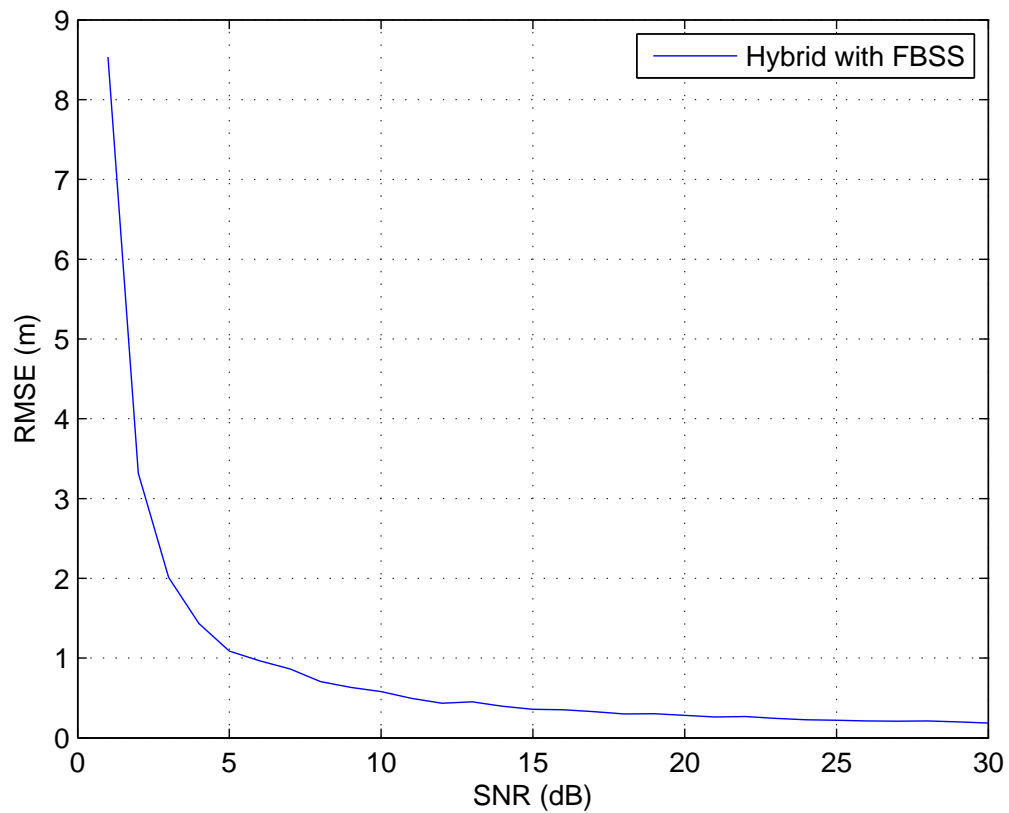}
        \caption[RMSE for different values of SNR using RSS and DOA Hybrid technique with Spatial Smoothing]{RMSE for different values of SNR using RSS and DOA Hybrid technique with Spatial Smoothing}
		 \label{Fig:hybridspatial}
\end{figure}

In our simulation for the hybrid technique, we will consider the same environment depicted in Figure \ref{Fig:hybridspatial}. In order to make the environment correlated, the unknown node will experience a scenario where it received three correlated signals simultaneously. The known node will receive three correlated signals with the angles $53.13^o$, $116.57^o$, $32^o$ in respect to the horizontal line. Also, frequency of the received signals are set to 1 GHz and the number of snapshot taken for them is 100.  Both the unknown and anchor nodes are assumed to possess a UCA configuration with 8 antennas and RSS capability. The behavior of the hybrid technique is showing in Figure \ref{Fig:hybridspatial}.By analyzing Figure \ref{Fig:hybridspatial}, we observe that as the SNR increases, the value of RMSE decreases. In addition, as the value of SNR grows larger, the RMSE, the RMSE follows a plateau of around 0. From this results, we conclude that the Spatial Smoothing was capable of restoring the operation of hybrid system in correlated environment.

\section{Hybrid technique using Least Square Based Techniques}
However, when the RSS technique is used to estimate the position, the noise in the environment affect the intersection of the circles and the estimation will not be in single point .Therefore, LS is method to minimize the effect of the noise and give one point. The LS is a close form and easy to compute but its best performance appears when the noise power is small \cite{chan_hybrid_2014}.   
In this algorithm, the LS approach is used to find the location of the unknown node. In this scheme, two RSS and one Hybrid nodes are used to find the location of the unknown node. Initially, the RSS technique is used to find the measurements. Then, the results are used to estimate the location of the unknown node using LS approach as in equation \ref{3.18}. 

Then, with this result a circle is drawn using this estimated position using LS and its center is the hybrid node. The radius of this circle is the difference between the center and estimated point, as shown in equation \ref{5.10}.

Assume the hybrid node $p_{hyb}$  and for the LS point $p_{LS}$. Then, the difference in the $x$-coordinates is noted as  $x_{diff}$ and the equation is the following:

\begin{equation}%
x_{diff}=\left| x_{hyb}-x_{LS}\right|
\label{5.8}
\end{equation}%

The same thing goes for the y-coordinates and the results are the following:

\begin{equation}%
y_{diff}=\left| y_{hyb}-y_{LS}\right|
\label{5.9}
\end{equation}%

Using the results from equations \ref{5.8} and \ref{5.9} and substitute them in equation \ref{5.10} which represents the radius.

\begin{equation}%
r=\sqrt{(x_{diff})^2+(y_{diff})^2}
\label{5.10}
\end{equation}%

DOA will estimate the position by measuring the angle ($\theta_{m}$) and drawing a line. This will result in another point. The point  $p_{DOA}$   is found by the following equations:  

\begin{equation}%
x_{DOA}=x_{hyb}+r\cos(\theta_{m})
\label{5.11}
\end{equation}%

\begin{equation}%
y_{DOA}=y_{hyb}+r\sin(\theta_{m})
\label{5.12}
\end{equation}%

These two points are averaged and the location is estimated. 
\begin{equation}%
\boldsymbol{\mathrm{p}}_s=\frac{1}{2}\left[\begin{array}{l}x_{LS}+x_{DOA}\\y_{LS}+y_{DOA}\end{array}\right]
\label{5.13}
\end{equation}%

However, the estimated position using LS includes high error because the least square does not provide information about the different distances between different anchors to the unknown node or simply the “link quality”.  Thus, the hybrid system will produce better results if this information is utilized.  The link which has high noise variance can be given less weight compared to that with low noise. This is what basically the WLS algorithm does; it introduces the weighting matrix \cite{salman_enhanced_2014}. 

In this hybrid algorithm, the WLS estimator is used instead of LS. The same method that was used in Hybrid LS is used for the fusing technique. The results are represented in the next section.  

\subsection{Simulation Results}

\begin{figure} [h]
       \centering
        \includegraphics[width=1\textwidth,clip]{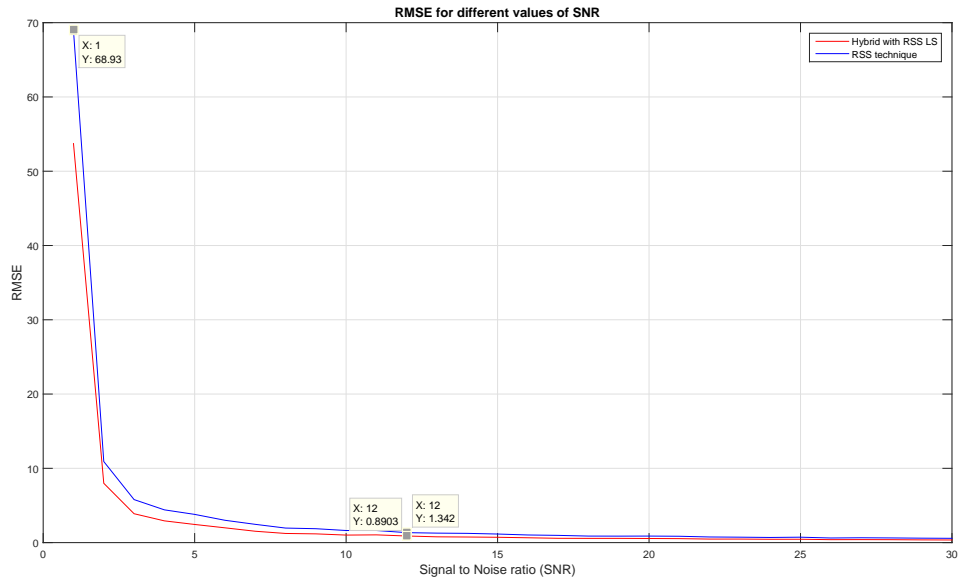}
        \caption[RMSE for different values of SNR using Hybrid technique RSS LS and RSS technique]{RMSE for different values of SNR using Hybrid technique RSS LS and RSS technique}
		 \label{Fig:hybrid2}
\end{figure}

The environment that was used in the first hybrid technique is used in testing the LS and WLS hybrid algorithm. The Hybrid anchor node contains 4 antennas placed in circular configuration. Each RSS nodes contain one antenna. The result is in Figure \ref{Fig:hybrid2}. The Hybrid technique shows less RMSE values than RSS with LS technique alone. However, in a published paper, it states that the hybrid technique with LS has no improvement over the RSS LS technique because according to method the RSS dominates and the estimator works only on the RSS measurements \cite{wang_hybrid_2012}. Thus, by using this technique we were able to overcome this problem.  

\begin{figure} [h]
       \centering
        \includegraphics[width=1\textwidth,clip]{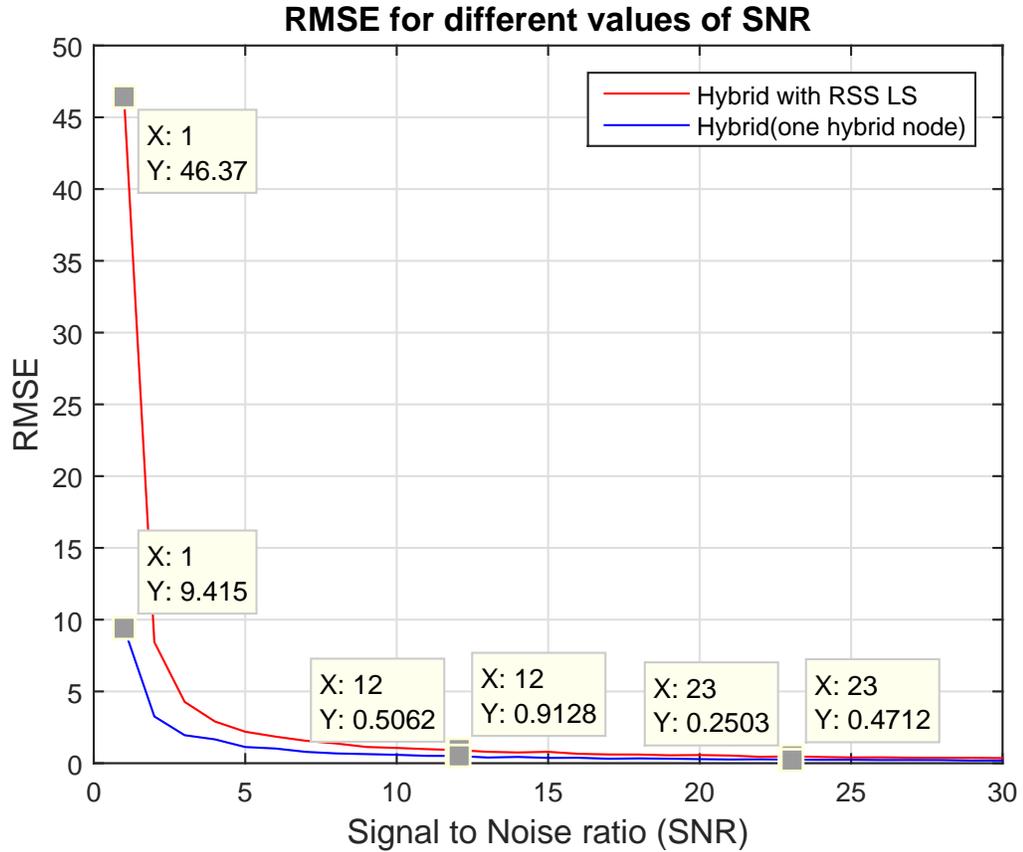}
        \caption[RMSE for different values of SNR using Hybrid technique RSS LS and first proposed Hybrid technique]{RMSE for different values of SNR using Hybrid technique RSS LS and first proposed Hybrid technique}
		 \label{Fig:h1_HLS}
\end{figure}

Comparing these results with the results in the first algorithm, we concluded that the first technique produces better results than the technique with LS. This because that there are more RSS nodes involved in the measurements so more errors will be introduced to the system, hence, low accuracy. However, the first technique uses only one hybrid node, thus, less error in the measurements compared to the Hybrid LS technique. The results are shown in Figure \ref{Fig:h1_HLS}. 
   
\begin{figure} [h]
       \centering
        \includegraphics[width=1\textwidth,clip]{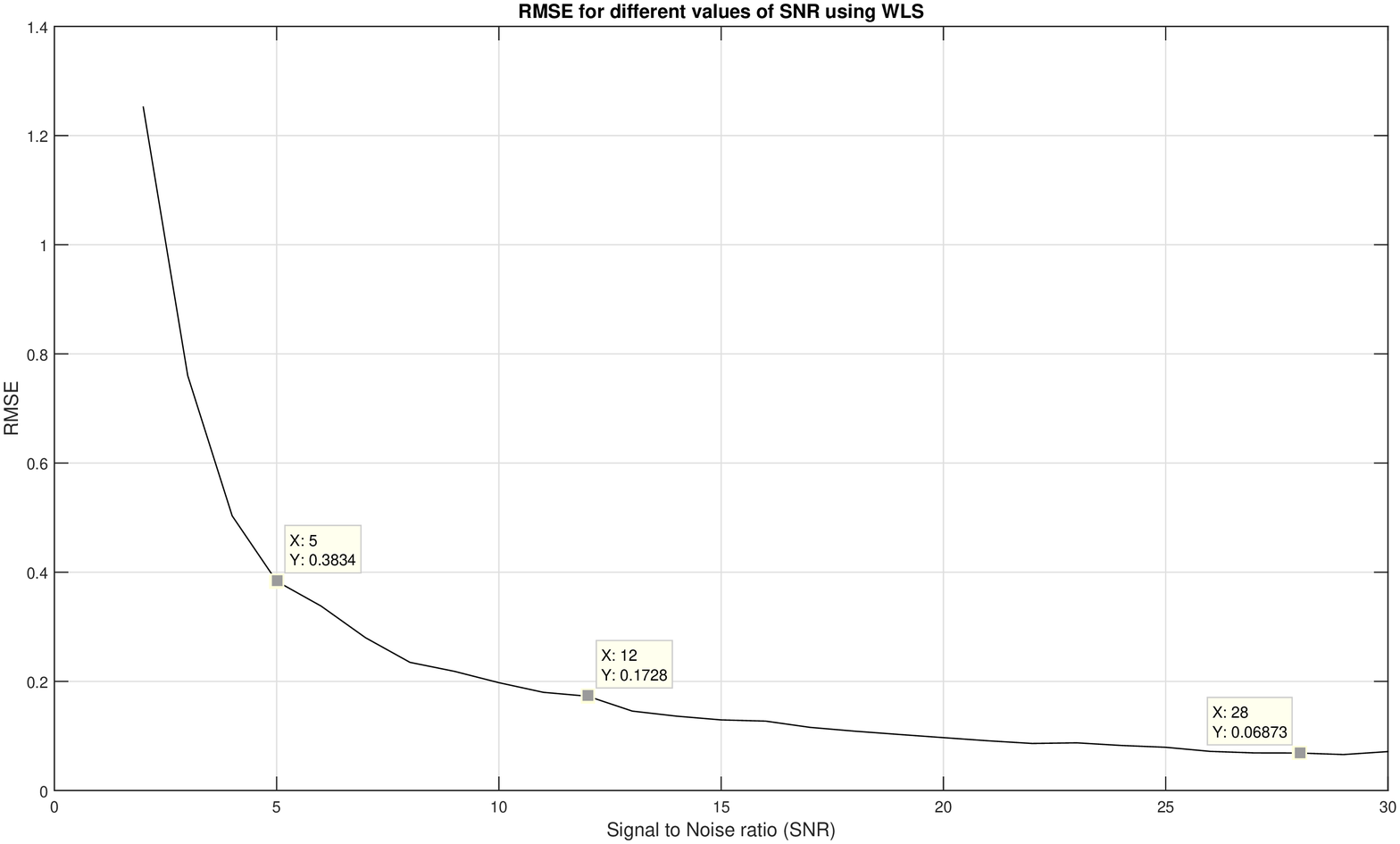}
        \caption[RMSE for different values of SNR using Hybrid technique RSS WLS]{RMSE for different values of SNR using Hybrid technique RSS WLS}
		 \label{Fig:hybridWLS}
\end{figure}

The third comparison was conducted in Hybrid LS and WLS and RSS technique. From the analysis, it is concluded that the WLS will outperform all the other technique. In Figure \ref{Fig:hybridWLS}, the results prove what it is in the theory. The WLS is the best performance because it gives more weight for the shorter distance with respect to the unknown node. At higher SNR the WLS shows dramatic reduction in error compared to the other techniques.

\section{Hybrid Technique using Two Lines}
After investigating different techniques that use the RSS technique to draw circles to corresponds to its measurements and the DOA to plot line that indicates the computed angle. It is worth to investigate the RSS measurements to be represented as a line. Thus, in this algorithm, both RSS and DOA will be used to draw two lines and the intersection for these two lines will be the estimated position($p_s$). To approach this technique, one RSS ($p_1$) and one hybrid nodes ($p_{hyp}$) are used. These nodes form two circles and each one is a center for each circle. These two circles intersect at two points to form the LOP. The equation of this line is represented as the following:    

\begin{equation}%
(x_{hyp}-x_1)x_s+(y_{hyp}-y_1)y_s=\frac{1}{2}(\left\Vert x_{hyp}\right\Vert^2-\left\Vert x_1\right\Vert^2+D^2_1-D^2_{hyp})
\label{5.14}
\end{equation}%

The second line is found from the DOA measurements which represented using the equations \ref{5.11} and \ref{5.12}. In this case $x_{DOA}=x_s$  and  $y_{DOA}=y_s$ . Thus, by rearranging the preceding two equations

\begin{equation}%
\frac{x_s-x_{hyp}}{\cos(\theta_{m})}=\frac{y_s-y_{hyp}}{\sin(\theta_{m})}
\label{5.15}
\end{equation}%

\begin{equation}%
(\sin(\theta_{m}))x_s-(\cos(\theta_{m}))y_s=(\sin(\theta_{m}))x_{hyp}-(\cos(\theta_{m}))y_{hyp}
\label{5.16}
\end{equation}%
 
Combing equations \ref{5.14} and \ref{5.16} will result in the following: 

\begin{equation}%
\left[\begin{array}{ll}(x_{hyp}-x_1)&(y_{hyp}-y_1)\\\sin(\theta_{m})&-\cos(\theta_{m})\end{array}\right]\left[\begin{array}{l}x_s\\y_s\end{array}\right]=\left[\begin{array}{l}\frac{1}{2}(\left\Vert x_{hyp}\right\Vert^2-\left\Vert x_1\right\Vert^2+D^2_1-D^2_{hyp})\\\sin(\theta_{m})x_{hyp}-\cos(\theta_{m})y_{hyp}\end{array}\right]
\label{5.17}
\end{equation}%

By assigning name for each matrix, the result is the following:

\begin{equation}%
\mathbf{\mathrm{Cp_s}}=\mathbf{\mathrm{D}}
\label{5.18}
\end{equation}%

Taking the inverse of both sides the results is the following:

\begin{equation}%
\mathbf{p}_s=\mathbf{\mathrm{C^{-1}D}}
\label{5.19}
\end{equation}%

The simulation results are presented in the following section. 

\subsection{Simulation Results}
\begin{figure} [h]
       \centering
        \includegraphics[width=1\textwidth,clip]{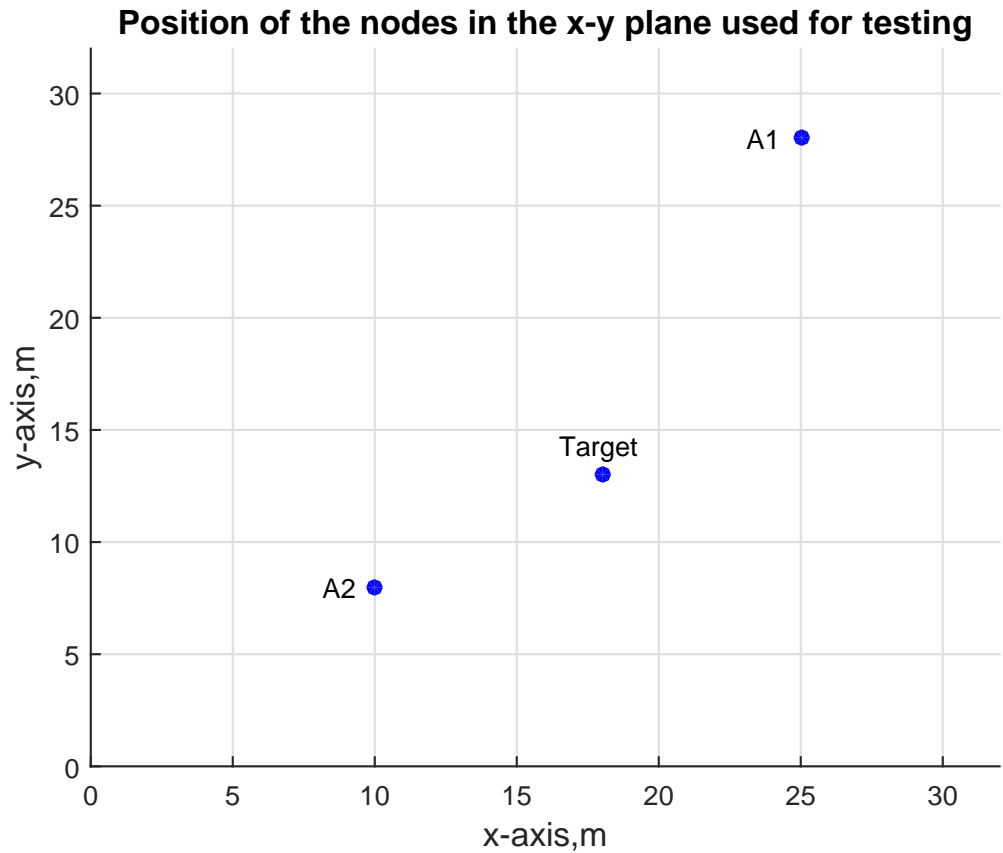}
        \caption[Position of the nodes in the $x$-$y$ plane used for testing the least hybrid ]{Position of the nodes in the $x$-$y$ plane used for testing the least hybrid}
		 \label{Fig:env3_2linehyb}
\end{figure}

\begin{figure} [h]
       \centering
        \includegraphics[width=1\textwidth,clip]{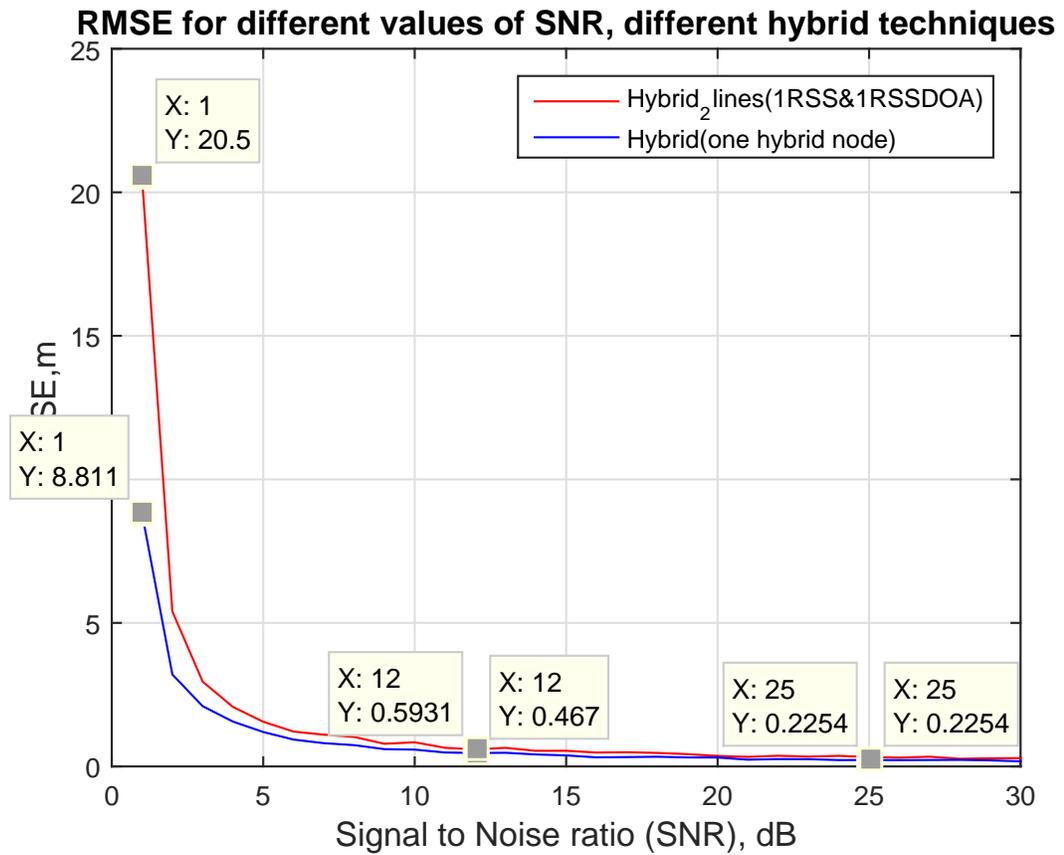}
        \caption[RMSE for different values of SNR using two Hybrid techniques, the last hybrid and hybrid with one hybrid node techniques ]{RMSE for different values of SNR using two Hybrid techniques, the last hybrid and hybrid with one hybrid node techniques}
		 \label{Fig:2hybf&1}
\end{figure}

In this technique, one hybrid and one RSS nodes are used along with the same specification that for the previous environment. The environment is shown in Figure \ref{Fig:env3_2linehyb}. From Figure \ref{Fig:2hybf&1}, the performance of this technique is less than the first proposed hybrid with one hybrid node at low SNR. However, at higher SNR the two techniques produce the same error, thus they have the same performance. The reason is that with the hybrid node, the all measurements from any technique will have approximately the same factors that affect the results. However, with many nodes that have different capabilities, the measurements may have different factor that affect the results and may produce higher errors. 

\begin{figure} [h]
       \centering
        \includegraphics[width=1\textwidth,clip]{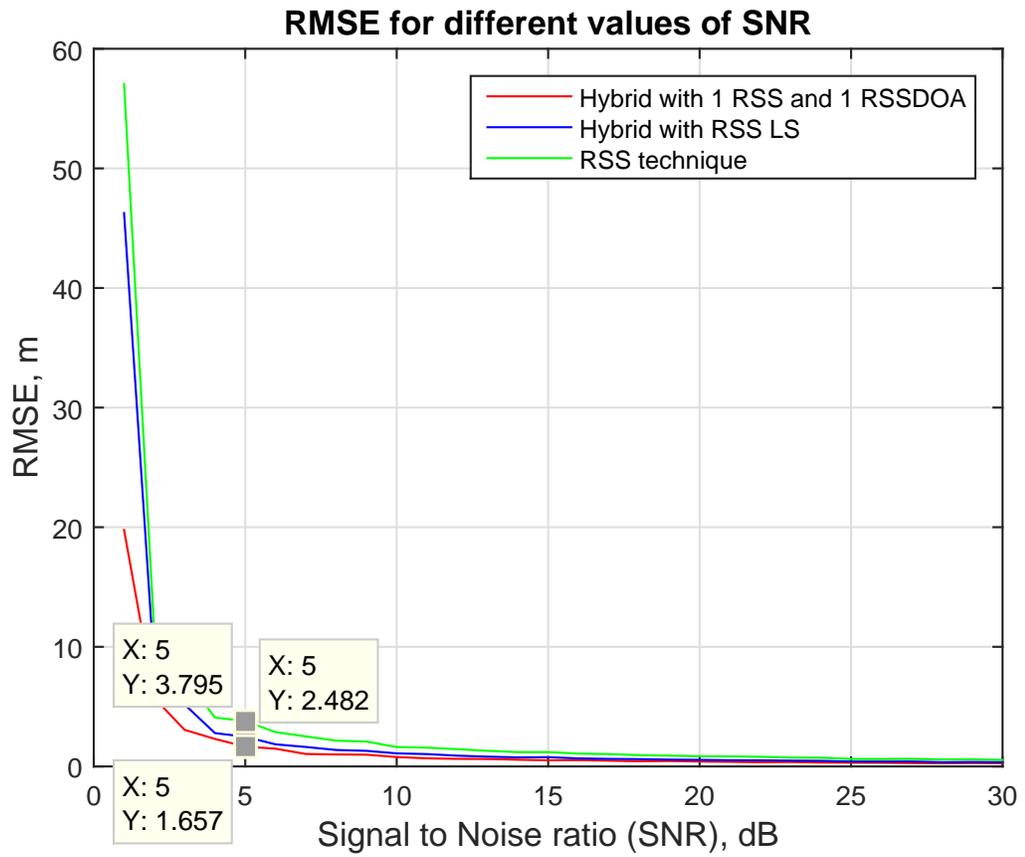}
        \caption[RMSE for different values of SNR using two Hybrid techniques, the last hybrid and hybrid LS and RSS techniques]{RMSE for different values of SNR using two Hybrid techniques, the last hybrid and hybrid LS and RSS techniques}
		 \label{Fig:LHyb_HybL}
\end{figure}

Also, comparison between this method and the hybrid method with LS and RSS technique is shown in Figure \ref{Fig:LHyb_HybL}. This technique outperforms the Hybrid LS and RSS technique alone because it uses only one RSS and one hybrid nodes this will introduce less error compare to other techniques.

\chapter{Conclusion}
\label{Conclusion}

\section{Summary of work done}

In this project, several tasks were achieved. These tasks are summarized as follows:

\begin{itemize}
\item Discussion regarding the methodologies behind localization discovery techniques which includes trilateration, triangulation and multilateration.
\item Investigation of the techniques used in ranging estimation which based on either distances or angles. Some of these techniques include TOA, RSS, Radio Hop Count, and DOA.
\item Detailed study about the classifications of localization algorithms in WSNs which is divided into two main branches; centralized and distributed localization algorithms.
\item Comprehension of the RSS model in theory and simulation using LS, WLS, and Huber robustness.
\item A high apprehension of DOA method of localization with a proper grasp and awareness of the various algorithms used to simulate the difference configuration models along with innovative solutions for problems that face the field such as the problem of correlated environments.
\item Investigation and simulation of different hybrid techniques to ripe the best performance from both techniques.
\end{itemize}

\section{Conclusion}

At the first stage of the project, we carried out a generalized survey about the methods, techniques, and algorithms used to achieve localization. This survey stage was necessary to provide us with the experience and the background in the localization field. To design a high-accuracy localization system, two techniques were selected, RSS and DOA, to be fused into one system called a Hybrid system. 

The RSS model is based on estimating the distance between the unknown node and several reference nodes.  To estimate the location of the unknown node, a minimum of 3 anchor nodes are needed which will form three corresponding circles and each anchor is at the center of its circle. The intersection of these three circles represents the location of the unknown node. To improve the estimation of the unknown node location, estimators such as the LS and WLS are used. The $\ell_2$-norm is investigated through these two estimators. The WLS outperforms the LS as the square error is less compared to the LS. This performance is clear in the case where the anchor nodes are at different distances with respect to the unknown nodes. Also, the $\ell_1$-norm or Huber robustness shows the same results compared to WLS. This is because the environment is Gaussian and the best estimator is WLS.  

The DOA model is based on determining the DOA of incoming signal between the unknown node and a reference node. To be able to locate the unknown node, a minimum of two reference nodes are needed. This will provide us with two bearing lines and the unknown node will lies on the intersection of these two bearing lines. In this model, the incident signal is detected using antenna array instead of single antenna to allow the capability of detecting more than one incident signal impinging at the same time. Two popular antenna array geometries are investigated which are ULA and UCA. Simulation results shows that the performance of UCA exceed ULA as it provides equal power distribution in all direction and resolve ambiguity due to $180^o$ coverage of ULA. 

Different types of signals can be detected, using DOA techniques, such as correlated and uncorrelated signals. In the case of uncorrelated signals, they can be detected directly using subspace algorithms namely MUSIC, Root-MUSIC, and ESPRIT. As the signal becomes correlated, the standard algorithms will fall to detect them. To resolve this problem, pre-processing schemes must be initially employed to remove the correlation between the received signals. The pre-processing schemes used in our projects involves phase mode excitation (PME), Spatial Smoothing (SS) and Toeplitz algorithm.

In the PME, UCA elements is mapped into converted to virtual ULA. With the use of PME, linear operations like Spatial Smoothing can now be employed indirectly with the UCA.  In the Spatial smoothing technique, two main methods can be used which are FSS and FBSS. FSS divides the array into subarrays in the forward direction allowing $N/2$ correlated signal to be detected. FBSS extends the capacity of detected signals to $2N/3$ through dividing the main array into subarrays in both forward and backward direction.  In Toeplitz algorithm, the correlated signals are fully de-correlate which, in terms, permits the full $N$-1 detection by subspace algorithms. Also, this technique surpasses spatial smoothing by offering more robust performance coupled with less computational load. 

Three main DOA subspace algorithms were used in this project to provide high accuracy in detecting the angles of incident signals impinging on an array. These techniques are MUSIC, Root-MUSIC and ESPRIT.  MUSIC algorithm estimate the DOA by employing an exhaustive search through all possible steering vectors that are orthogonal to the noise vectors. However, this methodology made MUSIC algorithm a high computational-load method. In Root-MUSIC, the DOA is estimated via the zeros of a polynomial allowing less computation compared to MUSIC algorithm. The third technique is the ESPRIT where the main subarray is divided into two identical doublets. The main feature of ESPRIT technique is the fact that the corresponding signal eigenvectors of doublets are related by rotational matrix, where DOA is embodied in that matrix. The application Root-MUSIC and ESPRIT techniques are limited to ULA as it possess a vandermonde structure. However, they can be applicable with UCA after converting the UCA into VULA using PME.    
 
The performance of DOA subspace algorithms was evaluated based on different parameters in both ULA and UCA. These parameters includes the number of antenna elements, angular separation between incident singles, SNR, number of samples and signal correlation. In the case of the signal correlation, our simulation results proves the superiority Toeplitz algorithm in case of ULA while FBSS in case of UCA.

Finally, the hybrid model is used to make the best of both techniques.  Different hybrid methods are implemented to achieve the best performance. The first method is to use one hybrid node. This node provides us with the RSS and DOA measurements. DOA measurement is represented by parametric line that crosses implicit circle which represents the RSS measurements.  This technique gives better results than the RSS technique alone. The preceding algorithm addresses the environment of uncorrelated signals. However, to consider the case of correlated signals, spatial smoothing specifically FBSS is used. The second method is to use the two estimators which are the LS and WLS. In this technique, one hybrid node and two RSS nodes are used to estimate the unknown node.  The fusing method is averaging the two points that result from the RSS and DOA measurements. The simulation results for the LS Hybrid outperform the performance the of the RSS technique. The same technique that was used in the LS hybrid is applied in WLS hybrid.  The WLS performance is the best compared to RSS technique and LS hybrid.  The Final technique is to use one hybrid and one RSS nodes. In which the two measurements are presented using two lines and their intersection is the unknown node. This technique shows a better performance than the LS hybrid and the RSS techniques. However, the hybrid technique that uses one hybrid node gives better results than the hybrid with two lines technique. 

\section{Critical Appraisal}

In this project many techniques are investigated and different algorithms are simulated and tested using MATLAB. Also, wide ranges of topics are discussed. The entire survey that was carried out at the initial stage of the project was sufficient to choose the two techniques that we worked on. Also, we investigated the advantages and disadvantages for many techniques and we chose the best two techniques to be able to develop Hybrid technique. Also, we investigated different estimators to improve the RSS accuracy.  We investigated the idea of the correlated signals that to be detected using RSS technique and we concluded that it will be advanced for senior design project.  Also, in the DOA, we studied different techniques to address the two types of signal which are correlated and uncorrelated signals. Furthermore, we come up with different methods for the fusing the RSS and DOA techniques to form the hybrid technique. To implement this project, we even read in various fields that are outside localization and WSN field. 

However, sometimes it was difficult to carry out the Matlab simulation because our laptops cannot handle the high computation complexity. Also, sometimes the database that is provided by the library was down and could not finish the required research. In addition, there are some resources that are not available in the university library or online which made the researches slow a little bit.

If time allows us, we would like to implement RSS and DOA techniques using hardware. This is because the real implementation will make the reader to appreciate the benefits of different simulators and algorithm that was used in this project. 

\section{Recommendations}
There are two potential topics that we come up with during our investigation in the SDP project. Due to the limited period of SDP project, we would like to leave these topics as a future tasks.  The first topic is about applying Toeplitz with UCA after it is converted into VULA. This procedure will greatly improve the performance of DOA algorithms with UCA in coherent environment. The second topic is changing the channel model and applying the $\ell_1$-norm to achieve the optimum performance in a more realistic environment.


\renewcommand{\chaptermark}[1]{%
\markboth{\appendixname\ \thechapter.\ #1}{}}
\appendix

\chapter{}
\label{Maximum Likelihood}

The least square is one technique from the well-known estimator the Maximum Likelihood techniques under the condition of Gaussian distribution, with mean ($\mu$) equals to zero and a variance ($\sigma^2$) \cite{duda_pattern_2004}. Hence, assume the equation for the straight line is:

\begin{equation}%
y_i=\alpha x_i + \epsilon_i
\label{B.1}
\end{equation}%

where $\alpha$ is the slope of the line and $\epsilon_i$ is the square error.

The Gaussian probability density function (pdf) is given by the following function: 

\begin{equation}%
f(x)=\dfrac{1}{\sqrt{2\pi\sigma^2}}e^{-\frac{(x-\mu)^2}{2\sigma^2}}
\label{B.2}
\end{equation}%
where $x$ is random variable 

Suppose that $x_i$  is fixed and $y_i$  is a random variable and $\epsilon$ is independent for $\alpha$, $m$ =0 and
$\epsilon_i$=$y_i$-$\alpha x_i$. Hence, 

\begin{equation}%
f(x~\vdots ~\alpha,\sigma^2)=\frac{1}{\sqrt{2\pi\sigma^2}}e^{-\frac{(y_i-\alpha x_i)^2}{2\sigma^2}}
\label{B.3}
\end{equation}%

\begin{equation}%
L=\prod\limits^{N}_{i=1}\dfrac{1}{\sqrt{2\pi\sigma^2}}e^{-\frac{(y_i-\alpha x_i)^2}{2\sigma^2}}=(\dfrac{1}{\sqrt{2\pi\sigma^2}})^N\prod\limits^{N}_{i=1}e^{-\dfrac{(y_i-\alpha x_i)^2}{2\sigma^2}}
\label{B.4}
\end{equation}%

Take the normal log for both sides:

\begin{equation}%
l=Nln(\dfrac{1}{\sqrt{2\pi\sigma^2}})-\dfrac{1}{2\sigma^2}\sum\limits^{N}_{i=1}(y_i-\alpha x_i)^2
\label{B.5}
\end{equation}%

Then differentiate and set the term to zero:

\begin{equation}%
\dfrac{dl}{d\alpha}=\dfrac{1}{\sigma^2}\sum\limits^{N}_{i=1}x_i(y_i-\alpha x_i)=0
\label{B.6}
\end{equation}%

Hence, the estimator $\widehat{\alpha}$  is:

\begin{equation}%
\widehat{\alpha_{ML}}=\dfrac{\sum\limits^{N}_{i=1}x_iy_i}{\sum\limits^{N}_{i=1}x^2_i}
\label{B.7}
\end{equation}%
 
\chapter{}
\label{LOP}

The Line of position (LOP) between $p_1$ and $p_2$

Starting from the formula of the distance between two points as shown below:

\begin{equation}%
D_i=\left\Vert p_i-p_s\right\Vert=\sqrt{(x_i-x_s)^2+(y_i-y_j)^2}
\label{C.1}
\end{equation}%

where $i$=1,2, Now, by squaring and taking the difference between $D_2^2$ and $D_1^2$:

\begin{equation}%
D_1=\left\Vert p_1-p_s\right\Vert=\sqrt{(x_1-x_s)^2+(y_1-y_j)^2}
\label{C.2}
\end{equation}%

\begin{equation}%
D_2=\left\Vert p_2-p_s\right\Vert=\sqrt{(x_2-x_s)^2+(y_2-y_j)^2}
\label{C.3}
\end{equation}%

\begin{equation}%
D^2_2-D^2_1=(x_2-x_s)^2+(y_2-y_s)^2-(x_1-x_s)^2-(y_1-y_s)^2
\label{C.4}
\end{equation}%

\begin{equation}%
(x_2-x_1)x_s+(y_2-y_1)y_s=\dfrac{1}{2}(x^2_2+y^2_2)-\dfrac{1}{2}(x^2_1+y^2_1)+\dfrac{1}{2}(D^2_1-D^2_2)
\label{C.5}
\end{equation}%

It can be simplified using the formula $\left\Vert p_i\right\Vert^2=x^2_i+y^2_i$

\begin{equation}%
(x_2-x_1)x_s+(y_2-y_1)y_s=\dfrac{1}{2}(\left\Vert p_2\right\Vert^2-\left\Vert p_1\right\Vert^2+D^2_1-D^2_2)
\label{C.6}
\end{equation}%

\chapter{}
\label{WLS}

The weighting matrix \textbf{W} is the inverse of the covariance matrix S of vector \textbf{b}. Assuming that the measurements of the distances are indepdent and $x_i$ and $y_i$ are constants, the matrix \textbf{S} can be easily calculated:

\begin{equation}%
\boldsymbol{\mathrm{S}}=E\{\boldsymbol{\mathrm{bb^T\}}}=\left[\begin{array}{cccc}Var(d^2_1)+Var(d^2_2)&Var(d^2_1)&\cdots&Var(d^2_1)\\Var(d^2_1)&Var(d^2_1)+Var(d^2_3)&\cdots&Var(d^2_1)\\\vdots&\vdots&\ddots&\vdots\\Var(d^2_1)&Var(d^2_1)&\ldots&Var(d^2_1)+Var(d^2_N)\end{array}\right]
\label{E.1}
\end{equation}%

Assuming that the channel is lognormal, it can be derived from equation \ref{3.4} that the estimated distance is a random variable defined by:

\begin{equation}%
\widetilde{d_i}=d_i 10^\frac{N(0,\sigma)}{10\eta}=10^{N(\log_{10}(d_i),\frac{\sigma}{10\eta})}=e^{N(\log_{10}(d_i),\frac{\sigma}{10\eta})\ln(10)}=e^{N(\ln(d_i),\frac{\sigma \ln(10)}{10\eta})}
\label{E.2}
\end{equation}%

That is $\widetilde{d_i}$ is a lognormal random variable with parameter \cite{devore_probability_2011}

\begin{equation}%
\mu_d=\ln(d_i);\sigma_d=\dfrac{\sigma \ln(10)}{10\eta}
\label{E.3}
\end{equation}%

The variance is calculated and subsituted into the covariance matrix \cite{tarrio_weighted_2011}

\begin{equation}%
Var(d^2_i)=e^{4\mu_d}(e^{8\sigma_d^2}-e^{4\sigma_d^2})
\label{E.4}
\end{equation}%

\backmatter
\renewcommand{\chaptermark}[1]{%
\markboth{#1}{}}

\addcontentsline{toc}{chapter}{References}
\begin{onehalfspace}
\bibliography{main1}
\end{onehalfspace}

\end{document}